\documentclass[superscriptaddress,nofootinbib,11pt,preprintnumbers]{revtex4-2}
\usepackage{packages}

\begin{document}

\pagenumbering{gobble}

\preprint{CALT-TH-2023-001}
\preprint{DESY-23-054}

\title{Energetic long-lived particles in the CMS muon chambers}
\begin{abstract}
{We present a recast in different benchmark models of the recent CMS search that uses the endcap muon detector system to identify displaced showers produced by decays of long-lived particles (LLPs). The exceptional shielding provided by the steel between the stations of the muon system drastically reduces the Standard Model background that limits other existing ATLAS and CMS searches. At the same time, by using the muon system as a sampling calorimeter, the search is sensitive to LLPs energies rather than masses. We show that, thanks to these characteristics, this new search approach is sensitive to LLPs masses even lighter than a GeV, and can be complementary to proposed and existing dedicated LLP experiments.}
\end{abstract}

\author{Andrea Mitridate}
\affiliation{Walter Burke Institute for Theoretical Physics, California Institute of Technology, Pasadena, CA 91125, USA}
\affiliation{Deutsches Elektronen-Synchrotron DESY, Notkestr. 85, 22607 Hamburg, Germany}
\author{Michele Papucci}
\affiliation{Walter Burke Institute for Theoretical Physics, California Institute of Technology, Pasadena, CA 91125, USA}
\author{Christina W. Wang}
\affiliation{California Institute of Technology, Pasadena, CA 91125, USA}
\author{Cristi\'an Pe\~na}
\affiliation{Fermi National Accelerator Laboratory, Batavia, IL 60510, USA}
\author{Si Xie}
\affiliation{California Institute of Technology, Pasadena, CA 91125, USA}
\affiliation{Fermi National Accelerator Laboratory, Batavia, IL 60510, USA}

\maketitle
\newpage

\renewcommand*{\thefootnote}{\arabic{footnote}}

\pagenumbering{arabic}

\tableofcontents

\newpage

\section{Introduction}\label{sec:intro}
Despite its successes, the Standard Model (SM) of particle physics has several shortcomings. Most notably, it fails to explain the nature of Dark Matter, the baryon asymmetry in the Universe, and the origin of neutrino masses. Because of this, several extensions of the SM have been proposed, a common byproduct of which are long-lived particles (LLPs). This is especially true for \emph{dark sectors} comprised of new neutral particles with masses below the electroweak scale that interact with the SM through suppressed renormalizable couplings, heavy mediators, or both. In recent years, several searches for neutral LLPs have been conducted by ATLAS, CMS, LHCb, and other dedicated LLP experiments and a vibrant search program is being developed~\cite{ATLAS:2017tny, LHCb:2017xxn, Lee:2018pag, ATLAS:2018rjc, Alimena:2019zri,Beacham:2019nyx, CMS:2019zxa, CMS:2019qjk, CMS:2020atg, CMS:2020iwv, ATLAS:2020xyo, ATLAS:2020wjh, Agrawal:2021dbo, Alimena:2021mdu,Borsato:2021aum, CMS:2021kdm, CMS:2021sch, CMS:2021tkn, LHCb:2021dyu, LHCb:2020akw, Knapen:2022afb, ATLAS:2022vhr, ATLAS:2022zhj, ATLAS:2022gbw, CMS:2022fut, CMS:2022qej, ATLAS:2023oti}. However, most ATLAS and CMS searches are known to face challenges for LLPs with masses at or below a GeV. While some room for improvement may be present in existing analyses~\cite{Gershtein:2017tsv,Gershtein:2019dhy,Alimena:2021mdu,Bernreuther:2020xus, Alimena:2021mdu, Bhattacherjee:2020nno, Bhattacherjee:2021qaa,Bhattacherjee:2021rml}, looking for tracks from displaced vertices will ultimately be limited by irreducible SM backgrounds. This is especially true in the long lifetime regime, where the few tracks produced by the decays of light LLPs are not enough to discriminate them against SM LLP decays such as those of $K_L$.

However, recently the CMS collaboration published a search~\cite{CMS:2021juv} for neutral LLPs which uses the CMS endcap muon detectors (EMDs) as a sampling calorimeter. Thanks to the unique design of the CMS EMDs (made of stations of cathode strip chambers (CSCs) interleaved with steel return-yoke), LLPs decay products can induce hadronic and electromagnetic showers that give rise to high hit multiplicity in localized detector regions. Because of this, the LLPs signature tracks the LLP energy instead of its mass. This, together with the exceptional shielding provided by the CMS calorimeters and the steel in the front layers of the EMD, allows this search to be sensitive to LLPs with masses smaller than a GeV. Despite this search currently lacking a dedicated trigger to maximize its potential, its reach is  competitive with many proposed dedicated LLP experiments~\cite{Chou:2016lxi,Curtin:2018mvb,MATHUSLA:2018bqv,Feng:2017uoz,FASER:2018eoc,Gligorov:2017nwh,SHiP:2015vad,Bauer:2019vqk,Feng:2022inv,Gligorov:2018vkc,Aielli:2019ivi,Dreyer:2021aqd,Haas:2014dda,Cerci:2021nlb,SHiP:2020sos,Boyarsky:2021moj,Acharya:2022nik} due to its large geometric acceptance.

The original CMS paper presented the results on a benchmark model motivated by the twin Higgs scenario, where the SM Higgs boson decays to a pair of neutral long-lived scalars, each of which decays in turn to a pair of bottom quarks, down quarks, or $\tau$ leptons. Masses of the scalar LLPs were probed down to 7~GeV. In this paper, we use the parameterized reconstruction and selection efficiency functions provided in the HEPData entry~\cite{hepdata.104408.v2} of the CMS paper to recast the analysis in different benchmark models and explore its strengths and weaknesses. The goal of this recast is to inform future iterations of the CMS analysis and inform the choice of benchmarks for other proposed LLP experiments away from already covered regions of the parameter space. Although we expect the barrel muon detector to also produce the shower signature seen in the CSCs, in this study, we will not consider the barrel region. The barrel muon detector is made of drift tubes, which are a different detector technology and system from the CSCs. Therefore, the response to LLPs will be different, and we will wait for CMS to release efficiency functions for the barrel muon detector for future studies.

The paper is structured as follows. In Section~\ref{sec:analysis}, we briefly review the CMS analysis, we outline the generation and simulation framework, as well as the validation of the framework against the CMS result. In Section~\ref{sec:benchmarks}, we discuss all the benchmark models that will be considered. Finally, in Section ~\ref{sec:results}, we discuss the results of this paper. 

\section{Analysis and recast strategy}\label{sec:analysis}
In this section, we describe the details of the analysis. We start by summarizing the CMS analysis~\cite{CMS:2021juv} and then we discuss and validate the recast procedure, including the event generation and detector simulation framework, signal selections, signal and background yield estimate, and statistical analysis to evaluate the upper limits.

\subsection{CMS endcap muon system analysis}\label{subsec:cms_summary}
The missing transverse momentum, $\ptmiss$, is calculated by CMS by using only tracker and calorimeter information. Therefore, LLPs decaying beyond the calorimeters can produce large $\ptmiss$ when they are produced in association with other prompt, visible objects or recoiled against an initial state radiation (ISR) jet. Due to this feature and the lack of a dedicated trigger, events are triggered by requiring $\ptmiss>120$~GeV. A further requirement of $\ptmiss>200$~GeV is then applied offline. To further suppress the background from $W$ and top quark production, the search also requires at least one jet from ISR with transverse momentum $\pt >50$~GeV and pseudorapidity $|\eta|< 2.4$, and no leptons in the event. 

 By clustering CSC hits using the DBSCAN algorithm~\cite{dbscan}, the search then identifies displaced showers produced by LLPs decaying in the endcap muon system. To suppress background from punch-through jets and muon bremsstrahlung, clusters that are geometrically matched ({\it i.e.} within a cone of radius $\Delta R = \sqrt{(\Delta\phi)^2+(\Delta\eta)^2}<0.4$) to jets (muons) with $p_T>10\,(20)$ GeV are rejected. Furthermore, a number of active vetoes are applied to veto clusters with hits or track segments in the muon detector stations with the least amount of shielding. Finally, to suppress the muon bremsstrahlung background originating from muons beyond the detector acceptance, clusters with $|\eta|>2.0$ are vetoed. Clusters are required to be consistent with an in-time interaction by restricting the cluster time ($ -5.0< t_{cluster} <12.5$~ns). To reject clusters composed of hits from multiple bunch crossings, the root mean square spread of a cluster’s hit times is required to be less than $20$~ns. 
 
 A cut-based ID that distinguishes signal from background clusters is defined by using several features, including the cluster $\eta$ position and the number of stations occupied by the cluster. The details required to simulate the cut-based ID efficiency have been provided in the HEPData entry of the CMS paper~\cite{hepdata.104408.v2}. 

Finally, the number of hits in the cluster, $\nhits$, and the azimuthal angle between $\ptmiss$ and the cluster location, $\Delta\phi$, are used to make the final discrimination between signal and background. The signal is required to have large $\nhits >130$ and small $\Delta\phi < 0.75$. For the chosen signal the bulk of the $\ptmiss$ is produced by the LLP when they are produced in association with other prompt, visible objects or recoiled against an ISR jet, while for the backgrounds, $\Delta\phi$ is independent of $\nhits$. The independence of the two variables for the background enables the use of the matrix (ABCD) method to predict the background yield. Using the matrix method and assuming no signal contribution, 2.0 $\pm$ 1.0 background events were predicted in the signal-enriched region, and 3 events were observed. No excess of events above the SM background was observed.

\subsection{Event generation}

We generated signal events using \madgraph~v2.9.3~\cite{Alwall:2014hca}, and performed the parton shower and hadronization with \pythia~v8.244~\cite{Sjostrand:2014zea}, while keeping the LLP stable. 
Samples with different jet multiplicities were merged according to the MLM algorithm~\cite{Mangano:2002ea,Alwall:2008qv}. 
Generator-level cuts were applied to the events in order to increase the statistics in the phase space regions selected by $\ptmiss$ cut in the CMS analysis. 
Additional details on the samples, including the specific generator-level cuts, are given in Sec.~\ref{sec:benchmarks} for each of the benchmark models considered in this work. 

To efficiently decay the LLP, we used the fact that the reconstruction efficiency parameterization provided by the CMS search is spatially binned in a small number of regions with simple shapes ( which are defined by the intersection of ranges in the radial direction, $r$, longitudinal direction, $z$, and pseudorapidity, $\eta$), for which the probability of decaying inside a region can be computed analytically.
For a region determined by the conditions
\begin{equation}
    \eta_0 \leq \eta \leq \eta_1, \qquad r_0 \leq r \leq r_1, \qquad z_0 \leq z \leq z_1,
\end{equation}
the probability, $P$, to decay inside the region for a particle traveling with momentum $p^\mu$ and proper decay length $c\tau$ is
\begin{equation}
    P = e^{-y_{min}} - e^{-y_{max}}
\end{equation}
where we have defined 
\begin{align}
     y_{min} &=\frac{1}{\beta\gamma c\tau}{\rm max}\Big(\min{\left(z_0 \coth{\eta}, z_1 \coth{\eta}\right)}, r_0 \cosh{\eta}\Big), \\
     y_{max} &=\frac{1}{\beta\gamma c\tau}{\rm min}\Big(\max{\left(z_0 \coth{\eta}, z_1 \coth{\eta}\right)}, r_1 \cosh{\eta}, \beta\gamma c\, t_{cut}\left(\beta\gamma+\sqrt{1+(\beta\gamma)^2}\right)\Big),
\end{align}
with $\eta$ being the pseudorapidity of the particle, and $\beta\gamma = |\vec p|/m$. We have further introduced a timing requirement such that $t_{decay} - d_{decay}/c < t_{cut}$ where $t_{decay}$ and $d_{decay}$ are the decay time and distance from the origin respectively.

The LLPs generated in the events were made to decay at fixed positions within each region using \pythia, as the presence of the decay vertex in a given region is the only geometrical information used by the detector simulation. For a given proper decay length $c\tau$, the probability for the LLP to decay in a given region was assigned to the event in the form of an event weight. The decay probability for each LLP in an event is independent, so the event-level probability is the product of the decay probability of the LLPs in an event. Therefore, for each input event with an undecayed LLP, the decay program generated as many decayed events as the number of regions intersected by the LLP trajectory. We used the multi-weight capabilities of the HepMC event format to perform a scan in $c\tau$ without having to reprocess the events. In case there were multiple LLPs in the same event (as in the case of Higgs decays to pairs of dark vectors or scalars), we also included decays outside the signal regions. In the case of LLP decays in the inner detector, where precise knowledge of the decay vertex position is used by the detector simulation, we generated the decay vertex position according to the decay probability distribution instead of keeping it fixed.

The events were subsequently passed to a simplified detector simulation based on \textsc{Delphes} v3.4.2~\cite{deFavereau:2013fsa} using the publicly available CMS configuration card for the reconstruction of prompt objects supplemented by a dedicated module, discussed in subsection~\ref{delphesModule}, simulating the LLP decay reconstruction and selection using the information provided by the CMS search.

\subsection{Detector simulation with dedicated \delphes modules}
\label{delphesModule}
We based the fast simulation of the detector response to standard particle flow (PF) candidates on the CMS detector response provided by the CMS configuration card in the parametric \delphes framework. The detector simulation of the hit clusters in the CSC endcap muon detector was based on a dedicated \delphes module and class for the CSC cluster objects that we developed~\cite{delphes_pr} based on the parameterized detector response functions provided in the HEPData entry~\cite{hepdata.104408.v2}. Based on the recasting instructions provided in the HEPData entry, the simulation of cluster-level selection efficiencies was divided into three components. 

The first component is cluster efficiency; which includes the cluster reconstruction efficiency, muon veto, active veto, time spread, and $\nhits$ cut efficiency. This cluster efficiency is provided as a function of the LLP electromagnetic and hadronic energy in two separate LLP decay regions in the CSC detector. Building upon the existing \code{Efficiency} module that is already used by all other PF candidates, we implemented a dedicated \code{CscClusterEfficiency} module in \delphes which encodes this parameterized function.

The second component is cluster identification efficiency. We implemented the \code{CscClusterID} module following the code function provided by the CMS HEPData entry.

The third and last component includes the model-dependent cluster time requirement, jet veto, and the $\Delta\phi$ requirement. The values of the three variables affected by these cuts were calculated using generator-level information and saved as part of the \code{CscCluster} class. Specifically, the cluster time was determined by calculating the LLP travel time from the production to the decay vertex in the lab frame. The jet veto was implemented by requiring no PF jets with $p_{T}>10$~GeV within a cone of $\Delta=0.4$ around the LLP. Finally, $\Delta\phi$ was computed as the azimuthal angle difference between the LLP momentum and missing transverse momentum (MET) using the result from the standard \delphes simulation modules. All these requirements were then made at a later stage of the analysis workflow. 

Finally, we modified the standard CMS configuration card from \delphes to include the \code{CSCClusterEfficiency} and \code{CSCClusterID} module in the processing sequence. The modules require only generator-level LLP information and can be used for the recasting of the result for any other model. The implementation can be found in~\cite{delphes_pr}.

\subsection{Analysis strategy}\label{subsec:strategy}
In this section, we discuss the procedure used to recast the CMS Run 2 results and to make projections for high-luminosity LHC (HL-LHC). 

For the recasting of the Run 2 results, we used the exact same selection and cuts used in the CMS paper and summarized in Sec.~\ref{subsec:cms_summary}. For all the standard particle flow objects, we used the default CMS configuration card from \delphes, which has been validated~\cite{deFavereau:2013fsa} to reproduce the object resolutions from Run 2. The \ptmiss calculation implemented in \delphes is accurate when the LLP decays outside of the calorimeters that it's treated as invisible and when the LLP decays sufficiently close to the interaction point that the energy of the decay particles are measured by the calorimeters. In models where there is only one LLP in the event the LLP is required to decay in the muon detector, so the LLPs are treated consistently as invisible in both \delphes and CMS simulations. For models with two or more LLPs per event, in the parameter space explored in this reinterpretation study, this approximation for the \ptmiss calculation leads to a systematic error of 20\% or less on the selection efficiency. For the CSC cluster objects, we ran the \code{CSCClusterEfficiency} and \code{CscClusterID} modules to select clusters that would pass the corresponding selections. We then applied the $\Delta\phi<0.75$ selection, jet veto, and time cut for CSC clusters. We used the number of signal events passing these signal selections as our estimate of the signal yield. Finally, by using this estimate for the signal yield together with the background yield ($2\pm 1$ background event) and observed data (3 observed events) obtained in the CMS analysis, we derived our constraints.

To further inform experimental studies and compare to other proposed LLP experiments, we projected the sensitivity of this analysis to Phase 2 conditions. To simulate the effect on signal yield from the increased number of pileup interactions during Phase 2, we modified the mean pileup number in the CMS configuration card from 32 to 200. We observed that, due to the larger number of jets from pileup interactions, the probability that a CSC cluster in the signal region is accidentally matched to and vetoed by a pileup jet with $p_{T}> 10$~GeV is 20\% higher. Signal region CSC clusters are concentrated in the region with $|\eta|<$ 1.6, while most pileup jets are concentrated in the high $\eta$ region, so the increase in the probability of accidental matching is only 20\%. Furthermore, the resolution of \ptmiss is degraded by the large number of pileup jets, such that the MET cut efficiency increased by a factor of 2 due to fake MET. However, a more realistic assumption would be that with the help of the new MIP Timing Layer (MTD) and pileup removal algorithms, the MET resolution will be kept at the same level. Similarly, we assumed that with the help of the MTD and new reconstruction algorithms, the efficiency and resolution would be kept at the same level for all PF candidates. Therefore, a simple projection for Phase 2 constraints can be derived by scaling the signal (and background) yield by the increased integrated luminosity, and applying an $80\%$ correction to the signal yield per cluster due to larger number of pileup jets while assuming the same efficiency and resolution for all PF candidates.

However, this simple recasting strategy significantly underestimates the potential sensitivity of a Phase 2 analysis. Realistically, given the larger dataset, we would apply tighter cuts to achieve near zero background. Therefore, we considered a second recasting strategy where we apply a tighter $\nhits$ selection, the main discriminator of the analysis, until the expected background reaches zero. To estimate the signal and background yield with a tighter $\nhits$ cut, we used the $\nhits$ distributions in the auxiliary materials from the CMS analysis. We fitted the $\nhits$ distribution for the background with an exponential function to extrapolate the background yield at higher $\nhits$ cuts. We found that a $\nhits$ of 200 would suppress the background yield to below 1 for the expected Phase 2 integrated luminosity. Similarly, we found that increasing the cut from 130 to 210 would give a signal efficiency of about $80\%$. Therefore, for this recast strategy, we scaled the signal yield by an additional $0.8$ with respect to the simple recasting strategy previously described and assumed a background yield of $0.2$. The majority of the background comes from low $p_{T}$ kaons and pions from the recoil, pileup, or underlying events that produce the clusters. The rate of low $p_{T}$ particles from pileup would increase linearly with the number of pileup, but the background rate from the recoil in the main collision wouldn't change with higher pileup. Therefore, we produce a conservative estimate of the effect of higher pileup by increasing the normalization of the $\nhits$ distribution by the increase in number of pileup and recomputing the $\nhits$ threshold. The estimated effect on the signal yield is at most 20\%.

Finally, we considered a search strategy that would be enabled by a new dedicated Level-1 and High Level Trigger targeting this signature starting from the beginning of Run 3~\cite{CMS-DP-2022-062}. The trigger selects for events with at least one CSC cluster with a large number of hits and is already in operation and validated. In this strategy, we remove the high $\ptmiss$ selection (which is necessary during Run 2 to trigger the events) and require at least two CSC clusters. In addition, we remove the requirement of at least one jet with $p_{T}>50$~GeV and $\Delta\phi(\textrm{cluster, MET})<0.75$ that has high signal efficiency only in the high MET phase space. For this strategy, due to the double cluster requirement, we assumed that zero background can be achieved.

For all the analyses discussed here, we assigned $20\%$ signal systematic uncertainty which is of the same order of signal systematic for the CMS result. There it is dominated by missing higher order QCD corrections, which have a size of 21\% for the gluon fusion production mode. The background uncertainty is dominated by statistical uncertainty, and we assign no additional background systematic uncertainty.

The result of our recast, for both Run 2 and the projection for Phase 2, are shown for all the benchmark models in Section~\ref{sec:results}. Unless differently stated, the first recasting strategy for Phase 2 is indicated with solid lines, and the second (third) scenario with dot-dashed (dashed) lines.
\begin{figure}[t!]
    \centering
    \includegraphics[width=\textwidth]{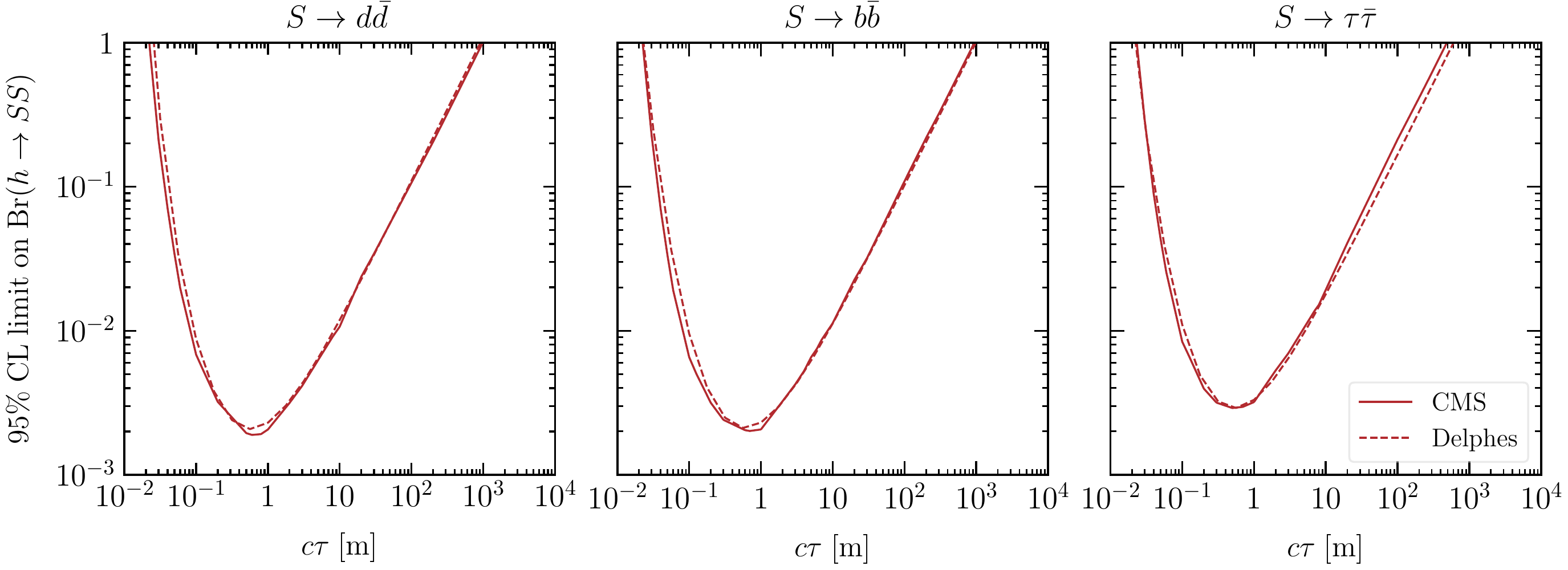}
    \caption{\small{Comparison of the 95\% CL upper limits on the branching fraction ${\rm Br}(h\to SS)$ as functions of $c\tau$ derived with the standalone workflow (dashed lines) and the CMS search (solid lines). In deriving these limits we have considered a $15\;\si{GeV}$ LLP decaying into $d$-quark pairs (left), $b$-quark pairs (center), and $\tau$ pairs (right). The limits from this work are shown to agree with the CMS search to within $30\%$.}}
    \label{fig:limit_validation}
\end{figure}

\subsection{Limit calculation and validation}
\begin{figure}[t!]
    \centering
    \includegraphics[width=0.8\textwidth]{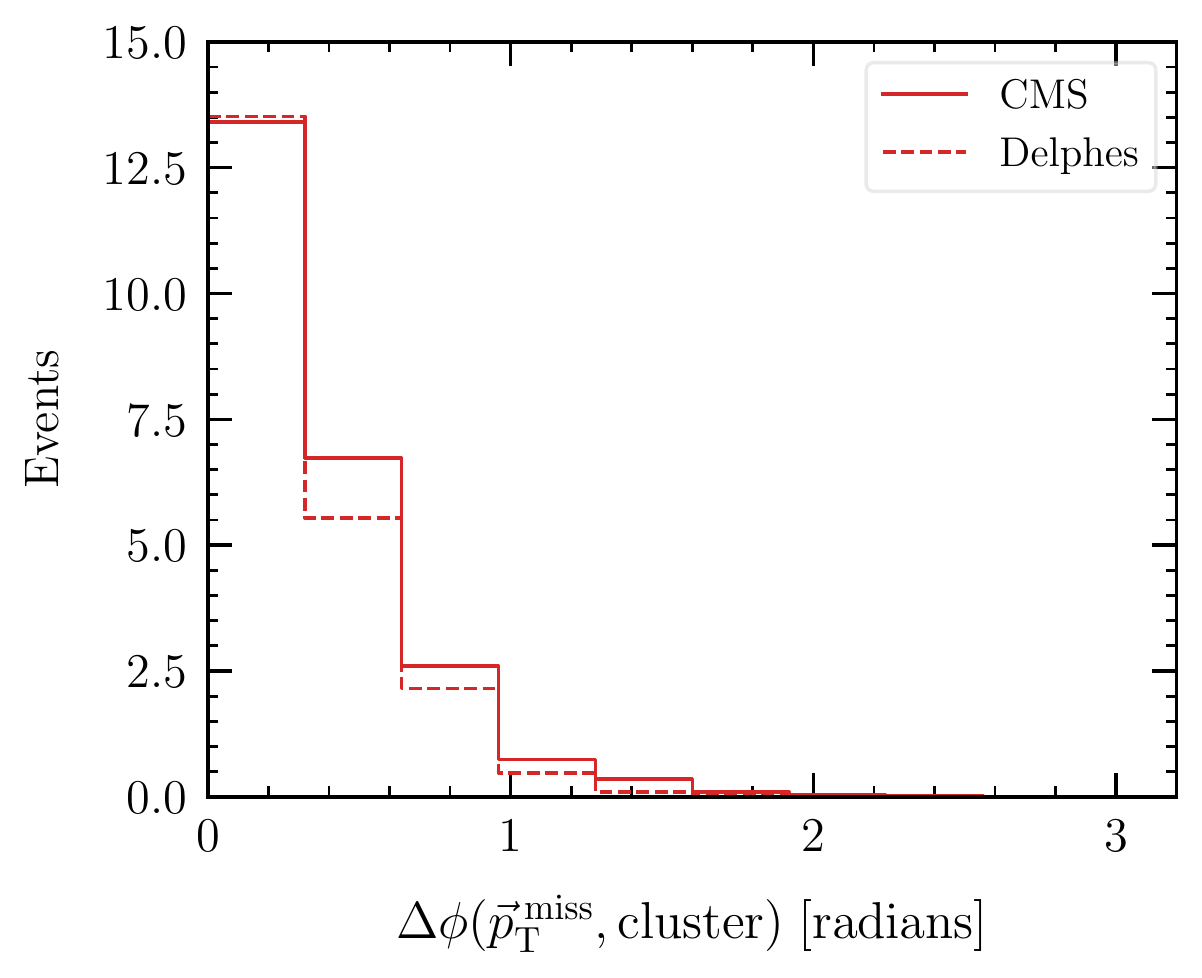}
    \caption{\small{Comparison of the $\Delta\phi(\textrm{cluster, MET})$ distributions derived with the standalone workflow (dashed line) and the CMS search (solid line) for a $15\;\si{GeV}$ and 1 m proper decay length LLP decaying into $d$-quark pairs with ${\rm Br}(h\to SS) = 0.01$ . The signal yield when requiring $\Delta\phi(\textrm{cluster, MET})<$1 agrees within 7\%. 
}}
    \label{fig:deltaphi_validation}
\end{figure}

Before moving to the discussion of the benchmark models, we present a validation of our recast analysis. Specifically, we derived the 95\% confidence level (CL) limits on the branching fraction ${\rm Br}(h\to SS)$ for different scenarios and compared them to the one derived in the CMS analysis. The observed $95\%$ CL upper limits on the branching fraction ${\rm Br}(h\to SS)$ for $15\;\si{GeV}$ LLP as functions of $c\tau$ for the $S\to d\bar d$, $S\to b\bar b$, and $S\to\tau\bar\tau$ decay modes were compared against the CMS results, as shown in Fig.~\ref{fig:limit_validation}. The limits evaluated using the fast simulation from \delphes agree with the CMS result to within $30\%$ for all lifetimes evaluated. Additionally, we also validate the $\Delta\phi(\textrm{cluster, MET})$ distributions for $15\;\si{GeV}$ and 1~m proper decay length LLP decaying to $d\bar d$ against the auxiliary materials provided by CMS, as shown in Fig.~\ref{fig:deltaphi_validation}. The signal yield when requiring $\Delta\phi(\textrm{cluster, MET})<$1 agrees within 7\%. 

\section{Benchmark models}\label{sec:benchmarks}
In this section, we briefly describe the benchmark models considered in this work.
Each of these models has been chosen to showcase the strengths and limitations of the current analysis in concrete examples exhibiting different kinematics and signal topologies. 
Specifically, we want to investigate what happens to the analysis reach when lower values of LLP masses are chosen, when the LLP energy, $E_{LLP}$, is reduced, or when the LLP momentum is not correlated in magnitude or direction with the missing transverse energy.
We also want to investigate what happens when there are multiple LLPs produced roughly in the same direction, potentially leading to failed isolation cuts in a non-trivial way.

Concretely, the models we consider are:
\begin{itemize}
    \item {\bf Exotic Higgs decays} into dark photons or light scalars. These are the closest models to the one considered in the original CMS analysis and are characterized by a production rate decoupled from the exclusive decay channels and LLP lifetime.
    Besides being commonly chosen benchmarks to compare the performance of different experiments in LLP searches, these benchmarks will allow us to probe the reach for LLP masses lighter than those presented in the CMS analysis, for a fixed production rate and using more realistic decay modes.
    \item {\bf Axion-like particles (ALPs)} coupled to SM gauge bosons. In this model, the coupling to the SM is provided by a dimension five operator. A single parameter (the ALP decay constant) controls the production rate and lifetime. These models are characterized by a production cross section which is enhanced for energetic LLPs, irrespective of the light LLP mass.
    \item {\bf Inelastic Dark Matter (DM)}. In these models, the LLP is provided by an almost degenerate partner of the DM, and the amount of energy carried out by the LLP is controlled by the DM-LLP mass splitting and decoupled from the missing transverse energy (MET). This allows us to probe the reach in the low $E_{LLP}$ region while allowing the other selection requirements to be passed without much of a penalty.
    \item {\bf Confining Hidden Valley models} where jets of LLPs are produced in perturbative hidden showers, analogously to the case of QCD. This benchmark allows us to study the impact of the jet veto in models where multiple LLPs are produced in the same detector region.
\end{itemize}

In addition to the models considered in this paper, the CMS analysis has also been shown to have non-trivial reach for Heavy Neutral Leptons (HNL)~\cite{Cottin:2022nwp}.

\subsection{Light scalar singlet}\label{subsec:scalar}
The most minimal extension of the SM is provided by adding a real scalar singlet ($S$) that mixes with the SM Higgs through renormalizable operators. The Lagrangian for this  model reads~\cite{OConnell:2006rsp}:
\begin{equation}
        \mathcal{L}_{SH}=\mathcal{L}_{\rm SM}+\mathcal{L}_{\rm DS}-\left(A_{HS}\,\hat{S}+\lambda_{HS}\,\hat{S}^2\right) \hat{H}^\dagger \hat{H}
\end{equation}
where $\mathcal{L}_{\rm SM}$ is the SM Lagrangian, $H$ is the complex Higgs doublet, and the dark sector Lagrangian is given by
\begin{equation}\label{eq:scal_potential}
    \begin{split}
        \mathcal{L}_{\rm DS}=\frac{1}{2}\partial_\mu \hat{S}\,\partial^\mu \hat{S}-\frac{\mu_S^2}{2}\hat{S}^2+...
     \end{split}
\end{equation}
where we have omitted possible self-interactions of the scalar singlet, which we assume have been chosen in such a way that $S$ does not have a vacuum expectation value. Here and in the following, we indicate with a hat the original fields with non-canonical kinetic terms, before any field redefinition is performed.

After electroweak symmetry breaking the Higgs scalar, $\hat{h}$, mixes with the singlet $\hat{S}$. The resulting physical states, $h$ and $S$, obtained by diagonalizing the mass matrix, are given by the linear combination
\begin{equation}\label{eq:scalar_mixing}
    \left(\begin{array}{c} h \\ S \end{array}\right)=
    \left(\begin{array}{cc} \cos\theta & \sin\theta\\ -\sin\theta & \cos\theta\end{array}\right)
    \left(\begin{array}{c} \hat{h}\\\hat{S}\end{array}\right)\,,
\end{equation}
where the mixing angle is controlled by the parameters $A_{HS}$, and explicitly given by
\begin{equation}
    \tan\theta=\frac{x}{1+\sqrt{1+x^2}}     \qquad\qquad      x=\frac{2vA_{HS}}{\mu_H^2-\mu_S^2-\lambda_{HS}\,v^2}\,,
\end{equation}
with $v$ being the Higgs vacuum expectation value (vev), and $\mu_H^2=\lambda_H v^2$ with $\lambda_H$ the Higgs quartic coupling. The mass eigenvalues can also be expressed in terms of the small parameter $x$ as 
\begin{equation}
    m^2_{h, S}=\left(\frac{\mu_H^2+\mu_S^2+\lambda_{HS}\,v^2}{2}\right)\pm\left(\frac{\mu_H^2-\mu_S^2-\lambda_{HS}\,v^2}{2}\right)\sqrt{1+x^2}\,,
\end{equation}
which for $x\ll1$ reduce to $m^2_h\simeq \mu_H^2$ and $m^2_s\simeq \mu_S^2+\lambda_{HS}\,v^2$.

Due to the mixing in Eq.~\eqref{eq:scalar_mixing}, $S$ inherits all the couplings of the SM Higgs, modulo a suppression factor, $\sin\theta$, which is controlled by the parameter $A_{HS}$. Therefore, the decay width of the singlet can be obtained by rescaling the one of a SM Higgs of the same mass. Specifically, we follow references~\cite{Winkler:2018qyg, Gershtein:2020mwi} to derive the singlet branching ratios used in this work. 

In general, the production cross section is fixed by a combination of the parameters $A_{HS}$ and $\lambda_{HS}$. The former controls the production via the $b\to s$ penguin diagram (allowed for $m_S<m_B-m_K$)~\cite{Willey:1982ti,Chivukula:1988gp,Grinstein:1988yu} and $s\to d$ penguins for (allowed for $m_S < m_K-m_\pi$); while the latter fixes the double $S$ production via the $b\to s$ penguin diagram with an off-shell SM Higgs~\cite{Batell:2009jf}, or through direct Higgs decay. Indeed, in presence of a non-vanishing $\lambda_{HS}$ (and for $2m_S<m_h$) the Higgs can decay into a couple of $S$ with a width given by~\cite{OConnell:2006rsp}
\begin{equation}
    \Gamma_{h\to S S}=\frac{\lambda_{HS}^2v^2}{8\pi\,m_h}\sqrt{1-4{m_S^2/m_h^2}}\,.
\end{equation}

When $b\to s$ transitions dominate the production channel, decay and production are controlled by the same parameter, $\theta$, and the model parameter space is given by $\{\sin\theta,\,m_S\}$. However, the analysis discussed in this work has no reach for the products of $b\to s$ transitions, as the LLPs would be mostly produced inside (or near) $b$-jets and fail isolation cuts. Therefore, we will concentrate on the limit where the production is dominated by Higgs decays to two $S$, which is controlled by the parameter $\lambda_{HS}$. Therefore, production and decay channels will be decoupled and the model parameter space given by $\{\lambda_{HS},\,\sin\theta,\,m_S\}$.

Concretely, we generated events for Higgs production from gluon fusion in association with up to two jets and decayed the Higgs to two scalars. No generator level cuts are imposed and the Higgs $p_T$ distribution is reweighed to the NNLO prediction.

We conclude by noticing that, given $m^2_S\simeq \mu_S^2+\lambda_{HS}\,v^2$, some level of fine-tuning is required for $m^2_S<\lambda_{HS}\,v^2 $. Measuring the degree of fine-tuning in terms of the parameter $\Delta\equiv m_S^2/(\lambda_{HS}\,v^2)$, we can write the branching ratio for the exotic decay $h\to SS$ as 
\begin{equation}\label{eq:fine_tuning}
    {\rm Br}\left(h\to SS\right)\simeq\frac{\Gamma_{h\to SS}}{\Gamma^{\rm SM}_h}\simeq 6\cdot 10^{-3}\bigg(\frac{m_S}{2.5\,{\rm GeV}}\bigg)^4\bigg(\frac{0.1}{\Delta}\bigg)^2\,,
\end{equation}
where $\Gamma^{\rm SM}_h$ is the total SM Higgs width. 

\subsection{Abelian hidden sector}\label{subsec:dark_vector}
The next benchmark model that we consider consists in extending the SM by adding a \emph{dark} $\rm{U}(1)$ gauge symmetry which is spontaneously broken by a dark Higgs field, $S$. The dark ${\rm U}(1)$ is mediated by a \emph{dark} photon, $X$, which kinetically mixes with the SM hypercharge as:
\begin{equation}
    \mathcal{L}_{SH}=\mathcal{L}_{\rm SM}+\mathcal{L}_{\rm DS}-\lambda_{HS}\,\hat S^2\hat H^\dagger \hat H-\frac{\epsilon}{2\cos\theta_W}\hat{X}_{\mu\nu}\hat{B}^{\mu\nu}\,,
\end{equation}
where $\hat{B}_{\mu\nu}$ and $\hat{X}_{\mu\nu}$ are the field strengths of the hypercharge and the new $\rm{U}(1)$ gauge group respectively. The dark sector Lagrangian is 
\begin{equation}
    \mathcal{L}_{\rm DS} = -\frac{1}{4}\hat{X}_{\mu\nu}\hat{X}^{\mu\nu}+\mu_S^2\hat{S}^2-\lambda_S \hat{S}^4+|(\partial_\mu+ig_D\hat{X}_\mu)\hat{S}|^2\,.
\end{equation}
As before, we indicate with a hat the original fields with non-canonical kinetic terms, before any field redefinition is performed. The dark $\rm{U}(1)$ is spontaneously broken by the vev of the dark Higgs, $\langle S \rangle= v_S/\sqrt{2}$, which generate a mass for the dark photon $m_{X,0}=g_D v_S$.

After electroweak symmetry breaking the kinetic mixing between the dark photon and the hypercharge induces a coupling of the dark photon to the SM fermions which, in the $m_X^2<<m_Z^2$ limits, reads
\begin{equation}
    \mathcal{L}_{Xf\bar f}=\epsilon e\,Q_{f}X_\mu\bar{f}\gamma^\mu f\,,
\end{equation}
where $Q_f$ is the fermion electric charge. This coupling, controlled by the small parameter $\epsilon$, provides the decay channel in visible states for the dark photon. Specifically, we compute the dark photon branching ratios by using the package provided in~\cite{Ilten:2018crw}.

The diagonalization of the scalar sector proceeds similarly to what was discussed in the previous section, with the only difference that now we are interested in the regime where $m_S\gg m_h$, so that the dark Higgs decouples from the phenomenology of the model. Given the non vanishing coupling between $S$ and $X$, the mixing between the SM and dark Higgs generates a non-zero $hXX$ coupling which gives rise to the exotic Higgs decay $h\to XX$, with a width given by
\begin{equation}
    \Gamma(h\to XX)=\frac{\lambda_{HS}^2}{32 \pi}\frac{m_h v^2}{m_S^2}\sqrt{1-\frac{4m_X^2}{m_h^2}}\frac{(m_h^2+2m_X^2)^2-8(m_h^2-m_X^2)m_X^2}{m_h^4}\,.
\end{equation}
In the limit of small $\epsilon$ (which will be the relevant limit for our analysis), this dominates over Drell-Yan and $h\to ZX$ production and becomes the dominant dark photon production channel. In this limit the decay channel, controlled by $\epsilon$, and the production channel, controlled by $\lambda_{HS}$, are decoupled; and the model parameter space is given by $\{\epsilon,\lambda_{HS},m_X\}$. The event generation for this benchmark was performed similarly to the light scalar singlet case.

\subsection{Inelastic Dark Matter}\label{subsec:inel}
Inelastic Dark Matter (iDM) models are characterized by a DM candidate that couples with the SM only through interactions with a nearly degenerate state. A simple realization of this scenario can be obtained by adding to the model discussed in the previous section a Dirac pair of Weyl fermions, $\eta$ and $\xi$, that couple to the dark photon, $X$, with opposite charges. As before, the Higgs provides a source of U(1) breaking, generating a mass for the dark photon and a Majorana mass, $\delta$, for the two Weyl fermions. A Dirac mass, $m_D$, involving the two Weyl fermions is also allowed, so that at energies below the dark ${\rm U}(1)$ breaking scale, the mass terms for the dark fermions are 
\begin{align}
    \mathcal{L}\supset -m_D\,\eta\xi-\frac{\delta}{2}(\eta^2+\xi^2)\,+ {\rm h.c.}.
\end{align}
For a technically natural small Majorana mass, these mass terms can be perturbatively diagonalized to give the physical states
\begin{align}
    \chi_1 \simeq\frac{i}{\sqrt{2}}(\eta-\xi) \qquad\qquad\chi_2\simeq \frac{1}{\sqrt 2}(\eta+\xi)\,,
\end{align}
which have nearly degenerate masses $m_{1,2}\simeq m_D\pm \delta$. These mass eigenstates couple off-diagonally to the dark photon, {\it i.e.}
\begin{align}
    \mathcal{L}\supset i e_D \hat{X}_\mu\,\bar\chi_1\gamma^\mu\chi_2+\mathcal{O}\left(\frac{\delta}{m_D}\right)\,,
\end{align}
where we have written $\chi_{1,2}$ as Majorana spinors using four-component notation. Therefore, if $m_{X}>m_1+m_2$ and $\alpha_D\gg\epsilon\alpha_{\rm em}$, once produced dark photons decay into $\chi_1\chi_2$ pairs with a rate given by $\Gamma_{X\to \chi_1\chi_2}\simeq\alpha_D m_X$, and provide the dominant production channel for $\chi_1\chi_2$ pairs at LHC. For the values of $\epsilon$ we are interested in this analysis, the dominant production channel for dark photons is provided by Drell-Yan processes and scales as $\epsilon^2$.

The lightest state, $\chi_1$, is stable and once produced leaves the detector as missing energy; $\chi_2$ can decay into $\chi_1$ plus a pair of SM particles through an off-shell dark photon, possibly leaving a detectable signature. The rate for decays with leptonic final states is given by~\cite{Duerr:2020muu}:
\begin{align}\label{eq:iDM_decay}
    \Gamma_{\chi_2\to\chi_1l\bar l}=\epsilon^2\alpha_{\rm em}\alpha_D\int_{4 m_l^2}^{(m_1\Delta)^2}ds\frac{|\vec{p}_1|(m_1^2\Delta^2-s)(2s+m_1^2(2+\Delta)^2)(s+2m_l^2)(s-4m_l^2)^{1/2}}{6\pi m_2^2 s^{3/2}(s-m_X^2)^2}
\end{align}
where $s$ is the invariant mass of the lepton pair, $\vec{p}_1$ is the momentum of $\chi_1$ in the rest frame of $\chi_2$, and we have introduced the dimensionless parameter $\Delta\equiv(m_2-m_1)/m_1$. The rate for decays involving hadronic final states can be derived by setting $m_l=m_\mu$ and multiplying the integrand of Eq.~\eqref{eq:iDM_decay} by the experimentally measured quantity $R(s)\equiv\sigma(e^+e^-\to{\rm hadrons})/\sigma(e^+e^-\to\mu^+\mu^-)$. 

For this benchmark, events were generated using a \madgraph $Z'$ model for $X$ via production in association with up to three jets. A generator level cut $p_T > 100$~GeV was applied on the $X$, as the truth-level $\ptmiss$ is given by $p_T$ of the $Z'$ (its decay products are one DM particle and an LLP decaying in the muon chambers).

\subsection{Axion-like particles}
Axion-like particles (ALPs) extend the axion scenario to include any pseudoscalar particle that couples to the SM through dimension five operators. The naturally suppressed couplings make them a natural candidate for LLP searches. The general Lagrangian for these kinds of models is given by 
\begin{align}\label{eq:alp}
    {\cal L} = & {\cal L}_{SM} + \frac{1}{2}\left(\partial_\mu a\right)^2 -\frac{1}{2}m_a^2 a^2 + \frac{c_q^{ij}}{2 f}(\partial_\mu a)\bar q_i \gamma^\mu \gamma^5 q_j  + \frac{c_\ell^{ij}}{2 f}(\partial_\mu a)\bar \ell_i \gamma^\mu \gamma^5 \ell_j \nn \\ 
    & + \frac{a}{4\pi f}\left(\alpha_s c_{GG}\, G_{\mu\nu}^a\, \widetilde G^{a,\mu\nu}+ \alpha_2 c_{WW} \, W_{\mu\nu}^a \widetilde W^{a,\mu\nu} + \alpha_1 c_{BB}\, B_{\mu\nu} \widetilde B^{\mu\nu}\right) + \cdots
\end{align}
where $\widetilde{G}_{\mu\nu}=1/2\,\epsilon_{\mu\nu\rho\sigma}G^{\rho\sigma}$ where $G^{\rho\sigma}$ is the gluon field strength, and similarly for $\widetilde W$ and $\widetilde B$. In the broken, phase the couplings to $W$ and $B$ bosons induce couplings to photons and $Z$-bosons which are given by:
\begin{align}
    c_{ZZ} = c_{WW} + c_{BB}\qquad c_{\gamma Z} = c_w^2 c_{WW} -s_w^2 c_{BB} \qquad c_{\gamma\gamma}=c_w^4 c_{WW} + s_w^4c_{BB}\,.
\end{align}
In this work, we will focus on benchmark models in which the ALP couples only to gauge bosons $(c_q^{ij}=c_{\ell}^{ij}=0)$. 
Since the focus is on the production of energetic, isolated LLPs, this choice is sufficient to capture most of the dominant production modes at the LHC. 
Specifically, we will consider the three following scenarios: ALP coupled to $W$ ($c_{WW}\ne0,\,c_{GG}=c_{BB}=0$), {\it photophilic} ALP ($c_{\gamma\gamma}\ne0,\,c_{\gamma Z}=c_{GG}=0$), and ALP coupled to gluons ($c_{GG}\ne0,\, c_{BB}=c_{WW}=0$). The latter is a well-studied benchmark model in the context of light LLP searches, yielding the highest production rate at the LHC. The photophilic model chosen here is one of the (infinite) possible choices of UV-completion at LHC energies of the well-studied low-energy benchmark of ``ALP coupled to photons''. The conservative choice $c_{\gamma Z}=0$ is to focus on the parameter region where the existing LEP bounds are the weakest. Finally, the $c_W \neq 0$ benchmark provides a better UV-motivated benchmark than the photophilic choice, where associated ALP production with all the gauge bosons is allowed.

For the ALP coupled to gluons, we generated events where the LLP is produced in association with up to 3 jets, and imposed a $p_T > 100$~GeV and a $|\eta| < 3$ generator-level cuts on the transverse momentum and pseudorapidity of the ALP. The \madgraph model used here has been described in~\cite{Brivio:2017ije}, and we have only adapted the normalization of the couplings to the one used above. We did not include ALP production in the shower (i.e. where the ALP is produced at intermediate scales between the hard process collision and the QCD confinement scale) which was first estimated in~\cite{Aielli:2019ivi}, as there are not yet reliable event generators that can be used to keep track of the angular separation between the ALP and QCD jets (necessary for the jet veto requirements of the analysis). Therefore, for this benchmark, our limits should be considered conservative estimates for the reach of this CMS analysis, as they miss an important production channel. Production from meson mixing and meson decay was also neglected because it yields softer and non-isolated ALPs, for which this analysis has no sensitivity. For the lifetime and exclusive decay branching ratios of this benchmark, we used the estimates of~\cite{Aloni:2018vki}.

For the case of the other two ALP benchmarks, we considered ALP production in association with either a $W$, a $Z$, or a photon and up to 2 extra jets. We kept the same generator-level pseudorapidity cut but lowered the $p_T$ cut to $50$~GeV as some of the missing transverse energy can be produced by the decay products of the $W$ and $Z$ bosons. In these two benchmarks, the ALP decays predominantly into two photons.

\subsection{Hidden Valley}\label{sec:hv}

Confining Hidden Valleys (HV)~\cite{Strassler:2006im}, with a perturbative evolution below the scale mediating the interactions producing hidden sector particles, are a generic hidden sector extension of the SM on which we can have some theoretical control based on our knowledge of QCD-like theories. In general, one expects jets of hidden sector partons to hadronize in HV particles, some of which may decay back into SM final states, potentially as LLPs. Still, large freedom exists in defining a specific model. From the field content of the hidden sector and its symmetries, to the portal interactions mediating both the production of HV states and their decay back to the SM~\cite{Knapen:2021eip}. Many studies of search strategies at the LHC have been performed for different incarnations of this paradigm~\cite{Knapen:2022afb}.

In the context of this reinterpretation study, we choose one particular realization as an example model generating  the LLP-jet signature, aiming at maximizing the multiplicity of LLP produced in a jet, while keeping a high level of simplicity of reinterpretation. Therefore the example chosen is by no means generic \emph{per se}, although the lessons learned about the CMS analysis are.
Specifically, we used the Hidden Valley module~\cite{Carloni:2010tw} implemented in \pythia to generate events and choose a perturbative hidden sector with an $SU(N_c)$ asymptotically free gauge group with $N_f$ hidden quark flavors, fixing $N_c = 3$ and $N_f = 1$. The choice of $N_f = 1$ is to guarantee the absence of stable hidden mesons, therefore reducing the amount of collider stable particles produced and maximizing the number of LLPs in a hidden jet. This has a drawback, namely the lack of knowledge of the mass spectrum of such a theory as it lacks chiral symmetry breaking which is an important handle used in lattice simulation. In particular, the mass ratio between the first (pseudo-)scalar, $\eta_V$, and vector, $\omega_V$, resonances are poorly known, but expected to be ${\cal O}(1)$~\cite{Farchioni:2007dw,Creutz:2006ts,Armoni:2003fb}. Again, motivated by maximizing LLP multiplicity, we choose $m_{\omega_V} = 2.5 m_{\eta_V}= \Lambda_{HV}$ and assumed that the lowest scalar state (which \pythia will not use in the hadronization of the HV partons) is also able to decay to pairs of $\eta_V$. In this way, vector resonances can promptly decay to pseudoscalar mesons $\eta_V$, which will be the LLPs. 
For portals, we decide to decouple production and HV meson decays so that we can study the effects of varying the LLP lifetime on the limits for a fixed production rate. Specifically, we will produce hidden quarks in Higgs decays and will decay back the hidden spin-0 mesons $100\%$ into pair of photons. The latter choice is purely driven by the fact that the CMS analysis is not too sensitive to the relative amount of hadronic vs electromagnetic energy in LLP decays. At the same time, existing limits on light LLPs decaying to pair of photons are quite weak, so we can focus on reinterpreting this analysis without worrying about recasting other existing searches\footnote{The case of a recent CMS search for trackless jets~\cite{CMS:2021rwb} provides likely the strongest constraint for low values of $c\tau$ where a significant fraction of LLPs decays in the inner detector. However, that analysis explicitly vetoes signatures compatible with loose photons and photon conversion. The efficiency for one or more light LLPs decaying into pairs of photons being identified as loose photons is hard to recast, therefore we do not consider such analysis when presenting our limits.}. From a model building point of view, these portals can be easily generated by introducing a heavy scalar and pseudoscalar states $S$ and $A$, having Yukawa interactions with the HV vector-like quark $q_V$. The scalar $S$ can then interact with the SM Higgs via a $|H|^2 S$ cubic interaction, generating a Yukawa coupling between $q_V$ and the SM Higgs and a $q_V$ mass after electroweak symmetry breaking. At the same time, the pseudoscalar $A$ can have a coupling to the SM photons $A F \tilde F$ which in turn will induce a small decay width for $\eta_V$ via $\eta_V - A$ mixing.

\section{Results}
\label{sec:results}
In this section, we present the results for the benchmark models discussed in the previous section. We present both the current constraints, derived from data collected from 2016 to 2018, corresponding to an integrated luminosity of 137~${\mathrm{fb}}^{-1}$ and the projected constraints for Phase 2. The different projections for Phase 2 are derived by using the three different search strategies discussed in Section \ref{subsec:strategy}. 
Specifically, solid lines correspond to the search with the same selections as the CMS paper and a background rescaled according to the higher luminosity, dot-dashed lines correspond to the search with a higher $\nhits$ cut and zero background, and the dashed lines to the search with a dedicated trigger (that no longer require the MET and isolation cuts, but the presence of two separate LLP decays in the muon chambers) and zero background (see Section \ref{subsec:strategy} for a more detailed discussion of the search strategies).
Other existing and projected limits shown in the following plots are all taken from the literature, as referenced in the figure captions. The only exception is a limit originating from an ATLAS mono-jet search for the case of the gluon-coupled ALP, Fig.~\ref{fig:ggalp_results}, whose mass dependence was derived in this work as described in Appendix~\ref{app:atlas_monojet}.

\begin{figure}[t]
    \centering
    \includegraphics[width=\textwidth]{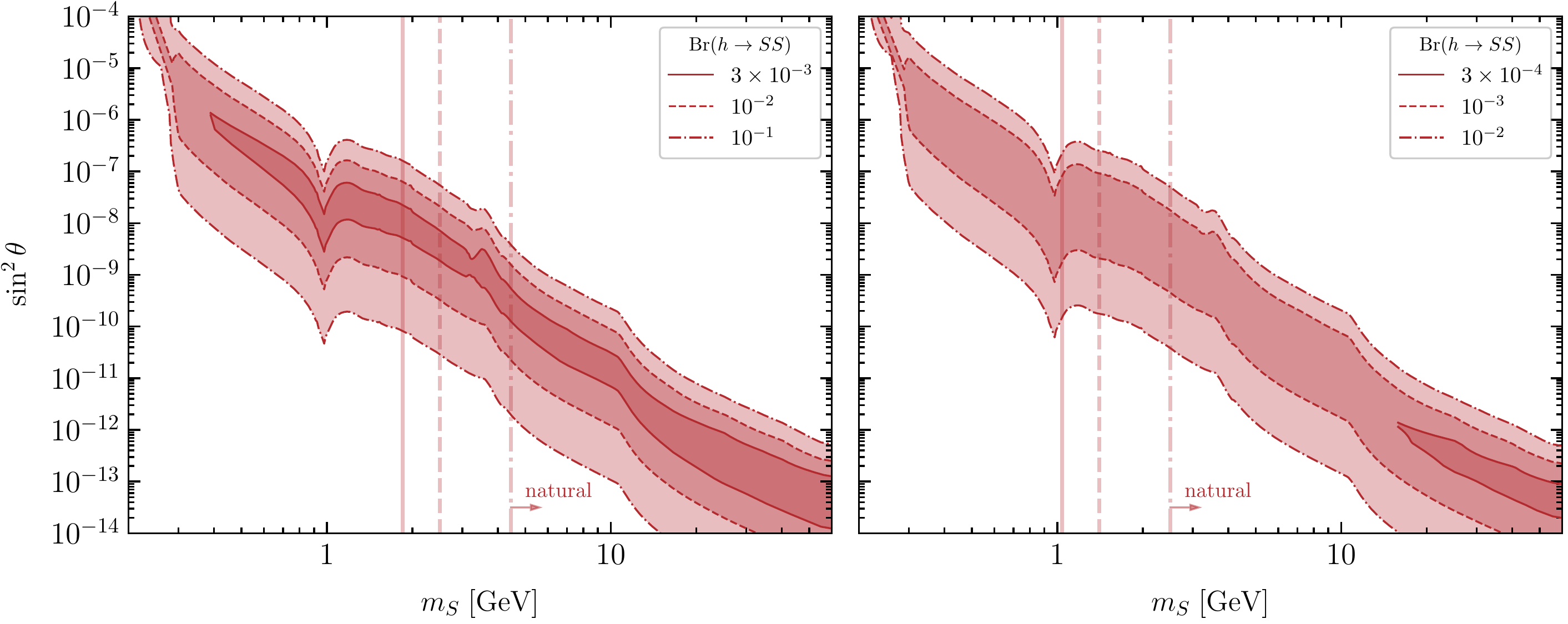}
    \caption{{\small Our limits for the light scalar model for different values of ${\rm Br}(h\to SS)$. In the left panel we show the current reach, while in the right panel we present the $3\;{\rm ab}^{-1}$ projections assuming that the same selections of the original CMS analysis are used. As in the previous plot, the vertical lines indicate the scalar mass below which the model needs to be fine-tuned (see discussion around Eq.~\eqref{eq:fine_tuning}).} \label{fig:scalar_br}}
    
\end{figure}

\begin{figure}[htbp]
    \centering
    \includegraphics[width=\textwidth]{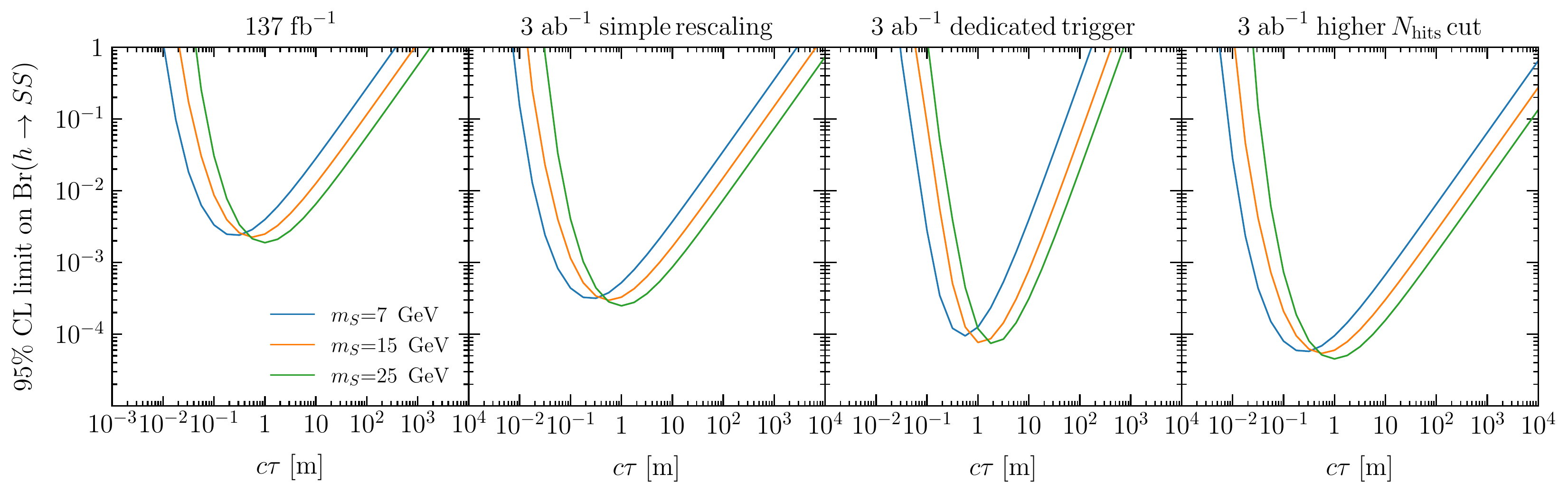}
    \caption{{\small Upper limits on the branching fraction ${\rm Br}(h\to SS)$ as functions of $c\tau$. In the left panel, we report the current constraints set by our analysis. In the other panels, we show the projected constraints for a luminosity of $3\;{\rm ab}^{-1}$ derived from the three different search strategies discussed in the main text.}}
    \label{fig:br_limits}
\end{figure}

In Fig.~\ref{fig:scalar_results} we show the reach for the light scalar model (discussed in Section~\ref{subsec:scalar}) with $\lambda=1.6\times10^{-3}$. This choice of $\lambda$ corresponds to an exotic Higgs branching fraction of ${\rm Br}(h\to SS)=0.01$, which is roughly the future reach for the Higgs branching into invisible final states. 
The present constraints are shown in the left panel, where we see that for low masses the analysis probes a previously unconstrained region of the parameter space, while at higher masses the constraints are similar to the ones of the ATLAS search for displaced vertices in the muon chambers (indicated as ATLAS mu-ROI in Fig.~\ref{fig:scalar_results}), whose reach was presented for $m_S > 5$~GeV. In the right panel, we show the projections for Phase 2 and compare them with the projected constraints from other future experiments and upgrades. We can see that, thanks to the different distance from the interaction point (IP), the projected results are complementary to dedicated LLPs experiments such as CODEX-b, FASER2, and MATHUSLA; all of which are positioned further away from the IP. To give an idea of how the constraints depend on the value of ${\rm Br}(h\to SS)$, in Fig.~\ref{fig:scalar_br} we show the same constraints of Fig.~\ref{fig:scalar_results} but for different values of ${\rm Br}(h\to SS)$. We see that the current search start to lose sensitivity for ${\rm Br}(h\to SS)\lesssim 3 \times 10^{-3}$, while for the future Phase 2 search we start to lose sensitivity for ${\rm Br}(h\to SS)\lesssim3\times10^{-4}$.
In all the plots we present also the values of the LLP mass (function of ${\rm Br}(h\to S S)$) below which tuning of more than 10\% is present.
Alternatively, in Fig.~\ref{fig:scalar_extra} we show the limits for a different slicing of the parameter space of this model, where the tree-level mass for $S$ is absent and the LLP mass is fully controlled by $\lambda_{HS}$ and therefore by ${\rm Br}(h\to S S)$.
In this case, there is no tuning even for lower masses, but the production rate varies with $m_S$ and searches for $H\to {\rm inv.}$ set an upper bound on $m_S$. Finally, to compare with the results of the CMS analysis, in Fig. \ref{fig:br_limits} we report the present and future limits on ${\rm Br}(h\to SS)$ as a function of the scalar lifetime.

\begin{figure}[htbp]
    \centering
    
    \includegraphics[width=\textwidth]{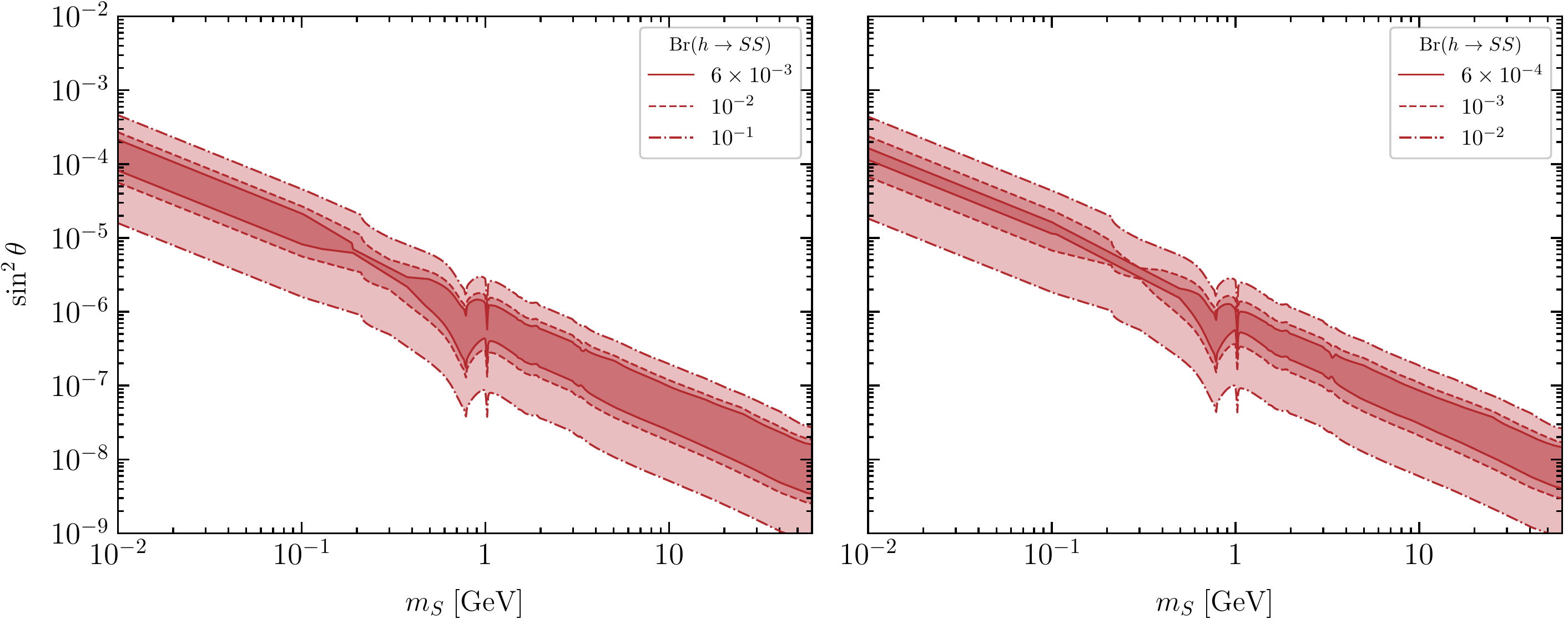}
    \caption{{\small Our limits for the dark photon model for different values of ${\rm Br}(h\to SS)$. In the left panel we show the current reach, while in the right panel we present the projections for $3\;{\rm ab}^{-1}$ assuming the same selections of the original CMS analysis are used.} \label{fig:dark_photon_br}}
    
\end{figure}

In Fig.~\ref{fig:dark_photon_results} we report the constraints for the Abelian hidden sector discussed in Section \ref{subsec:dark_vector}. As before, the value of the exotic Higgs branching ratio is fixed to ${\rm Br}(h\to A'A')=0.01$. We see that our current constraints (left panel) cover a mostly unconstrained region of the parameter space, except for the overlap with the ATLAS mu-ROI search at high masses. As for the scalar model, our projected constraints (right panel) well complement dedicated LLPs searches thanks to the different baselines. To investigate which is the lowest value of ${\rm Br}(h\to A'A')=0.01$ that we can probe, in Fig.~\ref{fig:dark_photon_br} we show present and future constraints for different values of the exotic Higgs branching. For the current search we see that we start to lose sensitivity for ${\rm Br}(h\to A'A')=3\times 10^{-3}$, while for the Phase 2 the constraints start to disappear for ${\rm Br}(h\to A'A')=3\times10^{-4}$.
This is consistent with what was found for the singlet scalar model and shows the relative insensitivity of the analysis to the specific exclusive decay modes. The only significant differences are around resonance mixing with hadronic resonances, which differ between the scalar and vector LLPs, and affect the LLP lifetime; and in the region between $200\,{\rm MeV} \lesssim m \lesssim 300\,{\rm MeV}$ where the $2\mu$ final state, to which this analysis is not sensitive to, contributes to ${\cal O}(50\%)$ of the dark photon branching ratios.

\begin{figure}[htbp]
    \centering
    \includegraphics[width=\textwidth]{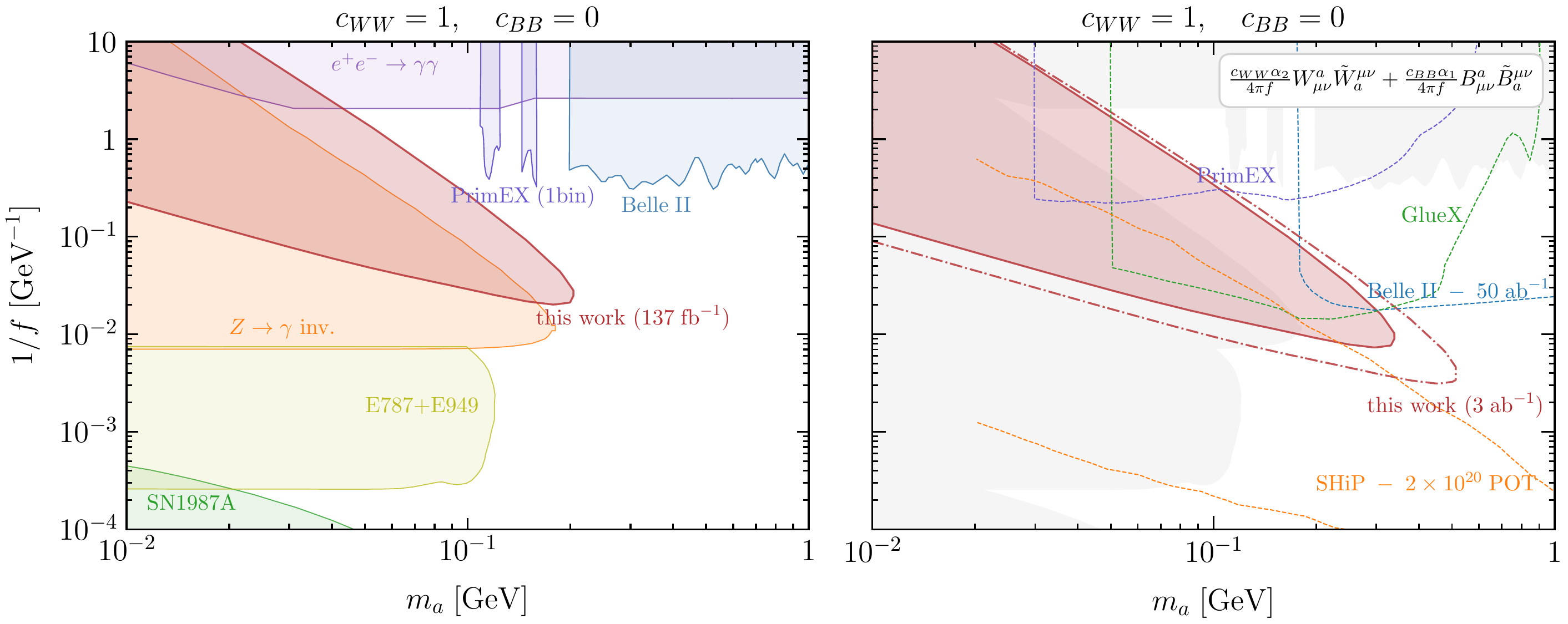}
    \caption{{\small Constraints on ALPs coupled to $W$ bosons. {\bf Left:} Comparison of our current reach (red region) with existing constraints from star cooling constraints (green)~\cite{Ertas:2020xcc}, beam dump experiments (yellow)~\cite{Ertas:2020xcc}, $Z$ invisible branching ratio (orange)~\cite{Dolan:2017osp, L3:1997exg}, limits on $e^+e^{-}\to\gamma\gamma$ from LEP data (violet)~\cite{Dolan:2017osp, Jaeckel:2015jla, DELPHI:1999fgt, L3:1995nbq}, PrimeEX (purple)~\cite{Aloni:2019ruo, PrimEx:2010fvg}, and Belle II (blue)~\cite{Belle-II:2020jti}.  {\bf Right:} Projections of our constraints for a luminosity of $3{\rm ab}^{-1}$ (red region). The solid and dot-dashed red contours correspond to the projections derived by using the same selections of the original CMS analysis, and the one derived by using a higher $\nhits$ cut and assuming zero background. We compare our results with current constraints (gray shaded region) and projections for SHiP (orange)~\cite{Beacham:2019nyx,SHiP:2015vad}, PrimEX (purple)~\cite{Aloni:2019ruo, PrimEx:2010fvg}, GlueX (violet)~\cite{Aloni:2019ruo, GlueX:2012idx}, , and Belle II (blue)~\cite{Beacham:2019nyx,Belle-II:2018jsg}.}}
    \label{fig:wwalp_results}
\end{figure}

The constraints for the three ALP models that we consider are shown in Fig.~\ref{fig:ggalp_results}$\,$-$\,$\ref{fig:wwalp_results}. For both the gluon (Fig.~\ref{fig:ggalp_results}) and electroweak (Fig.~\ref{fig:wwalp_results} and Fig.~\ref{fig:wwalp_tuned_results}) coupled scenarios, we find that the reinterpretation of the CMS analysis covers new territory beyond previous monojet~\cite{ATLAS:2021kxv} and fixed target~\cite{CHARM:1985anb} searches while being complementary to dedicated LLP experiments.
Moreover, one can expect the projections shown here to be underestimated, as dedicated searches using the fact the ALP is produced in association with a photon or a vector boson may allow us to relax some of the cuts and access softer LLPs that are produced with higher rates, pushing the estimated limits towards higher ALP masses and decay constants.

We now turn to the inelastic DM model results. The reinterpretation of this model is fairly sensitive to the LLP energy, $E_{LLP}$, via the mass splitting, $\Delta$. Unfortunately, the efficiency tables provided by the CMS Collaboration in HepData are not granular enough at low deposited energies $(E_{em}, E_{had})$ to resolve the turn-on shoulder of the 2D efficiency surface (the first bin is between 0 and 25~GeV). Therefore, our ability to reliably recast this model is hampered by the lack of knowledge about the minimal energy threshold for which the LLP visible decay products can produce ${\cal O}(20-30)$ charged particles emerging from a steel layer into the muon stations. To estimate this energy threshold, we impose an additional cut $E_{LLP}>5\,{\rm GeV}$ (which is approximately the energy needed for an electron to produce ${\cal O}(20)$ charged particles at the shower development maximum). The constraints for this choice of cut and using the model parameters $\Delta=0.005$, $\alpha_D=0.1$, and $m_{A'}=3m_1$, are reported in Fig.~\ref{fig:inel_results}. We see that the analysis covers previously unconstrained regions of the parameter space near the $Z$-resonance at $m_{A'}=3m_1=m_Z$. We have also derived the constraints for a model with $\Delta=0.01$ but decided not to show them since they are weaker than already existing limits, as larger mass splittings produce lifetimes too short to reach the CMS muon chambers. To further estimate the sensitivity of these results to the lower cut on the LLP energy, we show in Fig.~\ref{fig:inel_variation} the effect of varying it between 0 and 10~GeV. 

Finally, in Fig.~\ref{fig:hv_results} we report the limits on the exotic Higgs branching ratio ${\rm Br}(h\to Q\bar Q)$ for the hidden valley model discussed in Section~\ref{sec:hv}. We specifically choose a value for the HV confining scale $\Lambda_{HV} = 20\,{\rm GeV}$, which correspond to a pseudoscalar mass $m_{\eta_V}= 8\,{\rm GeV}$. Since in this model LLPs are produced within dark-showers in LLP jets, we expect the jet veto to reduce the sensitivity of the analysis. To quantify this effect, in the lower panels of Fig.~\ref{fig:hv_results} we show the ratio of the signal efficiency of the CMS analysis divided by the signal efficiency of the same analysis without the jet veto. As expected, this ratio rapidly approaches zero for small LLP lifetimes, when it is more likely for multiple LLPs to decay within the inner detector regions and the calorimeters in front of the cluster in the muon chambers selected as a signal by the analysis. Conversely, in the long lifetime area, the higher LLP multiplicity renders the limit more stringent than the case of Higgs decay to pairs of LLPs. Lowering the hidden confinement scale will increase the meson multiplicity inside hidden jets and therefore amplify this behavior.

\section{Discussion}
\label{sec:discussion}

We have explored some of the strengths and limitations of a recent search for LLPs using the muon chambers at CMS. The reinterpretation was made possible by the additional information provided by the collaboration in HEPData, which was embedded into \delphes modules.

\begin{figure}[t]
    \centering
    
    \includegraphics[width=\textwidth]{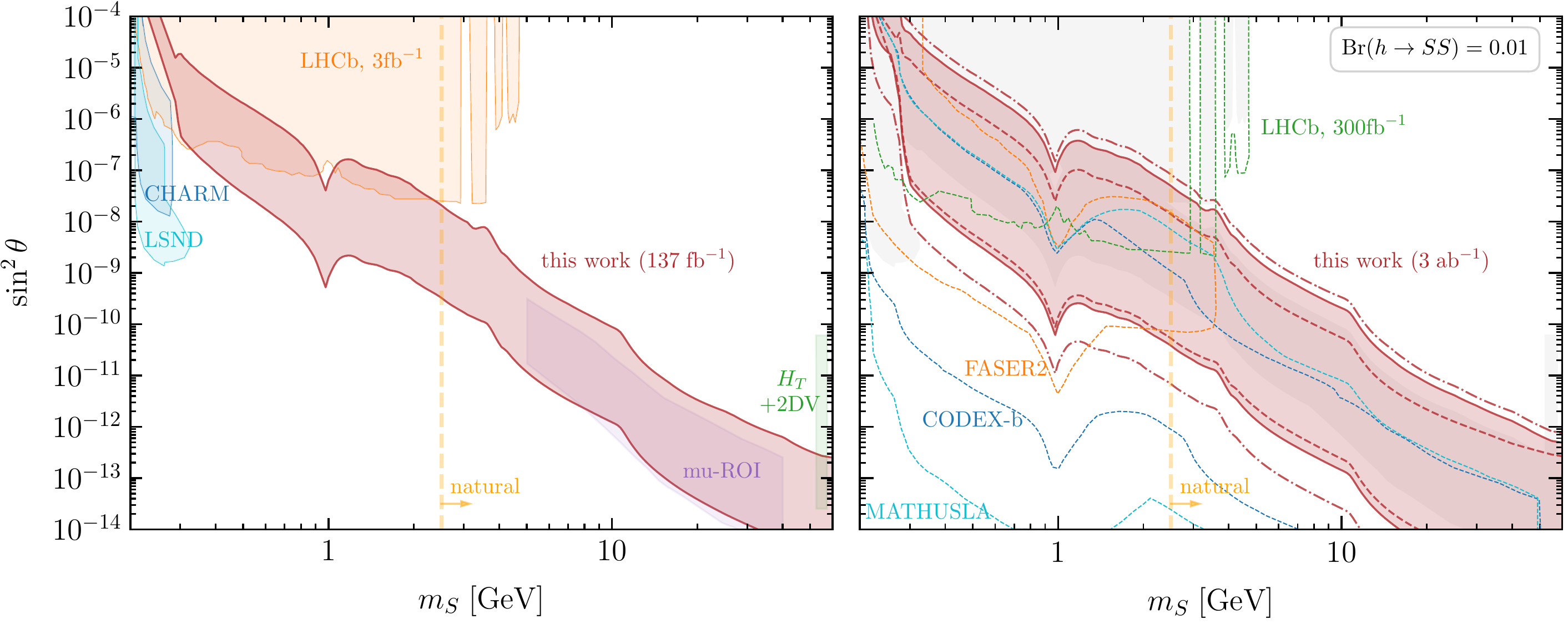}
    \caption{{\small Constraints on light scalars produced in Higgs decays for ${\rm Br}(h\to SS)=0.01$. {\bf Left:} Comparison of our current reach (red region) with existing limits from LHCb (orange)~\cite{LHCb:2016awg}, LSND (azure)~\cite{Foroughi-Abari:2020gju}, reinterpretation~\cite{Winkler:2018qyg} of the CHARM experiment (blue)~\cite{CHARM:1985anb}, CMS “HT +2DV” search (green)~\cite{Gershtein:2020mwi,CMS:2020iwv}, and reinterpretation of ATLAS mu-ROI (purple)~\cite{Gershtein:2020mwi, ATLAS:2018tup}. 
    {\bf Right:} Projections of our constraints for a luminosity of $3\,{\rm ab}^{-1}$ (red region). The three red contours (solid, dashed, and dot-dashed) correspond to the three search strategies discussed in the main text (rescaled CMS analysis, dedicated trigger, and higher $\nhits$). We compare our results with current constraints (gray shaded region) and projections for MATHUSLA~\cite{Alpigiani:2020tva}, CODEX-b~\cite{Aielli:2019ivi}, FASER2~\cite{Feng:2017vli}, and LHCb $300 {\rm fb}^{-1}$~\cite{Aaij:2016qsm}. The constraints for the projections are shown between the dashed lines with the corresponding colors. The vertical orange line indicates the scalar mass below which the model needs to be fine-tuned (see discussion around Eq.~\eqref{eq:fine_tuning}).} \label{fig:scalar_results}}
\end{figure}

\begin{figure}[htbp]
    \centering
    \includegraphics[width=0.5\textwidth]{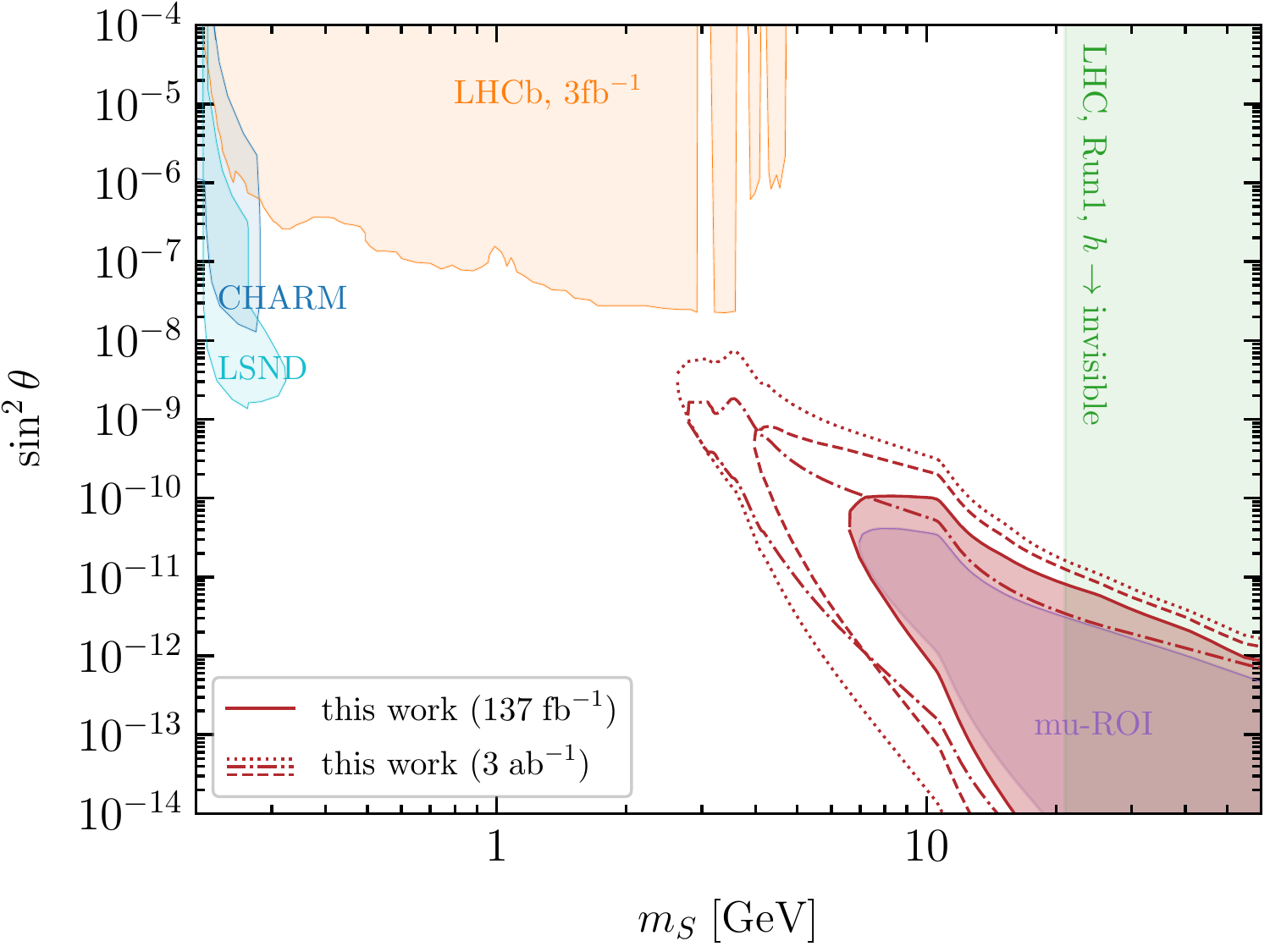}
    
    \caption{{\small Constraints on the singlet scalar model in absence of a tree-level mass for $S$ ($\mu_S=0$). The solid red line shows the current constraints, while the other contours (dashed, dot-dashed, and dotted) show $3\,{\rm ab}^{-1}$ projections derived by using the three recast strategies discussed in Section~\ref{subsec:strategy} (rescaled CMS analysis, dedicated trigger, and higher $\nhits$). The other existing constraints appearing on the plot are the same of Figure~\ref{fig:scalar_results}.}}
    \label{fig:scalar_extra}
\end{figure}

\begin{figure}[htbp]
    \centering
    
    \includegraphics[width=\textwidth]{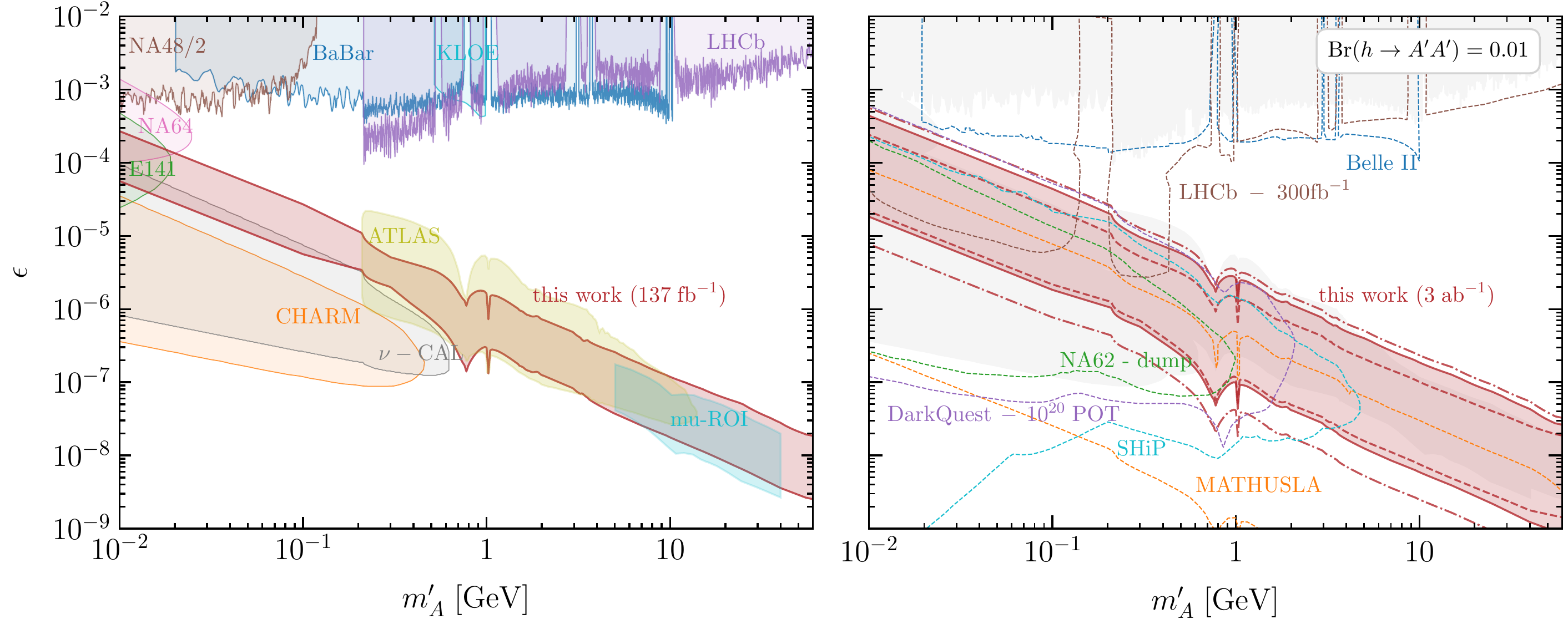}
    \caption{{\small Constraints on dark-photons produced in Higgs decays for ${\rm Br}(h\to SS)=0.01$. {\bf Left:} Comparison of our current reach (red region) with existing limits from BaBar (blue)~\cite{Lees:2014xha}, KLOE (azure)~\cite{Anastasi:2018azp}, LHCb (purple)~\cite{LHCb:2019vmc}, NA48 (brown)~\cite{Batley:2015lha}, reinterpretation of ATLAS $\mu$-ROI (yellow)~\cite{ATLAS:2018tup}, ATLAS search for displace dark-photon jets (yellow) \cite{ATLAS:2022izj}, and beam dump experiments (orange, gray, green, pink)~\cite{CHARM:1985anb, Blumlein:1991xh, Riordan:1987aw, NA64:2020qwq}. Most of the experimental constraints appearing in this plot have been digitized with the help of \texttt{darkcast}~\cite{Ilten:2018crw}.
    {\bf Right:} Projections of our constraints for a luminosity of $3\,{\rm ab}^{-1}$ (red region). The three red contours (solid, dashed, and dot-dashed) correspond to the three search strategies discussed in the main text (rescaled CMS analysis, dedicated trigger, and higher $\nhits$). We compare our results with current constraints (gray shaded region) and projections for MATHUSLA (orange)~\cite{Alpigiani:2020tva}, SHiP (azure)~\cite{SHiP:2015vad}, DarkQuest (purple)~\cite{Agrawal:2021dbo,Berlin:2018pwi}, NA62 in dump mode (green)~\cite{Agrawal:2021dbo,Collaboration:2691873}, LHCb upgrade (brown)~\cite{Agrawal:2021dbo,Ilten:2016tkc}, and Belle II (blue)~\cite{Kou:2018nap}.}\label{fig:dark_photon_results}}
\end{figure}

\begin{figure}[t]
    \centering
    \includegraphics[width=\textwidth]{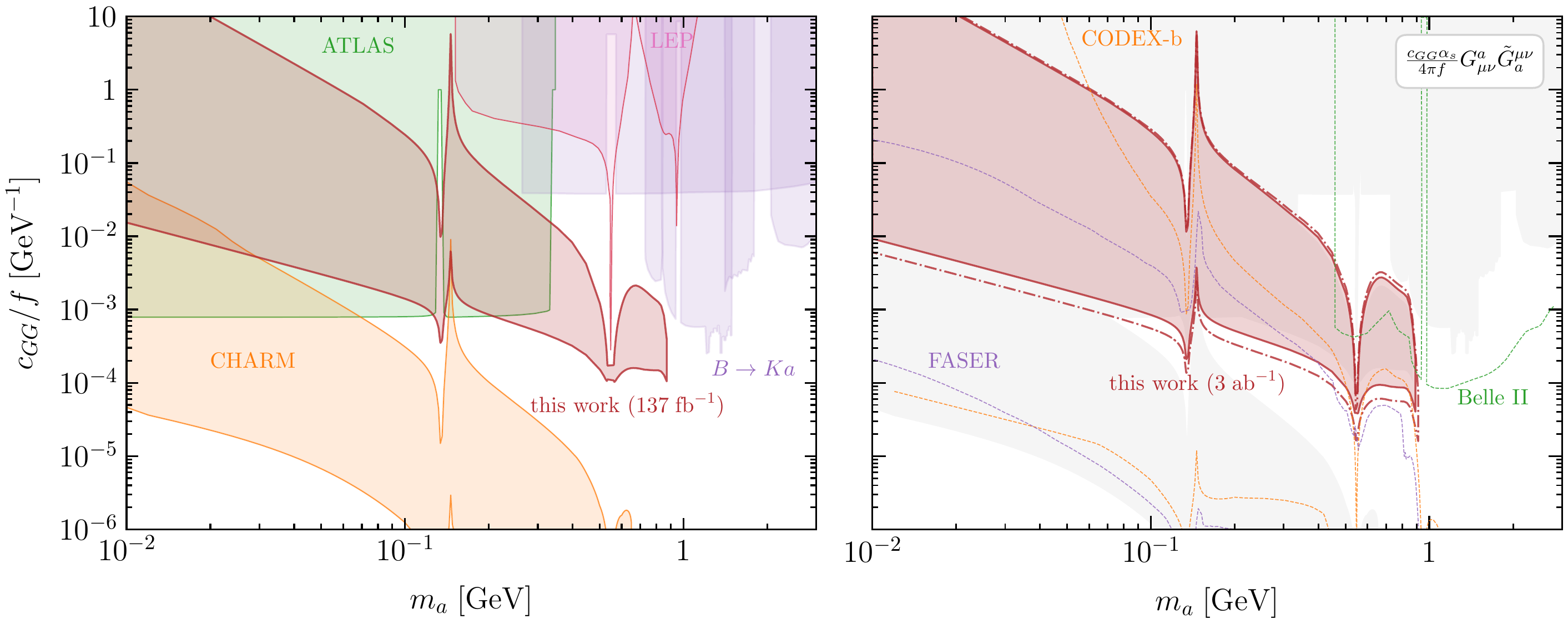}
    \caption{{\small Constraints on ALPs coupled to gluons. {\bf Left:} Comparison of our current reach (red region) with existing limits from CHARM (orange)~\cite{Aielli:2019ivi, CHARM:1985anb}, our reinterpretation of ATLAS (green)~\cite{ATLAS:2021kxv}, LEP~\cite{Aloni:2018vki}, and flavor probes (purple)~\cite{Chakraborty:2021wda, ParticleDataGroup:2020ssz, Belle:2013nby, BaBar:2008rth, BaBar:2011vod}. {\bf Right:} Projections of our constraints for a luminosity of $3\;{\rm ab}^{-1}$ (red region). The solid and dot-dashed red contours correspond to the projections derived by using the same selections of the original CMS analysis, and the one derived by using a higher $\nhits$ cut and assuming zero background. We compare our results with current constraints (gray shaded region) and projections for FASER (purple)~\cite{FASER:2018eoc}, CODEX-b (orange)~\cite{Aielli:2019ivi}, and Belle II (green)~\cite{Belle:2013nby, Chakraborty:2021wda}}}
    \label{fig:ggalp_results}
\end{figure}

\begin{figure}[htbp]
    \centering
     \includegraphics[width=\textwidth]{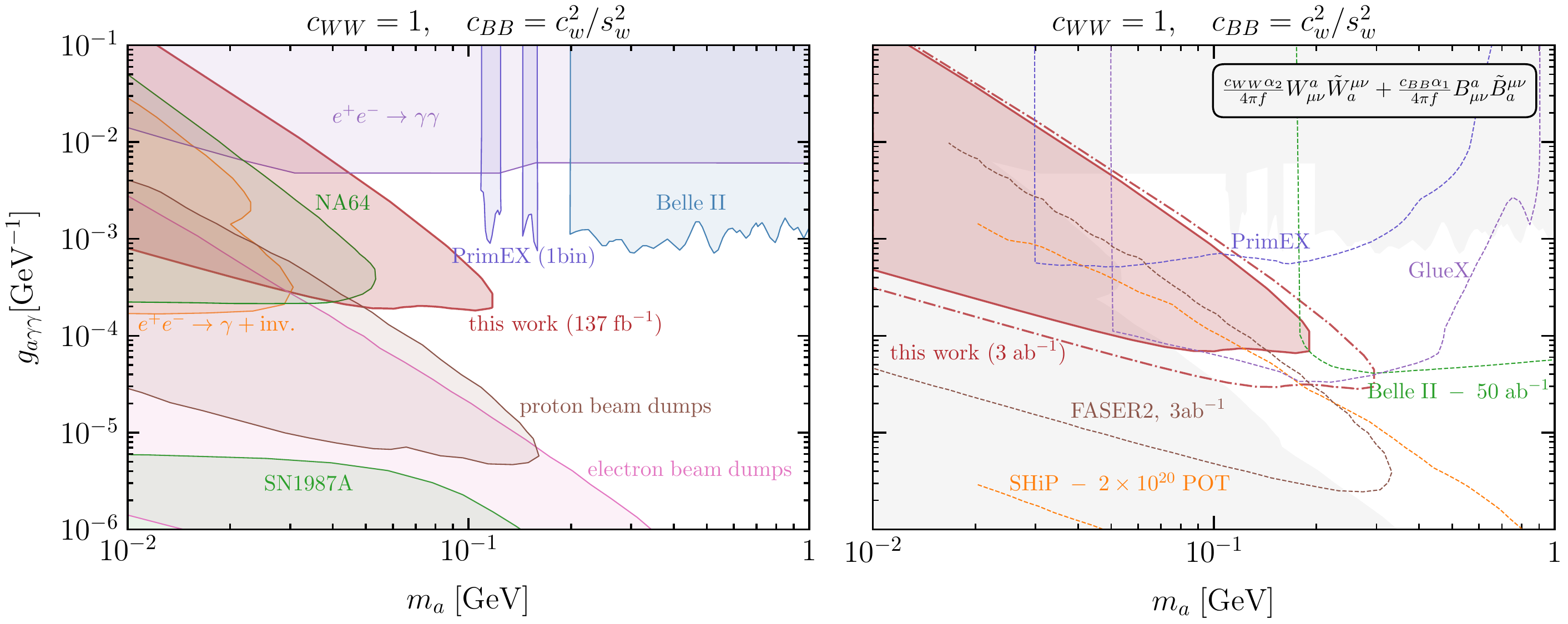}
    \caption{{\small Constraints on ALPs coupled to electroweak gauge bosons, and with $c_{\gamma Z}=0$. {\bf Left:} Comparison of our current reach (red region) with existing constraints from star cooling constraints (green)~\cite{Ertas:2020xcc}, electron~\cite{Dolan:2017osp, Bjorken:1988as, Riordan:1987aw} and proton~\cite{Beacham:2019nyx, Blumlein:1990ay, CHARM:1985anb} beam dump experiments (pink and brown), limits from mono-photon searches at LEP (orange)~\cite{Dolan:2017osp, DELPHI:2008uka}, NA64 (green)~\cite{NA64:2020qwq}, PrimEX (purple)~\cite{PrimEx:2010fvg, Aloni:2019ruo}, and Belle II (blue)~\cite{Belle-II:2020jti}. {\bf Right:} Projections of our constraints for a luminosity of $3{\rm ab}^{-1}$ (red region). The solid and dot-dashed red contours correspond to the projections derived by using the same selections of the original CMS analysis, and the one derived by using a higher $\nhits$ cut and assuming zero background. We compare our results with current constraints (gray shaded region) and projections for FASER (brown)~\cite{FASER:2018eoc}, SHiP (orange)~\cite{SHiP:2015vad}, PrimEX (purple)~\cite{PrimEx:2010fvg, Aloni:2019ruo}, GlueX (green)~\cite{Aloni:2019ruo, GlueX:2012idx}, and Belle II (green)~\cite{Belle-II:2018jsg}.}}
    \label{fig:wwalp_tuned_results}
\end{figure}

\begin{figure}[htbp]
    \centering
    \includegraphics[width=\textwidth]{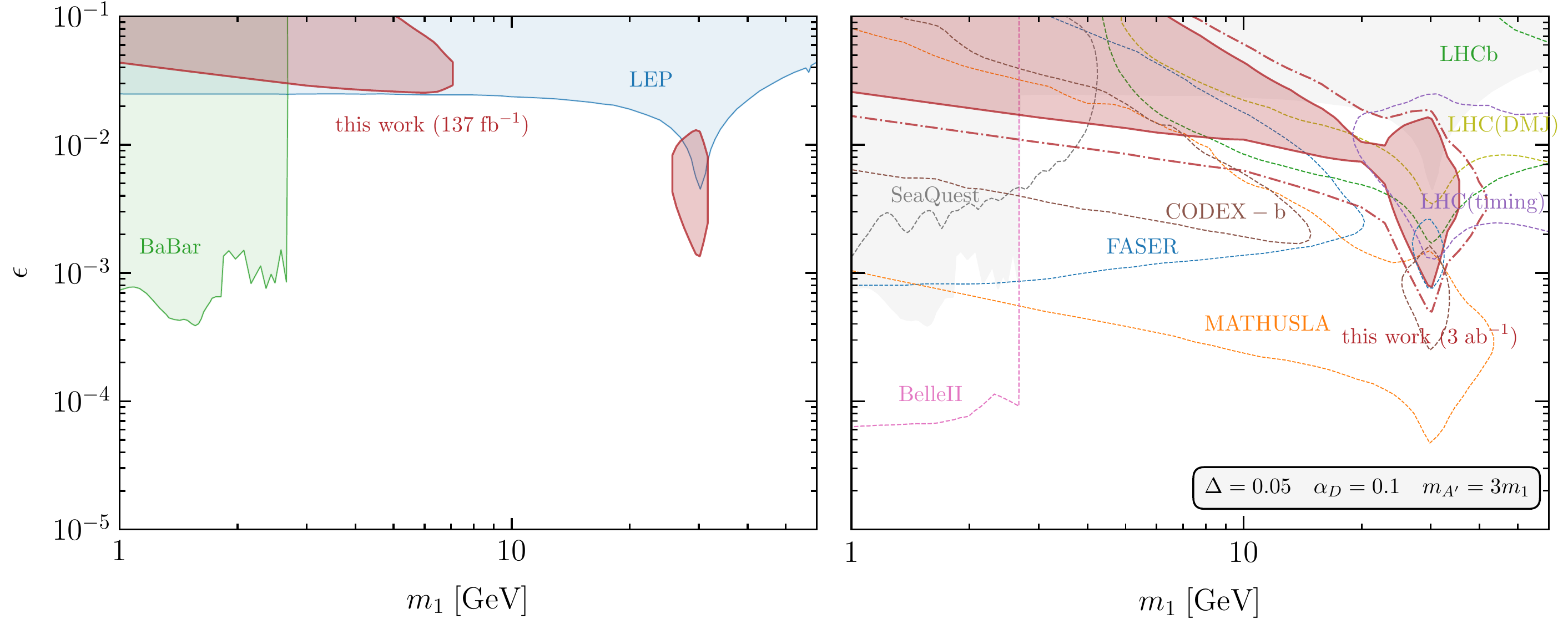}
    \caption{{\small Constraints on inelastic DM models, assuming a normalized mass splitting of $\Delta=0.05$, a dark coupling $\alpha_D=0.1$, and mediator mass given by $m_{A'}=3m_1$.
    {\bf Left:} Comparison of our current reach(red region) with existing constraints from BaBar (green)~\cite{Berlin:2018jbm, BaBar:2017tiz} and LEP (blue)~\cite{Berlin:2018jbm, Hook:2010tw, Ilten:2016tkc, Pierce:2017taw}. 
    {\bf Right:} Projections of our constraints for a luminosity of $3{\rm ab}^{-1}$ (red region). The solid and dot-dashed red contours correspond to the projections derived by assuming the same selections of the original CMS analysis, and the one derived by using a higher $\nhits$ cut and assuming zero background. We compare our results with current constraints (gray shaded region) and projections for BelleII (pink)~\cite{Belle-II:2018jsg}, SeaQuest (gray)~\cite{Berlin:2018pwi}, FASER (blue)~\cite{Feng:2017uoz}, MATHUSLA (orange)~\cite{Chou:2016lxi}, CODEX-b (brown)~\cite{Gligorov:2017nwh}, LHC (yellow and purple)~\cite{Izaguirre:2015zva, Liu:2018wte, Berlin:2018jbm}, and LHCb (green)~\cite{Berlin:2018jbm,LHCb:2017trq}.} \label{fig:inel_results}}
\end{figure}

\begin{figure}[htbp]
    \centering
    \includegraphics[width=\textwidth]{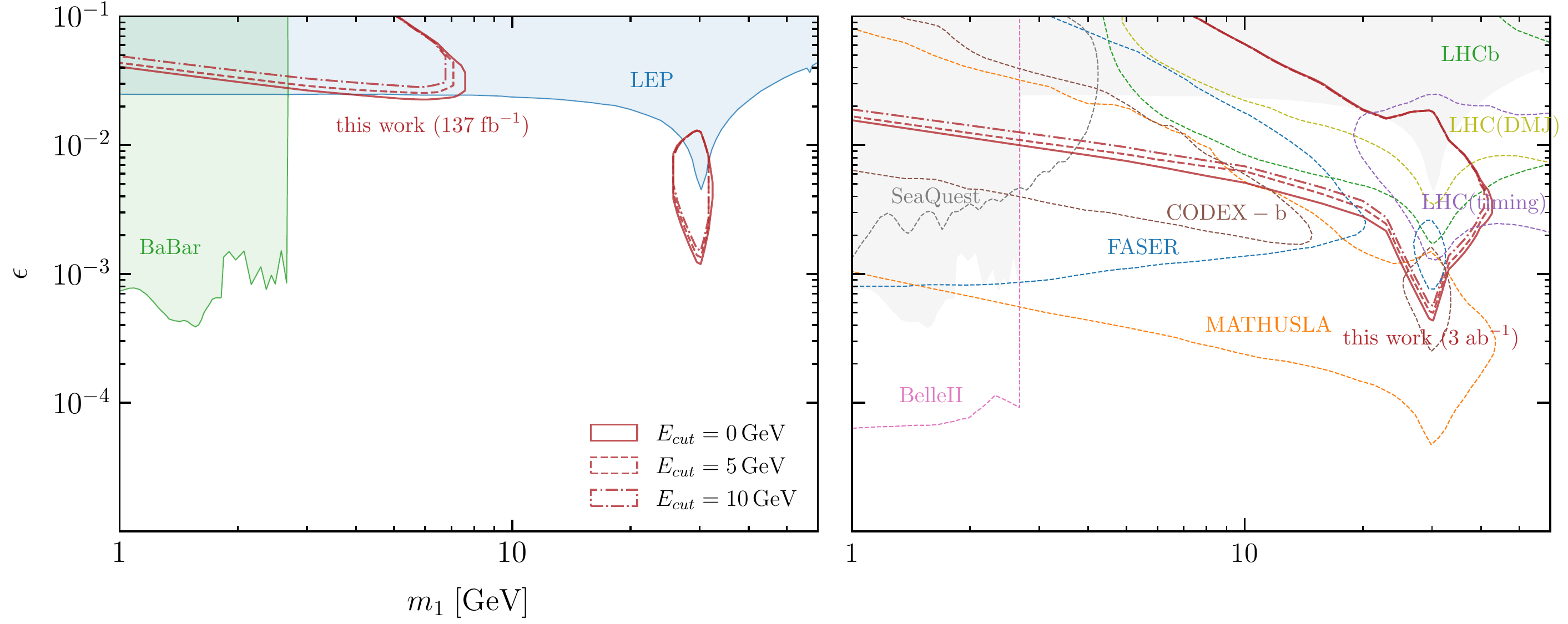}
    
    \caption{{\small Constraints on the inelastic DM model for different choices of the lower cut on the LLP energy. The other constraints appearing in the plot are the same reported in Figure~\ref{fig:inel_results}. The projections in the right panel are for a luminosity of $3{\rm ab}^{-1}$ luminosity, and assuming a tighter $N_{\rm hits}$ cut and zero background.}}
    \label{fig:inel_variation}
\end{figure}

\begin{figure}[htbp]
    \centering
    \includegraphics[width=\textwidth]{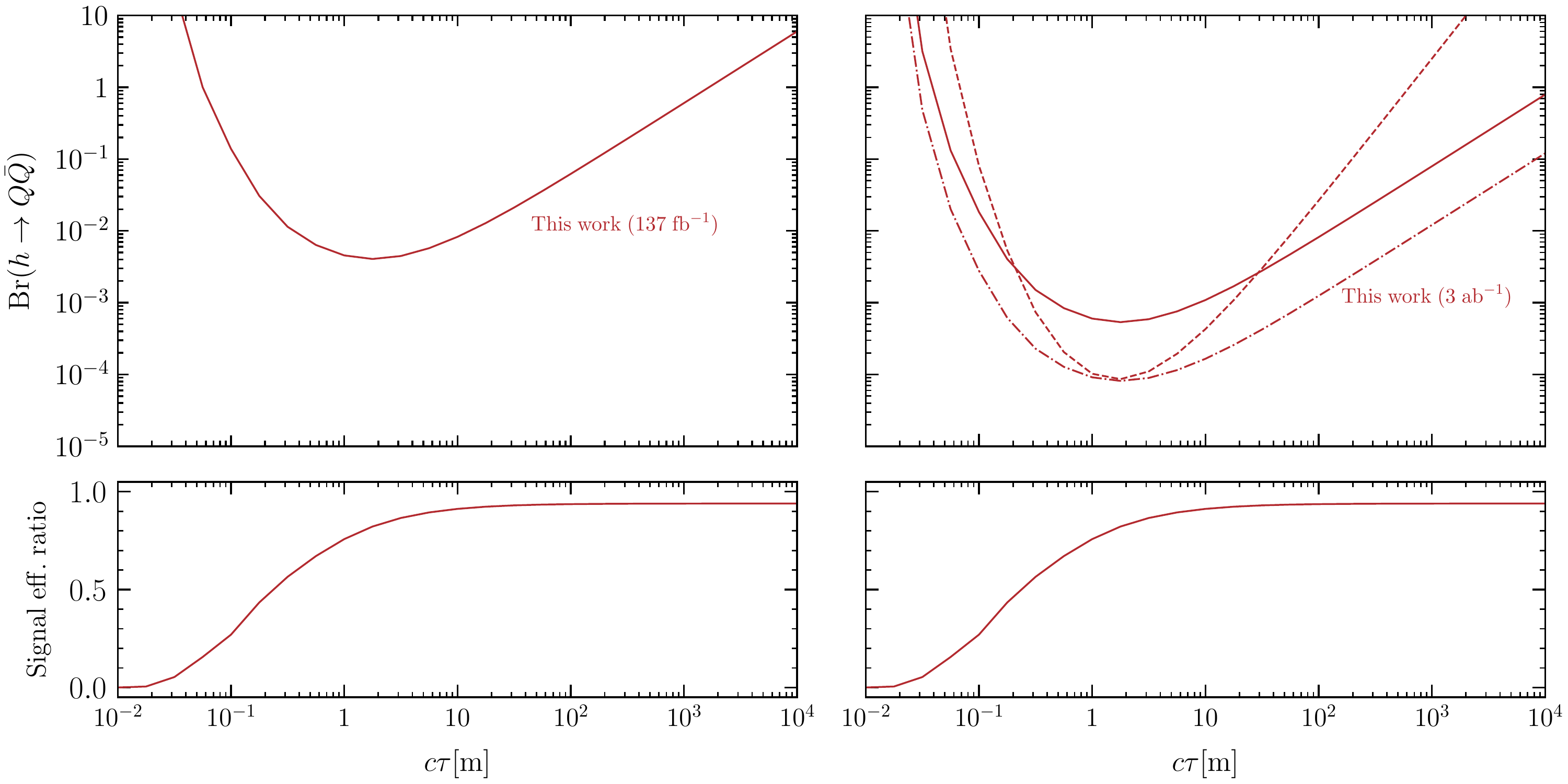}
    \caption{{\small Current (upper left panel) and projected (upper right panel) constraints on the Higgs exotic decay into dark quarks of a confining hidden valley model. For the projections we report (solid, dashed, and dot-dashed lines) the results obtained by using the three search strategies discussed in the main text (rescaled CMS analysis, dedicated trigger, and higher $\nhits$).} \label{fig:hv_results}}
\end{figure}

We have shown that this analysis proves very effective at constraining light LLPs, $m_{LLP} < {\cal O}({\rm GeV})$, as long as they can be produced energetically in LHC hadronic collisions and have $c\tau \lesssim {\cal O}({\rm m})$. In fact, we found that the current version of such a search strategy not only provides a counterexample to the lore that LLP searches at ATLAS and CMS are limited at low masses by irreducible SM backgrounds, but it is already able to cover previously unconstrained parameter space in many models, see Figs.~\ref{fig:scalar_results},~\ref{fig:dark_photon_results},~\ref{fig:ggalp_results},~\ref{fig:wwalp_tuned_results},~\ref{fig:inel_results},~\ref{fig:hv_results}, competing with and complementing the reach of dedicated LLP detectors. 

Still, various avenues for improvement exist. As mentioned in Sec.~\ref{sec:results}, producing signal categories with lower MET requirements but in association with another object such as a photon, lepton(s) or b-jet, may improve limits on specific models such as ALPs and HNLs. This will greatly increase the coverage of the search for many other models. This is especially true given the particular simplicity and reliability of the recasting provided by the publicly released information in HepData \cite{hepdata.104408.v2}. In this respect, we encourage the CMS Collaboration to provide more finely spaced efficiency maps at low $(E_{had},\,E_{em})$ to fully capture the turn-on shoulders, which is required in models where LLPs are producing less visible energy such as in the inelastic Dark Matter benchmark shown here.

Perhaps the most important avenue of improvement may be the investigation of how much the cluster isolation requirement can be relaxed. Many models, and production modes within a model, produce LLPs inside ($b$-)jets. Examples include the case of a light scalar model, where $S$ can be produced efficiently in $b$ decays and would yield muon chamber clusters not isolated from a $b$-jet; the case of ALPs produced in $b$-flavored hadron decays or in hadronic showers via $\pi^0$- $\eta^{(\prime)}$ mixing; or the case of emerging jets~\cite{Schwaller:2015gea} where showering within QCD and a Hidden Valley happens concurrently. Extending this kind of searches into the non-isolated regime will inevitably require some characterization and understanding of the origin of SM backgrounds mimicking clusters in the CSC. This effort has also implications and synergies beyond CMS itself. In fact, the amount of (instrumented) shielding provided by the calorimeters and the steel layers in the muon chambers is about $20-27$ nuclear interaction lengths, not far from the required shielding of other proposals for dedicated LLP experiments, such as e.g. the $30\lambda$ of active shield estimated to be necessary for CODEX-b~\cite{Gligorov:2017nwh}. Therefore, any characterization of SM backgrounds for CMS LLP searches would also benefit and inform the ongoing  shielding design and simulation of other experimental proposals such as CODEX-b.

Many benchmarks chosen here correspond to some of those selected to compare present and future efforts in the LLP search program such as within the CERN Physics Beyond Collider (PBC)~\cite{Beacham:2019nyx}. Given the relevance of this novel type of CMS search on the LHC reach for LLPs, we encourage the Collaboration to produce official limits that can be included in the PBC comparison plots. The capabilities of this kind of CMS search in probing light LLPs greatly extend what was considered possible for general-purpose existing LHC experiments to achieve and will undoubtedly complement and inform the broader future search program for LLPs beyond the Standard Model.

\acknowledgments

We would like to thank the CMS Collaboration and especially the CMS Exotica group, where part of this work was presented in 2021, for their positive feedback. 
CW and SX are partially supported by the U.S. Department of Energy, Office of Science, Office of High Energy Physics, under Award Number DE-SC0011925. This work  has been discussed in the relevant  Snowmass community study subgroups in 2022 and is part of CW's PhD thesis along with additional reinterpretation studies based solely on CMS published results and Delphes public codes \cite{CMS:2021juv,hepdata.104408.v2,delphes_pr}. 
CW, CP, and SX are grateful to the organizers and participants of the ``New ideas in detecting long-lived particles at the LHC'' workshop at LBNL in the Summer of 2018 where experimentalists and theorists gathered to generate new ideas on triggers and analysis strategies for long-lived particles searches at the LHC \cite{workshop} as well as the Fermilab LPC LLP group.   

AM was supported by the U.S. Department of Energy, Office of Science, Office of High Energy Physics, under Award No. DE-SC0021431, the Quantum Information Science Enabled Discovery (QuantISED) for High Energy Physics (KA2401032), and the Deutsche Forschungsgemeinschaft under Germany’s Excellence Strategy - EXC 2121 Quantum Universe - 390833306.

MP was supported by the U.S. Department of Energy, Office of Science, Office of High Energy Physics, under Award Number DE-SC0011632 and by the Walter Burke Institute for Theoretical Physics. MP would like to thank the Aspen Center for Physics where part of this work was performed.

\newpage
\FloatBarrier

\FloatBarrier
\appendix

\section{ATLAS mono-jet limit for Axion-like Particles coupled to gluons}\label{app:atlas_monojet}

Here we summarize the procedure used to reinterpret the ATLAS monojet limit on ALPs coupled to gluons~\cite{ATLAS:2021kxv}. The ATLAS collaboration already provides a lower limit on the ALP decay constant at a fixed ALP mass $m_a=1\,{\rm MeV}$ in this particular model and claims that such limit should hold for ALP masses up to approximately $1\,{\rm GeV}$. This claim is motivated from ALP literature prior to the improved estimates on ALP lifetimes and branching ratios provided in Ref.~\cite{Aloni:2018vki} and it is modified in the region $0.1-1\,{\rm GeV}$ due to the non-trivial behavior from ALP mixing with the neutral pseudoscalar mesons.
To estimate the limit curve in this region we use our ALP+jet simulation to extract the 2D LLP energy and pseudorapidity distributions, convolve that with the lifetime model of~\cite{Aloni:2018vki}, and require that the ALP does not decay in the ATLAS detector volume, for a fixed value of $m_a$ and $f$. We then rescale the ATLAS limit for the ratio of the two efficiencies described above computed at $m_a=1\,{\rm MeV}$ and at a different mass point. This produces a function of $(m_a/1\, {\rm MeV}, f/f_{\rm limit, 1 {\rm MeV}})$. We then invert this function to solve for the limit on $f$ as function of $m_a$ as shown in Fig.~\ref{fig:ggalp_results}. As expected the limit is fairly flat at low ALP masses but gets cut off earlier than 1~GeV due to the ALP lifetime significantly changing after the $m_{\eta}$ threshold. The steepness of the turn-off renders this limit curve a little sensitive to the specific geometric dimensions considered for the ATLAS detector.

\addtocontents{toc}{\protect\vspace*{10pt}}
\bibliographystyle{apsrev4-1}
\bibliography{bibliography-inspireized}

\begin{thebibliography}{127}%
\makeatletter
\providecommand \@ifxundefined [1]{%
 \@ifx{#1\undefined}
}%
\providecommand \@ifnum [1]{%
 \ifnum #1\expandafter \@firstoftwo
 \else \expandafter \@secondoftwo
 \fi
}%
\providecommand \@ifx [1]{%
 \ifx #1\expandafter \@firstoftwo
 \else \expandafter \@secondoftwo
 \fi
}%
\providecommand \natexlab [1]{#1}%
\providecommand \enquote  [1]{``#1''}%
\providecommand \bibnamefont  [1]{#1}%
\providecommand \bibfnamefont [1]{#1}%
\providecommand \citenamefont [1]{#1}%
\providecommand \href@noop [0]{\@secondoftwo}%
\providecommand \href [0]{\begingroup \@sanitize@url \@href}%
\providecommand \@href[1]{\@@startlink{#1}\@@href}%
\providecommand \@@href[1]{\endgroup#1\@@endlink}%
\providecommand \@sanitize@url [0]{\catcode `\\12\catcode `\$12\catcode
  `\&12\catcode `\#12\catcode `\^12\catcode `\_12\catcode `\%12\relax}%
\providecommand \@@startlink[1]{}%
\providecommand \@@endlink[0]{}%
\providecommand \url  [0]{\begingroup\@sanitize@url \@url }%
\providecommand \@url [1]{\endgroup\@href {#1}{\urlprefix }}%
\providecommand \urlprefix  [0]{URL }%
\providecommand \Eprint [0]{\href }%
\providecommand \doibase [0]{http://dx.doi.org/}%
\providecommand \selectlanguage [0]{\@gobble}%
\providecommand \bibinfo  [0]{\@secondoftwo}%
\providecommand \bibfield  [0]{\@secondoftwo}%
\providecommand \translation [1]{[#1]}%
\providecommand \BibitemOpen [0]{}%
\providecommand \bibitemStop [0]{}%
\providecommand \bibitemNoStop [0]{.\EOS\space}%
\providecommand \EOS [0]{\spacefactor3000\relax}%
\providecommand \BibitemShut  [1]{\csname bibitem#1\endcsname}%
\let\auto@bib@innerbib\@empty
\bibitem [{\citenamefont {Aaboud}\ \emph {et~al.}(2018)\citenamefont {Aaboud}
  \emph {et~al.}}]{ATLAS:2017tny}%
  \BibitemOpen
  \bibfield  {author} {\bibinfo {author} {\bibfnamefont {M.}~\bibnamefont
  {Aaboud}} \emph {et~al.} (\bibinfo {collaboration} {ATLAS}),\ }\href
  {\doibase 10.1103/PhysRevD.97.052012} {\bibfield  {journal} {\bibinfo
  {journal} {Phys. Rev. D}\ }\textbf {\bibinfo {volume} {97}},\ \bibinfo
  {pages} {052012} (\bibinfo {year} {2018})},\ \Eprint
  {http://arxiv.org/abs/1710.04901} {arXiv:1710.04901 [hep-ex]} \BibitemShut
  {NoStop}%
\bibitem [{\citenamefont {Aaij}\ \emph
  {et~al.}(2017{\natexlab{a}})\citenamefont {Aaij} \emph
  {et~al.}}]{LHCb:2017xxn}%
  \BibitemOpen
  \bibfield  {author} {\bibinfo {author} {\bibfnamefont {R.}~\bibnamefont
  {Aaij}} \emph {et~al.} (\bibinfo {collaboration} {LHCb}),\ }\href {\doibase
  10.1140/epjc/s10052-017-5178-x} {\bibfield  {journal} {\bibinfo  {journal}
  {Eur. Phys. J. C}\ }\textbf {\bibinfo {volume} {77}},\ \bibinfo {pages} {812}
  (\bibinfo {year} {2017}{\natexlab{a}})},\ \Eprint
  {http://arxiv.org/abs/1705.07332} {arXiv:1705.07332 [hep-ex]} \BibitemShut
  {NoStop}%
\bibitem [{\citenamefont {Lee}\ \emph {et~al.}(2019)\citenamefont {Lee},
  \citenamefont {Ohm}, \citenamefont {Soffer},\ and\ \citenamefont
  {Yu}}]{Lee:2018pag}%
  \BibitemOpen
  \bibfield  {author} {\bibinfo {author} {\bibfnamefont {L.}~\bibnamefont
  {Lee}}, \bibinfo {author} {\bibfnamefont {C.}~\bibnamefont {Ohm}}, \bibinfo
  {author} {\bibfnamefont {A.}~\bibnamefont {Soffer}}, \ and\ \bibinfo {author}
  {\bibfnamefont {T.-T.}\ \bibnamefont {Yu}},\ }\href {\doibase
  10.1016/j.ppnp.2019.02.006} {\bibfield  {journal} {\bibinfo  {journal} {Prog.
  Part. Nucl. Phys.}\ }\textbf {\bibinfo {volume} {106}},\ \bibinfo {pages}
  {210} (\bibinfo {year} {2019})},\ \bibinfo {note} {[Erratum:
  Prog.Part.Nucl.Phys. 122, 103912 (2022)]},\ \Eprint
  {http://arxiv.org/abs/1810.12602} {arXiv:1810.12602 [hep-ph]} \BibitemShut
  {NoStop}%
\bibitem [{\citenamefont {Aaboud}\ \emph
  {et~al.}(2019{\natexlab{a}})\citenamefont {Aaboud} \emph
  {et~al.}}]{ATLAS:2018rjc}%
  \BibitemOpen
  \bibfield  {author} {\bibinfo {author} {\bibfnamefont {M.}~\bibnamefont
  {Aaboud}} \emph {et~al.} (\bibinfo {collaboration} {ATLAS}),\ }\href
  {\doibase 10.1103/PhysRevD.99.012001} {\bibfield  {journal} {\bibinfo
  {journal} {Phys. Rev. D}\ }\textbf {\bibinfo {volume} {99}},\ \bibinfo
  {pages} {012001} (\bibinfo {year} {2019}{\natexlab{a}})},\ \Eprint
  {http://arxiv.org/abs/1808.03057} {arXiv:1808.03057 [hep-ex]} \BibitemShut
  {NoStop}%
\bibitem [{\citenamefont {Alimena}\ \emph {et~al.}(2020)\citenamefont {Alimena}
  \emph {et~al.}}]{Alimena:2019zri}%
  \BibitemOpen
  \bibfield  {author} {\bibinfo {author} {\bibfnamefont {J.}~\bibnamefont
  {Alimena}} \emph {et~al.},\ }\href {\doibase 10.1088/1361-6471/ab4574}
  {\bibfield  {journal} {\bibinfo  {journal} {J. Phys. G}\ }\textbf {\bibinfo
  {volume} {47}},\ \bibinfo {pages} {090501} (\bibinfo {year} {2020})},\
  \Eprint {http://arxiv.org/abs/1903.04497} {arXiv:1903.04497 [hep-ex]}
  \BibitemShut {NoStop}%
\bibitem [{\citenamefont {Beacham}\ \emph {et~al.}(2020)\citenamefont {Beacham}
  \emph {et~al.}}]{Beacham:2019nyx}%
  \BibitemOpen
  \bibfield  {author} {\bibinfo {author} {\bibfnamefont {J.}~\bibnamefont
  {Beacham}} \emph {et~al.},\ }\href {\doibase 10.1088/1361-6471/ab4cd2}
  {\bibfield  {journal} {\bibinfo  {journal} {J. Phys. G}\ }\textbf {\bibinfo
  {volume} {47}},\ \bibinfo {pages} {010501} (\bibinfo {year} {2020})},\
  \Eprint {http://arxiv.org/abs/1901.09966} {arXiv:1901.09966 [hep-ex]}
  \BibitemShut {NoStop}%
\bibitem [{\citenamefont {Sirunyan}\ \emph
  {et~al.}(2019{\natexlab{a}})\citenamefont {Sirunyan} \emph
  {et~al.}}]{CMS:2019zxa}%
  \BibitemOpen
  \bibfield  {author} {\bibinfo {author} {\bibfnamefont {A.~M.}\ \bibnamefont
  {Sirunyan}} \emph {et~al.} (\bibinfo {collaboration} {CMS}),\ }\href
  {\doibase 10.1103/PhysRevD.100.112003} {\bibfield  {journal} {\bibinfo
  {journal} {Phys. Rev. D}\ }\textbf {\bibinfo {volume} {100}},\ \bibinfo
  {pages} {112003} (\bibinfo {year} {2019}{\natexlab{a}})},\ \Eprint
  {http://arxiv.org/abs/1909.06166} {arXiv:1909.06166 [hep-ex]} \BibitemShut
  {NoStop}%
\bibitem [{\citenamefont {Sirunyan}\ \emph
  {et~al.}(2019{\natexlab{b}})\citenamefont {Sirunyan} \emph
  {et~al.}}]{CMS:2019qjk}%
  \BibitemOpen
  \bibfield  {author} {\bibinfo {author} {\bibfnamefont {A.~M.}\ \bibnamefont
  {Sirunyan}} \emph {et~al.} (\bibinfo {collaboration} {CMS}),\ }\href
  {\doibase 10.1016/j.physletb.2019.134876} {\bibfield  {journal} {\bibinfo
  {journal} {Phys. Lett. B}\ }\textbf {\bibinfo {volume} {797}},\ \bibinfo
  {pages} {134876} (\bibinfo {year} {2019}{\natexlab{b}})},\ \Eprint
  {http://arxiv.org/abs/1906.06441} {arXiv:1906.06441 [hep-ex]} \BibitemShut
  {NoStop}%
\bibitem [{\citenamefont {Sirunyan}\ \emph {et~al.}(2020)\citenamefont
  {Sirunyan} \emph {et~al.}}]{CMS:2020atg}%
  \BibitemOpen
  \bibfield  {author} {\bibinfo {author} {\bibfnamefont {A.~M.}\ \bibnamefont
  {Sirunyan}} \emph {et~al.} (\bibinfo {collaboration} {CMS}),\ }\href
  {\doibase 10.1016/j.physletb.2020.135502} {\bibfield  {journal} {\bibinfo
  {journal} {Phys. Lett. B}\ }\textbf {\bibinfo {volume} {806}},\ \bibinfo
  {pages} {135502} (\bibinfo {year} {2020})},\ \Eprint
  {http://arxiv.org/abs/2004.05153} {arXiv:2004.05153 [hep-ex]} \BibitemShut
  {NoStop}%
\bibitem [{\citenamefont {Sirunyan}\ \emph
  {et~al.}(2021{\natexlab{a}})\citenamefont {Sirunyan} \emph
  {et~al.}}]{CMS:2020iwv}%
  \BibitemOpen
  \bibfield  {author} {\bibinfo {author} {\bibfnamefont {A.~M.}\ \bibnamefont
  {Sirunyan}} \emph {et~al.} (\bibinfo {collaboration} {CMS}),\ }\href
  {\doibase 10.1103/PhysRevD.104.012015} {\bibfield  {journal} {\bibinfo
  {journal} {Phys. Rev. D}\ }\textbf {\bibinfo {volume} {104}},\ \bibinfo
  {pages} {012015} (\bibinfo {year} {2021}{\natexlab{a}})},\ \Eprint
  {http://arxiv.org/abs/2012.01581} {arXiv:2012.01581 [hep-ex]} \BibitemShut
  {NoStop}%
\bibitem [{\citenamefont {Aad}\ \emph {et~al.}(2020)\citenamefont {Aad} \emph
  {et~al.}}]{ATLAS:2020xyo}%
  \BibitemOpen
  \bibfield  {author} {\bibinfo {author} {\bibfnamefont {G.}~\bibnamefont
  {Aad}} \emph {et~al.} (\bibinfo {collaboration} {ATLAS}),\ }\href {\doibase
  10.1103/PhysRevD.102.032006} {\bibfield  {journal} {\bibinfo  {journal}
  {Phys. Rev. D}\ }\textbf {\bibinfo {volume} {102}},\ \bibinfo {pages}
  {032006} (\bibinfo {year} {2020})},\ \Eprint
  {http://arxiv.org/abs/2003.11956} {arXiv:2003.11956 [hep-ex]} \BibitemShut
  {NoStop}%
\bibitem [{\citenamefont {Aad}\ \emph {et~al.}(2021{\natexlab{a}})\citenamefont
  {Aad} \emph {et~al.}}]{ATLAS:2020wjh}%
  \BibitemOpen
  \bibfield  {author} {\bibinfo {author} {\bibfnamefont {G.}~\bibnamefont
  {Aad}} \emph {et~al.} (\bibinfo {collaboration} {ATLAS}),\ }\href {\doibase
  10.1103/PhysRevLett.127.051802} {\bibfield  {journal} {\bibinfo  {journal}
  {Phys. Rev. Lett.}\ }\textbf {\bibinfo {volume} {127}},\ \bibinfo {pages}
  {051802} (\bibinfo {year} {2021}{\natexlab{a}})},\ \Eprint
  {http://arxiv.org/abs/2011.07812} {arXiv:2011.07812 [hep-ex]} \BibitemShut
  {NoStop}%
\bibitem [{\citenamefont {Agrawal}\ \emph {et~al.}(2021)\citenamefont {Agrawal}
  \emph {et~al.}}]{Agrawal:2021dbo}%
  \BibitemOpen
  \bibfield  {author} {\bibinfo {author} {\bibfnamefont {P.}~\bibnamefont
  {Agrawal}} \emph {et~al.},\ }\href {\doibase 10.1140/epjc/s10052-021-09703-7}
  {\bibfield  {journal} {\bibinfo  {journal} {Eur. Phys. J. C}\ }\textbf
  {\bibinfo {volume} {81}},\ \bibinfo {pages} {1015} (\bibinfo {year}
  {2021})},\ \Eprint {http://arxiv.org/abs/2102.12143} {arXiv:2102.12143
  [hep-ph]} \BibitemShut {NoStop}%
\bibitem [{\citenamefont {Acosta}\ \emph {et~al.}(2021)\citenamefont {Acosta}
  \emph {et~al.}}]{Alimena:2021mdu}%
  \BibitemOpen
  \bibfield  {author} {\bibinfo {author} {\bibfnamefont {D.}~\bibnamefont
  {Acosta}} \emph {et~al.},\ }\href@noop {} {\  (\bibinfo {year} {2021})},\
  \Eprint {http://arxiv.org/abs/2110.14675} {arXiv:2110.14675 [hep-ex]}
  \BibitemShut {NoStop}%
\bibitem [{\citenamefont {Borsato}\ \emph {et~al.}(2022)\citenamefont {Borsato}
  \emph {et~al.}}]{Borsato:2021aum}%
  \BibitemOpen
  \bibfield  {author} {\bibinfo {author} {\bibfnamefont {M.}~\bibnamefont
  {Borsato}} \emph {et~al.},\ }\href {\doibase 10.1088/1361-6633/ac4649}
  {\bibfield  {journal} {\bibinfo  {journal} {Rept. Prog. Phys.}\ }\textbf
  {\bibinfo {volume} {85}},\ \bibinfo {pages} {024201} (\bibinfo {year}
  {2022})},\ \Eprint {http://arxiv.org/abs/2105.12668} {arXiv:2105.12668
  [hep-ph]} \BibitemShut {NoStop}%
\bibitem [{\citenamefont {Tumasyan}\ \emph
  {et~al.}(2022{\natexlab{a}})\citenamefont {Tumasyan} \emph
  {et~al.}}]{CMS:2021kdm}%
  \BibitemOpen
  \bibfield  {author} {\bibinfo {author} {\bibfnamefont {A.}~\bibnamefont
  {Tumasyan}} \emph {et~al.} (\bibinfo {collaboration} {CMS}),\ }\href
  {\doibase 10.1140/epjc/s10052-022-10027-3} {\bibfield  {journal} {\bibinfo
  {journal} {Eur. Phys. J. C}\ }\textbf {\bibinfo {volume} {82}},\ \bibinfo
  {pages} {153} (\bibinfo {year} {2022}{\natexlab{a}})},\ \Eprint
  {http://arxiv.org/abs/2110.04809} {arXiv:2110.04809 [hep-ex]} \BibitemShut
  {NoStop}%
\bibitem [{\citenamefont {Tumasyan}\ \emph
  {et~al.}(2022{\natexlab{b}})\citenamefont {Tumasyan} \emph
  {et~al.}}]{CMS:2021sch}%
  \BibitemOpen
  \bibfield  {author} {\bibinfo {author} {\bibfnamefont {A.}~\bibnamefont
  {Tumasyan}} \emph {et~al.} (\bibinfo {collaboration} {CMS}),\ }\href
  {\doibase 10.1007/JHEP04(2022)062} {\bibfield  {journal} {\bibinfo  {journal}
  {JHEP}\ }\textbf {\bibinfo {volume} {04}},\ \bibinfo {pages} {062} (\bibinfo
  {year} {2022}{\natexlab{b}})},\ \Eprint {http://arxiv.org/abs/2112.13769}
  {arXiv:2112.13769 [hep-ex]} \BibitemShut {NoStop}%
\bibitem [{\citenamefont {Sirunyan}\ \emph
  {et~al.}(2021{\natexlab{b}})\citenamefont {Sirunyan} \emph
  {et~al.}}]{CMS:2021tkn}%
  \BibitemOpen
  \bibfield  {author} {\bibinfo {author} {\bibfnamefont {A.~M.}\ \bibnamefont
  {Sirunyan}} \emph {et~al.} (\bibinfo {collaboration} {CMS}),\ }\href
  {\doibase 10.1103/PhysRevD.104.052011} {\bibfield  {journal} {\bibinfo
  {journal} {Phys. Rev. D}\ }\textbf {\bibinfo {volume} {104}},\ \bibinfo
  {pages} {052011} (\bibinfo {year} {2021}{\natexlab{b}})},\ \Eprint
  {http://arxiv.org/abs/2104.13474} {arXiv:2104.13474 [hep-ex]} \BibitemShut
  {NoStop}%
\bibitem [{\citenamefont {Aaij}\ \emph {et~al.}(2022)\citenamefont {Aaij} \emph
  {et~al.}}]{LHCb:2021dyu}%
  \BibitemOpen
  \bibfield  {author} {\bibinfo {author} {\bibfnamefont {R.}~\bibnamefont
  {Aaij}} \emph {et~al.} (\bibinfo {collaboration} {LHCb}),\ }\href {\doibase
  10.1140/epjc/s10052-022-10186-3} {\bibfield  {journal} {\bibinfo  {journal}
  {Eur. Phys. J. C}\ }\textbf {\bibinfo {volume} {82}},\ \bibinfo {pages} {373}
  (\bibinfo {year} {2022})},\ \Eprint {http://arxiv.org/abs/2110.07293}
  {arXiv:2110.07293 [hep-ex]} \BibitemShut {NoStop}%
\bibitem [{\citenamefont {Aaij}\ \emph {et~al.}(2021)\citenamefont {Aaij} \emph
  {et~al.}}]{LHCb:2020akw}%
  \BibitemOpen
  \bibfield  {author} {\bibinfo {author} {\bibfnamefont {R.}~\bibnamefont
  {Aaij}} \emph {et~al.} (\bibinfo {collaboration} {LHCb}),\ }\href {\doibase
  10.1140/epjc/s10052-021-08994-0} {\bibfield  {journal} {\bibinfo  {journal}
  {Eur. Phys. J. C}\ }\textbf {\bibinfo {volume} {81}},\ \bibinfo {pages} {261}
  (\bibinfo {year} {2021})},\ \Eprint {http://arxiv.org/abs/2012.02696}
  {arXiv:2012.02696 [hep-ex]} \BibitemShut {NoStop}%
\bibitem [{\citenamefont {Knapen}\ and\ \citenamefont
  {Lowette}(2022)}]{Knapen:2022afb}%
  \BibitemOpen
  \bibfield  {author} {\bibinfo {author} {\bibfnamefont {S.}~\bibnamefont
  {Knapen}}\ and\ \bibinfo {author} {\bibfnamefont {S.}~\bibnamefont
  {Lowette}},\ }\href@noop {} {\  (\bibinfo {year} {2022})},\ \Eprint
  {http://arxiv.org/abs/2212.03883} {arXiv:2212.03883 [hep-ph]} \BibitemShut
  {NoStop}%
\bibitem [{\citenamefont {{ATLAS
  Collaboration}}(2022{\natexlab{a}})}]{ATLAS:2022vhr}%
  \BibitemOpen
  \bibfield  {author} {\bibinfo {author} {\bibnamefont {{ATLAS
  Collaboration}}},\ }\href@noop {} {\  (\bibinfo {year}
  {2022}{\natexlab{a}})},\ \Eprint {http://arxiv.org/abs/2209.01029}
  {arXiv:2209.01029 [hep-ex]} \BibitemShut {NoStop}%
\bibitem [{\citenamefont {Aad}\ \emph {et~al.}(2022{\natexlab{a}})\citenamefont
  {Aad} \emph {et~al.}}]{ATLAS:2022zhj}%
  \BibitemOpen
  \bibfield  {author} {\bibinfo {author} {\bibfnamefont {G.}~\bibnamefont
  {Aad}} \emph {et~al.} (\bibinfo {collaboration} {ATLAS}),\ }\href {\doibase
  10.1007/JHEP06(2022)005} {\bibfield  {journal} {\bibinfo  {journal} {JHEP}\
  }\textbf {\bibinfo {volume} {06}},\ \bibinfo {pages} {005} (\bibinfo {year}
  {2022}{\natexlab{a}})},\ \Eprint {http://arxiv.org/abs/2203.01009}
  {arXiv:2203.01009 [hep-ex]} \BibitemShut {NoStop}%
\bibitem [{\citenamefont {Aad}\ \emph {et~al.}(2022{\natexlab{b}})\citenamefont
  {Aad} \emph {et~al.}}]{ATLAS:2022gbw}%
  \BibitemOpen
  \bibfield  {author} {\bibinfo {author} {\bibfnamefont {G.}~\bibnamefont
  {Aad}} \emph {et~al.} (\bibinfo {collaboration} {ATLAS}),\ }\href {\doibase
  10.1103/PhysRevD.106.032005} {\bibfield  {journal} {\bibinfo  {journal}
  {Phys. Rev. D}\ }\textbf {\bibinfo {volume} {106}},\ \bibinfo {pages}
  {032005} (\bibinfo {year} {2022}{\natexlab{b}})},\ \Eprint
  {http://arxiv.org/abs/2203.00587} {arXiv:2203.00587 [hep-ex]} \BibitemShut
  {NoStop}%
\bibitem [{\citenamefont {Tumasyan}\ \emph
  {et~al.}(2022{\natexlab{c}})\citenamefont {Tumasyan} \emph
  {et~al.}}]{CMS:2022fut}%
  \BibitemOpen
  \bibfield  {author} {\bibinfo {author} {\bibfnamefont {A.}~\bibnamefont
  {Tumasyan}} \emph {et~al.} (\bibinfo {collaboration} {CMS}),\ }\href
  {\doibase 10.1007/JHEP07(2022)081} {\bibfield  {journal} {\bibinfo  {journal}
  {JHEP}\ }\textbf {\bibinfo {volume} {07}},\ \bibinfo {pages} {081} (\bibinfo
  {year} {2022}{\natexlab{c}})},\ \Eprint {http://arxiv.org/abs/2201.05578}
  {arXiv:2201.05578 [hep-ex]} \BibitemShut {NoStop}%
\bibitem [{\citenamefont {{CMS
  Collaboration}}(2022{\natexlab{a}})}]{CMS:2022qej}%
  \BibitemOpen
  \bibfield  {author} {\bibinfo {author} {\bibnamefont {{CMS Collaboration}}},\
  }\href@noop {} {\  (\bibinfo {year} {2022}{\natexlab{a}})},\ \Eprint
  {http://arxiv.org/abs/2205.08582} {arXiv:2205.08582 [hep-ex]} \BibitemShut
  {NoStop}%
\bibitem [{\citenamefont {{ ATLAS Collaboration}}(2023)}]{ATLAS:2023oti}%
  \BibitemOpen
  \bibfield  {author} {\bibinfo {author} {\bibnamefont {{ ATLAS
  Collaboration}}},\ }\href@noop {} {\  (\bibinfo {year} {2023})},\ \Eprint
  {http://arxiv.org/abs/2301.13866} {arXiv:2301.13866 [hep-ex]} \BibitemShut
  {NoStop}%
\bibitem [{\citenamefont {Gershtein}(2017)}]{Gershtein:2017tsv}%
  \BibitemOpen
  \bibfield  {author} {\bibinfo {author} {\bibfnamefont {Y.}~\bibnamefont
  {Gershtein}},\ }\href {\doibase 10.1103/PhysRevD.96.035027} {\bibfield
  {journal} {\bibinfo  {journal} {Phys. Rev. D}\ }\textbf {\bibinfo {volume}
  {96}},\ \bibinfo {pages} {035027} (\bibinfo {year} {2017})},\ \Eprint
  {http://arxiv.org/abs/1705.04321} {arXiv:1705.04321 [hep-ph]} \BibitemShut
  {NoStop}%
\bibitem [{\citenamefont {Gershtein}\ and\ \citenamefont
  {Knapen}(2020)}]{Gershtein:2019dhy}%
  \BibitemOpen
  \bibfield  {author} {\bibinfo {author} {\bibfnamefont {Y.}~\bibnamefont
  {Gershtein}}\ and\ \bibinfo {author} {\bibfnamefont {S.}~\bibnamefont
  {Knapen}},\ }\href {\doibase 10.1103/PhysRevD.101.032003} {\bibfield
  {journal} {\bibinfo  {journal} {Phys. Rev. D}\ }\textbf {\bibinfo {volume}
  {101}},\ \bibinfo {pages} {032003} (\bibinfo {year} {2020})},\ \Eprint
  {http://arxiv.org/abs/1907.00007} {arXiv:1907.00007 [hep-ex]} \BibitemShut
  {NoStop}%
\bibitem [{\citenamefont {Bernreuther}\ \emph {et~al.}(2021)\citenamefont
  {Bernreuther}, \citenamefont {Mejia}, \citenamefont {Kahlhoefer},
  \citenamefont {Kr\"amer},\ and\ \citenamefont
  {Tunney}}]{Bernreuther:2020xus}%
  \BibitemOpen
  \bibfield  {author} {\bibinfo {author} {\bibfnamefont {E.}~\bibnamefont
  {Bernreuther}}, \bibinfo {author} {\bibfnamefont {J.~C.}\ \bibnamefont
  {Mejia}}, \bibinfo {author} {\bibfnamefont {F.}~\bibnamefont {Kahlhoefer}},
  \bibinfo {author} {\bibfnamefont {M.}~\bibnamefont {Kr\"amer}}, \ and\
  \bibinfo {author} {\bibfnamefont {P.}~\bibnamefont {Tunney}},\ }\href
  {\doibase 10.1007/JHEP04(2021)210} {\bibfield  {journal} {\bibinfo  {journal}
  {JHEP}\ }\textbf {\bibinfo {volume} {04}},\ \bibinfo {pages} {210} (\bibinfo
  {year} {2021})},\ \Eprint {http://arxiv.org/abs/2011.06604} {arXiv:2011.06604
  [hep-ph]} \BibitemShut {NoStop}%
\bibitem [{\citenamefont {Bhattacherjee}\ \emph {et~al.}(2020)\citenamefont
  {Bhattacherjee}, \citenamefont {Mukherjee}, \citenamefont {Sengupta},\ and\
  \citenamefont {Solanki}}]{Bhattacherjee:2020nno}%
  \BibitemOpen
  \bibfield  {author} {\bibinfo {author} {\bibfnamefont {B.}~\bibnamefont
  {Bhattacherjee}}, \bibinfo {author} {\bibfnamefont {S.}~\bibnamefont
  {Mukherjee}}, \bibinfo {author} {\bibfnamefont {R.}~\bibnamefont {Sengupta}},
  \ and\ \bibinfo {author} {\bibfnamefont {P.}~\bibnamefont {Solanki}},\ }\href
  {\doibase 10.1007/JHEP08(2020)141} {\bibfield  {journal} {\bibinfo  {journal}
  {JHEP}\ }\textbf {\bibinfo {volume} {08}},\ \bibinfo {pages} {141} (\bibinfo
  {year} {2020})},\ \Eprint {http://arxiv.org/abs/2003.03943} {arXiv:2003.03943
  [hep-ph]} \BibitemShut {NoStop}%
\bibitem [{\citenamefont {Bhattacherjee}\ \emph
  {et~al.}(2022{\natexlab{a}})\citenamefont {Bhattacherjee}, \citenamefont
  {Ghosh}, \citenamefont {Sengupta},\ and\ \citenamefont
  {Solanki}}]{Bhattacherjee:2021qaa}%
  \BibitemOpen
  \bibfield  {author} {\bibinfo {author} {\bibfnamefont {B.}~\bibnamefont
  {Bhattacherjee}}, \bibinfo {author} {\bibfnamefont {T.}~\bibnamefont
  {Ghosh}}, \bibinfo {author} {\bibfnamefont {R.}~\bibnamefont {Sengupta}}, \
  and\ \bibinfo {author} {\bibfnamefont {P.}~\bibnamefont {Solanki}},\ }\href
  {\doibase 10.1007/JHEP08(2022)254} {\bibfield  {journal} {\bibinfo  {journal}
  {JHEP}\ }\textbf {\bibinfo {volume} {08}},\ \bibinfo {pages} {254} (\bibinfo
  {year} {2022}{\natexlab{a}})},\ \Eprint {http://arxiv.org/abs/2112.04518}
  {arXiv:2112.04518 [hep-ph]} \BibitemShut {NoStop}%
\bibitem [{\citenamefont {Bhattacherjee}\ \emph
  {et~al.}(2022{\natexlab{b}})\citenamefont {Bhattacherjee}, \citenamefont
  {Matsumoto},\ and\ \citenamefont {Sengupta}}]{Bhattacherjee:2021rml}%
  \BibitemOpen
  \bibfield  {author} {\bibinfo {author} {\bibfnamefont {B.}~\bibnamefont
  {Bhattacherjee}}, \bibinfo {author} {\bibfnamefont {S.}~\bibnamefont
  {Matsumoto}}, \ and\ \bibinfo {author} {\bibfnamefont {R.}~\bibnamefont
  {Sengupta}},\ }\href {\doibase 10.1103/PhysRevD.106.095018} {\bibfield
  {journal} {\bibinfo  {journal} {Phys. Rev. D}\ }\textbf {\bibinfo {volume}
  {106}},\ \bibinfo {pages} {095018} (\bibinfo {year} {2022}{\natexlab{b}})},\
  \Eprint {http://arxiv.org/abs/2111.02437} {arXiv:2111.02437 [hep-ph]}
  \BibitemShut {NoStop}%
\bibitem [{\citenamefont {Tumasyan}\ \emph {et~al.}(2021)\citenamefont
  {Tumasyan} \emph {et~al.}}]{CMS:2021juv}%
  \BibitemOpen
  \bibfield  {author} {\bibinfo {author} {\bibfnamefont {A.}~\bibnamefont
  {Tumasyan}} \emph {et~al.} (\bibinfo {collaboration} {CMS}),\ }\href
  {\doibase 10.1103/PhysRevLett.127.261804} {\bibfield  {journal} {\bibinfo
  {journal} {Phys. Rev. Lett.}\ }\textbf {\bibinfo {volume} {127}},\ \bibinfo
  {pages} {261804} (\bibinfo {year} {2021})},\ \Eprint
  {http://arxiv.org/abs/2107.04838} {arXiv:2107.04838 [hep-ex]} \BibitemShut
  {NoStop}%
\bibitem [{\citenamefont {Chou}\ \emph {et~al.}(2017)\citenamefont {Chou},
  \citenamefont {Curtin},\ and\ \citenamefont {Lubatti}}]{Chou:2016lxi}%
  \BibitemOpen
  \bibfield  {author} {\bibinfo {author} {\bibfnamefont {J.~P.}\ \bibnamefont
  {Chou}}, \bibinfo {author} {\bibfnamefont {D.}~\bibnamefont {Curtin}}, \ and\
  \bibinfo {author} {\bibfnamefont {H.~J.}\ \bibnamefont {Lubatti}},\ }\href
  {\doibase 10.1016/j.physletb.2017.01.043} {\bibfield  {journal} {\bibinfo
  {journal} {Phys. Lett. B}\ }\textbf {\bibinfo {volume} {767}},\ \bibinfo
  {pages} {29} (\bibinfo {year} {2017})},\ \Eprint
  {http://arxiv.org/abs/1606.06298} {arXiv:1606.06298 [hep-ph]} \BibitemShut
  {NoStop}%
\bibitem [{\citenamefont {Curtin}\ \emph {et~al.}(2019)\citenamefont {Curtin}
  \emph {et~al.}}]{Curtin:2018mvb}%
  \BibitemOpen
  \bibfield  {author} {\bibinfo {author} {\bibfnamefont {D.}~\bibnamefont
  {Curtin}} \emph {et~al.},\ }\href {\doibase 10.1088/1361-6633/ab28d6}
  {\bibfield  {journal} {\bibinfo  {journal} {Rept. Prog. Phys.}\ }\textbf
  {\bibinfo {volume} {82}},\ \bibinfo {pages} {116201} (\bibinfo {year}
  {2019})},\ \Eprint {http://arxiv.org/abs/1806.07396} {arXiv:1806.07396
  [hep-ph]} \BibitemShut {NoStop}%
\bibitem [{\citenamefont {Alpigiani}\ \emph {et~al.}(2018)\citenamefont
  {Alpigiani} \emph {et~al.}}]{MATHUSLA:2018bqv}%
  \BibitemOpen
  \bibfield  {author} {\bibinfo {author} {\bibfnamefont {C.}~\bibnamefont
  {Alpigiani}} \emph {et~al.} (\bibinfo {collaboration} {MATHUSLA}),\
  }\href@noop {} {\  (\bibinfo {year} {2018})},\ \Eprint
  {http://arxiv.org/abs/1811.00927} {arXiv:1811.00927 [physics.ins-det]}
  \BibitemShut {NoStop}%
\bibitem [{\citenamefont {Feng}\ \emph
  {et~al.}(2018{\natexlab{a}})\citenamefont {Feng}, \citenamefont {Galon},
  \citenamefont {Kling},\ and\ \citenamefont {Trojanowski}}]{Feng:2017uoz}%
  \BibitemOpen
  \bibfield  {author} {\bibinfo {author} {\bibfnamefont {J.~L.}\ \bibnamefont
  {Feng}}, \bibinfo {author} {\bibfnamefont {I.}~\bibnamefont {Galon}},
  \bibinfo {author} {\bibfnamefont {F.}~\bibnamefont {Kling}}, \ and\ \bibinfo
  {author} {\bibfnamefont {S.}~\bibnamefont {Trojanowski}},\ }\href {\doibase
  10.1103/PhysRevD.97.035001} {\bibfield  {journal} {\bibinfo  {journal} {Phys.
  Rev. D}\ }\textbf {\bibinfo {volume} {97}},\ \bibinfo {pages} {035001}
  (\bibinfo {year} {2018}{\natexlab{a}})},\ \Eprint
  {http://arxiv.org/abs/1708.09389} {arXiv:1708.09389 [hep-ph]} \BibitemShut
  {NoStop}%
\bibitem [{\citenamefont {Ariga}\ \emph {et~al.}(2019)\citenamefont {Ariga}
  \emph {et~al.}}]{FASER:2018eoc}%
  \BibitemOpen
  \bibfield  {author} {\bibinfo {author} {\bibfnamefont {A.}~\bibnamefont
  {Ariga}} \emph {et~al.} (\bibinfo {collaboration} {FASER}),\ }\href {\doibase
  10.1103/PhysRevD.99.095011} {\bibfield  {journal} {\bibinfo  {journal} {Phys.
  Rev. D}\ }\textbf {\bibinfo {volume} {99}},\ \bibinfo {pages} {095011}
  (\bibinfo {year} {2019})},\ \Eprint {http://arxiv.org/abs/1811.12522}
  {arXiv:1811.12522 [hep-ph]} \BibitemShut {NoStop}%
\bibitem [{\citenamefont {Gligorov}\ \emph {et~al.}(2018)\citenamefont
  {Gligorov}, \citenamefont {Knapen}, \citenamefont {Papucci},\ and\
  \citenamefont {Robinson}}]{Gligorov:2017nwh}%
  \BibitemOpen
  \bibfield  {author} {\bibinfo {author} {\bibfnamefont {V.~V.}\ \bibnamefont
  {Gligorov}}, \bibinfo {author} {\bibfnamefont {S.}~\bibnamefont {Knapen}},
  \bibinfo {author} {\bibfnamefont {M.}~\bibnamefont {Papucci}}, \ and\
  \bibinfo {author} {\bibfnamefont {D.~J.}\ \bibnamefont {Robinson}},\ }\href
  {\doibase 10.1103/PhysRevD.97.015023} {\bibfield  {journal} {\bibinfo
  {journal} {Phys. Rev. D}\ }\textbf {\bibinfo {volume} {97}},\ \bibinfo
  {pages} {015023} (\bibinfo {year} {2018})},\ \Eprint
  {http://arxiv.org/abs/1708.09395} {arXiv:1708.09395 [hep-ph]} \BibitemShut
  {NoStop}%
\bibitem [{\citenamefont {Anelli}\ \emph {et~al.}(2015)\citenamefont {Anelli}
  \emph {et~al.}}]{SHiP:2015vad}%
  \BibitemOpen
  \bibfield  {author} {\bibinfo {author} {\bibfnamefont {M.}~\bibnamefont
  {Anelli}} \emph {et~al.} (\bibinfo {collaboration} {SHiP}),\ }\href@noop {}
  {\  (\bibinfo {year} {2015})},\ \Eprint {http://arxiv.org/abs/1504.04956}
  {arXiv:1504.04956 [physics.ins-det]} \BibitemShut {NoStop}%
\bibitem [{\citenamefont {Bauer}\ \emph {et~al.}(2019)\citenamefont {Bauer},
  \citenamefont {Brandt}, \citenamefont {Lee},\ and\ \citenamefont
  {Ohm}}]{Bauer:2019vqk}%
  \BibitemOpen
  \bibfield  {author} {\bibinfo {author} {\bibfnamefont {M.}~\bibnamefont
  {Bauer}}, \bibinfo {author} {\bibfnamefont {O.}~\bibnamefont {Brandt}},
  \bibinfo {author} {\bibfnamefont {L.}~\bibnamefont {Lee}}, \ and\ \bibinfo
  {author} {\bibfnamefont {C.}~\bibnamefont {Ohm}},\ }\href@noop {} {\
  (\bibinfo {year} {2019})},\ \Eprint {http://arxiv.org/abs/1909.13022}
  {arXiv:1909.13022 [physics.ins-det]} \BibitemShut {NoStop}%
\bibitem [{\citenamefont {Feng}\ \emph {et~al.}(2022)\citenamefont {Feng} \emph
  {et~al.}}]{Feng:2022inv}%
  \BibitemOpen
  \bibfield  {author} {\bibinfo {author} {\bibfnamefont {J.~L.}\ \bibnamefont
  {Feng}} \emph {et~al.},\ }\href@noop {} {\  (\bibinfo {year} {2022})},\
  \Eprint {http://arxiv.org/abs/2203.05090} {arXiv:2203.05090 [hep-ex]}
  \BibitemShut {NoStop}%
\bibitem [{\citenamefont {Gligorov}\ \emph {et~al.}(2019)\citenamefont
  {Gligorov}, \citenamefont {Knapen}, \citenamefont {Nachman}, \citenamefont
  {Papucci},\ and\ \citenamefont {Robinson}}]{Gligorov:2018vkc}%
  \BibitemOpen
  \bibfield  {author} {\bibinfo {author} {\bibfnamefont {V.~V.}\ \bibnamefont
  {Gligorov}}, \bibinfo {author} {\bibfnamefont {S.}~\bibnamefont {Knapen}},
  \bibinfo {author} {\bibfnamefont {B.}~\bibnamefont {Nachman}}, \bibinfo
  {author} {\bibfnamefont {M.}~\bibnamefont {Papucci}}, \ and\ \bibinfo
  {author} {\bibfnamefont {D.~J.}\ \bibnamefont {Robinson}},\ }\href {\doibase
  10.1103/PhysRevD.99.015023} {\bibfield  {journal} {\bibinfo  {journal} {Phys.
  Rev. D}\ }\textbf {\bibinfo {volume} {99}},\ \bibinfo {pages} {015023}
  (\bibinfo {year} {2019})},\ \Eprint {http://arxiv.org/abs/1810.03636}
  {arXiv:1810.03636 [hep-ph]} \BibitemShut {NoStop}%
\bibitem [{\citenamefont {Aielli}\ \emph {et~al.}(2020)\citenamefont {Aielli}
  \emph {et~al.}}]{Aielli:2019ivi}%
  \BibitemOpen
  \bibfield  {author} {\bibinfo {author} {\bibfnamefont {G.}~\bibnamefont
  {Aielli}} \emph {et~al.},\ }\href {\doibase 10.1140/epjc/s10052-020-08711-3}
  {\bibfield  {journal} {\bibinfo  {journal} {Eur. Phys. J. C}\ }\textbf
  {\bibinfo {volume} {80}},\ \bibinfo {pages} {1177} (\bibinfo {year}
  {2020})},\ \Eprint {http://arxiv.org/abs/1911.00481} {arXiv:1911.00481
  [hep-ex]} \BibitemShut {NoStop}%
\bibitem [{\citenamefont {Dreyer}\ \emph {et~al.}(2021)\citenamefont {Dreyer}
  \emph {et~al.}}]{Dreyer:2021aqd}%
  \BibitemOpen
  \bibfield  {author} {\bibinfo {author} {\bibfnamefont {S.}~\bibnamefont
  {Dreyer}} \emph {et~al.},\ }\href@noop {} {\  (\bibinfo {year} {2021})},\
  \Eprint {http://arxiv.org/abs/2105.12962} {arXiv:2105.12962 [hep-ph]}
  \BibitemShut {NoStop}%
\bibitem [{\citenamefont {Haas}\ \emph {et~al.}(2015)\citenamefont {Haas},
  \citenamefont {Hill}, \citenamefont {Izaguirre},\ and\ \citenamefont
  {Yavin}}]{Haas:2014dda}%
  \BibitemOpen
  \bibfield  {author} {\bibinfo {author} {\bibfnamefont {A.}~\bibnamefont
  {Haas}}, \bibinfo {author} {\bibfnamefont {C.~S.}\ \bibnamefont {Hill}},
  \bibinfo {author} {\bibfnamefont {E.}~\bibnamefont {Izaguirre}}, \ and\
  \bibinfo {author} {\bibfnamefont {I.}~\bibnamefont {Yavin}},\ }\href
  {\doibase 10.1016/j.physletb.2015.04.062} {\bibfield  {journal} {\bibinfo
  {journal} {Phys. Lett. B}\ }\textbf {\bibinfo {volume} {746}},\ \bibinfo
  {pages} {117} (\bibinfo {year} {2015})},\ \Eprint
  {http://arxiv.org/abs/1410.6816} {arXiv:1410.6816 [hep-ph]} \BibitemShut
  {NoStop}%
\bibitem [{\citenamefont {Cerci}\ \emph {et~al.}(2022)\citenamefont {Cerci}
  \emph {et~al.}}]{Cerci:2021nlb}%
  \BibitemOpen
  \bibfield  {author} {\bibinfo {author} {\bibfnamefont {S.}~\bibnamefont
  {Cerci}} \emph {et~al.},\ }\href {\doibase 10.1007/JHEP06(2022)110}
  {\bibfield  {journal} {\bibinfo  {journal} {JHEP}\ }\textbf {\bibinfo
  {volume} {2022}},\ \bibinfo {pages} {110} (\bibinfo {year} {2022})},\ \Eprint
  {http://arxiv.org/abs/2201.00019} {arXiv:2201.00019 [hep-ex]} \BibitemShut
  {NoStop}%
\bibitem [{\citenamefont {Ahdida}\ \emph {et~al.}(2020)\citenamefont {Ahdida}
  \emph {et~al.}}]{SHiP:2020sos}%
  \BibitemOpen
  \bibfield  {author} {\bibinfo {author} {\bibfnamefont {C.}~\bibnamefont
  {Ahdida}} \emph {et~al.} (\bibinfo {collaboration} {SHiP}),\ }\href@noop {}
  {\  (\bibinfo {year} {2020})},\ \Eprint {http://arxiv.org/abs/2002.08722}
  {arXiv:2002.08722 [physics.ins-det]} \BibitemShut {NoStop}%
\bibitem [{\citenamefont {Boyarsky}\ \emph {et~al.}(2022)\citenamefont
  {Boyarsky}, \citenamefont {Mikulenko}, \citenamefont {Ovchynnikov},\ and\
  \citenamefont {Shchutska}}]{Boyarsky:2021moj}%
  \BibitemOpen
  \bibfield  {author} {\bibinfo {author} {\bibfnamefont {A.}~\bibnamefont
  {Boyarsky}}, \bibinfo {author} {\bibfnamefont {O.}~\bibnamefont {Mikulenko}},
  \bibinfo {author} {\bibfnamefont {M.}~\bibnamefont {Ovchynnikov}}, \ and\
  \bibinfo {author} {\bibfnamefont {L.}~\bibnamefont {Shchutska}},\ }\href
  {\doibase 10.1007/JHEP03(2022)006} {\bibfield  {journal} {\bibinfo  {journal}
  {JHEP}\ }\textbf {\bibinfo {volume} {03}},\ \bibinfo {pages} {006} (\bibinfo
  {year} {2022})},\ \Eprint {http://arxiv.org/abs/2104.09688} {arXiv:2104.09688
  [hep-ph]} \BibitemShut {NoStop}%
\bibitem [{\citenamefont {Acharya}\ \emph {et~al.}(2022)\citenamefont {Acharya}
  \emph {et~al.}}]{Acharya:2022nik}%
  \BibitemOpen
  \bibfield  {author} {\bibinfo {author} {\bibfnamefont {B.}~\bibnamefont
  {Acharya}} \emph {et~al.},\ }in\ \href@noop {} {\emph {\bibinfo {booktitle}
  {{2022 Snowmass Summer Study}}}}\ (\bibinfo {year} {2022})\ \Eprint
  {http://arxiv.org/abs/2209.03988} {arXiv:2209.03988 [hep-ph]} \BibitemShut
  {NoStop}%
\bibitem [{\citenamefont {{CMS Collaboration}}(2021)}]{hepdata.104408.v2}%
  \BibitemOpen
  \bibfield  {author} {\bibinfo {author} {\bibnamefont {{CMS Collaboration}}},\
  }\href@noop {} {\  (\bibinfo {year} {2021})},\ \bibinfo {note}
  {\url{https://doi.org/10.17182/hepdata.104408.v2}}\BibitemShut {NoStop}%
\bibitem [{\citenamefont {Ester}\ \emph {et~al.}(1996)\citenamefont {Ester},
  \citenamefont {Kriegel}, \citenamefont {Sander},\ and\ \citenamefont
  {Xu}}]{dbscan}%
  \BibitemOpen
  \bibfield  {author} {\bibinfo {author} {\bibfnamefont {M.}~\bibnamefont
  {Ester}}, \bibinfo {author} {\bibfnamefont {H.-P.}\ \bibnamefont {Kriegel}},
  \bibinfo {author} {\bibfnamefont {J.}~\bibnamefont {Sander}}, \ and\ \bibinfo
  {author} {\bibfnamefont {X.}~\bibnamefont {Xu}},\ }\href@noop {} {\bibfield
  {journal} {\bibinfo  {journal} {Data Mining and Knowledge Discovery}\ ,\
  \bibinfo {pages} {226}} (\bibinfo {year} {1996})}\BibitemShut {NoStop}%
\bibitem [{\citenamefont {Alwall}\ \emph {et~al.}(2014)\citenamefont {Alwall},
  \citenamefont {Frederix}, \citenamefont {Frixione}, \citenamefont {Hirschi},
  \citenamefont {Maltoni}, \citenamefont {Mattelaer}, \citenamefont {Shao},
  \citenamefont {Stelzer}, \citenamefont {Torrielli},\ and\ \citenamefont
  {Zaro}}]{Alwall:2014hca}%
  \BibitemOpen
  \bibfield  {author} {\bibinfo {author} {\bibfnamefont {J.}~\bibnamefont
  {Alwall}}, \bibinfo {author} {\bibfnamefont {R.}~\bibnamefont {Frederix}},
  \bibinfo {author} {\bibfnamefont {S.}~\bibnamefont {Frixione}}, \bibinfo
  {author} {\bibfnamefont {V.}~\bibnamefont {Hirschi}}, \bibinfo {author}
  {\bibfnamefont {F.}~\bibnamefont {Maltoni}}, \bibinfo {author} {\bibfnamefont
  {O.}~\bibnamefont {Mattelaer}}, \bibinfo {author} {\bibfnamefont {H.~S.}\
  \bibnamefont {Shao}}, \bibinfo {author} {\bibfnamefont {T.}~\bibnamefont
  {Stelzer}}, \bibinfo {author} {\bibfnamefont {P.}~\bibnamefont {Torrielli}},
  \ and\ \bibinfo {author} {\bibfnamefont {M.}~\bibnamefont {Zaro}},\ }\href
  {\doibase 10.1007/JHEP07(2014)079} {\bibfield  {journal} {\bibinfo  {journal}
  {JHEP}\ }\textbf {\bibinfo {volume} {07}},\ \bibinfo {pages} {079} (\bibinfo
  {year} {2014})},\ \Eprint {http://arxiv.org/abs/1405.0301} {arXiv:1405.0301
  [hep-ph]} \BibitemShut {NoStop}%
\bibitem [{\citenamefont {Sj\"ostrand}\ \emph {et~al.}(2015)\citenamefont
  {Sj\"ostrand}, \citenamefont {Ask}, \citenamefont {Christiansen},
  \citenamefont {Corke}, \citenamefont {Desai}, \citenamefont {Ilten},
  \citenamefont {Mrenna}, \citenamefont {Prestel}, \citenamefont {Rasmussen},\
  and\ \citenamefont {Skands}}]{Sjostrand:2014zea}%
  \BibitemOpen
  \bibfield  {author} {\bibinfo {author} {\bibfnamefont {T.}~\bibnamefont
  {Sj\"ostrand}}, \bibinfo {author} {\bibfnamefont {S.}~\bibnamefont {Ask}},
  \bibinfo {author} {\bibfnamefont {J.~R.}\ \bibnamefont {Christiansen}},
  \bibinfo {author} {\bibfnamefont {R.}~\bibnamefont {Corke}}, \bibinfo
  {author} {\bibfnamefont {N.}~\bibnamefont {Desai}}, \bibinfo {author}
  {\bibfnamefont {P.}~\bibnamefont {Ilten}}, \bibinfo {author} {\bibfnamefont
  {S.}~\bibnamefont {Mrenna}}, \bibinfo {author} {\bibfnamefont
  {S.}~\bibnamefont {Prestel}}, \bibinfo {author} {\bibfnamefont {C.~O.}\
  \bibnamefont {Rasmussen}}, \ and\ \bibinfo {author} {\bibfnamefont {P.~Z.}\
  \bibnamefont {Skands}},\ }\href {\doibase 10.1016/j.cpc.2015.01.024}
  {\bibfield  {journal} {\bibinfo  {journal} {Comput. Phys. Commun.}\ }\textbf
  {\bibinfo {volume} {191}},\ \bibinfo {pages} {159} (\bibinfo {year}
  {2015})},\ \Eprint {http://arxiv.org/abs/1410.3012} {arXiv:1410.3012
  [hep-ph]} \BibitemShut {NoStop}%
\bibitem [{\citenamefont {Mangano}\ \emph {et~al.}(2003)\citenamefont
  {Mangano}, \citenamefont {Moretti}, \citenamefont {Piccinini}, \citenamefont
  {Pittau},\ and\ \citenamefont {Polosa}}]{Mangano:2002ea}%
  \BibitemOpen
  \bibfield  {author} {\bibinfo {author} {\bibfnamefont {M.~L.}\ \bibnamefont
  {Mangano}}, \bibinfo {author} {\bibfnamefont {M.}~\bibnamefont {Moretti}},
  \bibinfo {author} {\bibfnamefont {F.}~\bibnamefont {Piccinini}}, \bibinfo
  {author} {\bibfnamefont {R.}~\bibnamefont {Pittau}}, \ and\ \bibinfo {author}
  {\bibfnamefont {A.~D.}\ \bibnamefont {Polosa}},\ }\href {\doibase
  10.1088/1126-6708/2003/07/001} {\bibfield  {journal} {\bibinfo  {journal}
  {JHEP}\ }\textbf {\bibinfo {volume} {07}},\ \bibinfo {pages} {001} (\bibinfo
  {year} {2003})},\ \Eprint {http://arxiv.org/abs/hep-ph/0206293}
  {arXiv:hep-ph/0206293} \BibitemShut {NoStop}%
\bibitem [{\citenamefont {Alwall}\ \emph {et~al.}(2009)\citenamefont {Alwall},
  \citenamefont {de~Visscher},\ and\ \citenamefont {Maltoni}}]{Alwall:2008qv}%
  \BibitemOpen
  \bibfield  {author} {\bibinfo {author} {\bibfnamefont {J.}~\bibnamefont
  {Alwall}}, \bibinfo {author} {\bibfnamefont {S.}~\bibnamefont {de~Visscher}},
  \ and\ \bibinfo {author} {\bibfnamefont {F.}~\bibnamefont {Maltoni}},\ }\href
  {\doibase 10.1088/1126-6708/2009/02/017} {\bibfield  {journal} {\bibinfo
  {journal} {JHEP}\ }\textbf {\bibinfo {volume} {02}},\ \bibinfo {pages} {017}
  (\bibinfo {year} {2009})},\ \Eprint {http://arxiv.org/abs/0810.5350}
  {arXiv:0810.5350 [hep-ph]} \BibitemShut {NoStop}%
\bibitem [{\citenamefont {de~Favereau}\ \emph {et~al.}(2014)\citenamefont
  {de~Favereau}, \citenamefont {Delaere}, \citenamefont {Demin}, \citenamefont
  {Giammanco}, \citenamefont {Lema\^\i{}tre}, \citenamefont {Mertens},\ and\
  \citenamefont {Selvaggi}}]{deFavereau:2013fsa}%
  \BibitemOpen
  \bibfield  {author} {\bibinfo {author} {\bibfnamefont {J.}~\bibnamefont
  {de~Favereau}}, \bibinfo {author} {\bibfnamefont {C.}~\bibnamefont
  {Delaere}}, \bibinfo {author} {\bibfnamefont {P.}~\bibnamefont {Demin}},
  \bibinfo {author} {\bibfnamefont {A.}~\bibnamefont {Giammanco}}, \bibinfo
  {author} {\bibfnamefont {V.}~\bibnamefont {Lema\^\i{}tre}}, \bibinfo {author}
  {\bibfnamefont {A.}~\bibnamefont {Mertens}}, \ and\ \bibinfo {author}
  {\bibfnamefont {M.}~\bibnamefont {Selvaggi}} (\bibinfo {collaboration}
  {DELPHES 3}),\ }\href {\doibase 10.1007/JHEP02(2014)057} {\bibfield
  {journal} {\bibinfo  {journal} {JHEP}\ }\textbf {\bibinfo {volume} {02}},\
  \bibinfo {pages} {057} (\bibinfo {year} {2014})},\ \Eprint
  {http://arxiv.org/abs/1307.6346} {arXiv:1307.6346 [hep-ex]} \BibitemShut
  {NoStop}%
\bibitem [{\citenamefont {Wang}(2022)}]{delphes_pr}%
  \BibitemOpen
  \bibfield  {author} {\bibinfo {author} {\bibfnamefont {C.}~\bibnamefont
  {Wang}},\ }\href@noop {} {\enquote {\bibinfo {title} {{Dedicated Delphes
  Module: \url{https://github.com/delphes/delphes/pull/103}}},}\ } (\bibinfo
  {year} {2022})\BibitemShut {NoStop}%
\bibitem [{\citenamefont {{CMS
  Collaboration}}(2022{\natexlab{b}})}]{CMS-DP-2022-062}%
  \BibitemOpen
  \bibfield  {author} {\bibinfo {author} {\bibnamefont {{CMS Collaboration}}},\
  }\href@noop {} {\  (\bibinfo {year} {2022}{\natexlab{b}})},\ \bibinfo {note}
  {\url{https://cds.cern.ch/record/2842376}}\BibitemShut {NoStop}%
\bibitem [{\citenamefont {Cottin}\ \emph {et~al.}(2023)\citenamefont {Cottin},
  \citenamefont {Helo}, \citenamefont {Hirsch}, \citenamefont {Pe\~na},
  \citenamefont {Wang},\ and\ \citenamefont {Xie}}]{Cottin:2022nwp}%
  \BibitemOpen
  \bibfield  {author} {\bibinfo {author} {\bibfnamefont {G.}~\bibnamefont
  {Cottin}}, \bibinfo {author} {\bibfnamefont {J.~C.}\ \bibnamefont {Helo}},
  \bibinfo {author} {\bibfnamefont {M.}~\bibnamefont {Hirsch}}, \bibinfo
  {author} {\bibfnamefont {C.}~\bibnamefont {Pe\~na}}, \bibinfo {author}
  {\bibfnamefont {C.}~\bibnamefont {Wang}}, \ and\ \bibinfo {author}
  {\bibfnamefont {S.}~\bibnamefont {Xie}},\ }\href {\doibase
  10.1007/JHEP02(2023)011} {\bibfield  {journal} {\bibinfo  {journal} {JHEP}\
  }\textbf {\bibinfo {volume} {02}},\ \bibinfo {pages} {011} (\bibinfo {year}
  {2023})},\ \Eprint {http://arxiv.org/abs/2210.17446} {arXiv:2210.17446
  [hep-ph]} \BibitemShut {NoStop}%
\bibitem [{\citenamefont {O'Connell}\ \emph {et~al.}(2007)\citenamefont
  {O'Connell}, \citenamefont {Ramsey-Musolf},\ and\ \citenamefont
  {Wise}}]{OConnell:2006rsp}%
  \BibitemOpen
  \bibfield  {author} {\bibinfo {author} {\bibfnamefont {D.}~\bibnamefont
  {O'Connell}}, \bibinfo {author} {\bibfnamefont {M.~J.}\ \bibnamefont
  {Ramsey-Musolf}}, \ and\ \bibinfo {author} {\bibfnamefont {M.~B.}\
  \bibnamefont {Wise}},\ }\href {\doibase 10.1103/PhysRevD.75.037701}
  {\bibfield  {journal} {\bibinfo  {journal} {Phys. Rev. D}\ }\textbf {\bibinfo
  {volume} {75}},\ \bibinfo {pages} {037701} (\bibinfo {year} {2007})},\
  \Eprint {http://arxiv.org/abs/hep-ph/0611014} {arXiv:hep-ph/0611014}
  \BibitemShut {NoStop}%
\bibitem [{\citenamefont {Winkler}(2019)}]{Winkler:2018qyg}%
  \BibitemOpen
  \bibfield  {author} {\bibinfo {author} {\bibfnamefont {M.~W.}\ \bibnamefont
  {Winkler}},\ }\href {\doibase 10.1103/PhysRevD.99.015018} {\bibfield
  {journal} {\bibinfo  {journal} {Phys. Rev. D}\ }\textbf {\bibinfo {volume}
  {99}},\ \bibinfo {pages} {015018} (\bibinfo {year} {2019})},\ \Eprint
  {http://arxiv.org/abs/1809.01876} {arXiv:1809.01876 [hep-ph]} \BibitemShut
  {NoStop}%
\bibitem [{\citenamefont {Gershtein}\ \emph {et~al.}(2021)\citenamefont
  {Gershtein}, \citenamefont {Knapen},\ and\ \citenamefont
  {Redigolo}}]{Gershtein:2020mwi}%
  \BibitemOpen
  \bibfield  {author} {\bibinfo {author} {\bibfnamefont {Y.}~\bibnamefont
  {Gershtein}}, \bibinfo {author} {\bibfnamefont {S.}~\bibnamefont {Knapen}}, \
  and\ \bibinfo {author} {\bibfnamefont {D.}~\bibnamefont {Redigolo}},\ }\href
  {\doibase 10.1016/j.physletb.2021.136758} {\bibfield  {journal} {\bibinfo
  {journal} {Phys. Lett. B}\ }\textbf {\bibinfo {volume} {823}},\ \bibinfo
  {pages} {136758} (\bibinfo {year} {2021})},\ \Eprint
  {http://arxiv.org/abs/2012.07864} {arXiv:2012.07864 [hep-ph]} \BibitemShut
  {NoStop}%
\bibitem [{\citenamefont {Willey}\ and\ \citenamefont
  {Yu}(1982)}]{Willey:1982ti}%
  \BibitemOpen
  \bibfield  {author} {\bibinfo {author} {\bibfnamefont {R.~S.}\ \bibnamefont
  {Willey}}\ and\ \bibinfo {author} {\bibfnamefont {H.~L.}\ \bibnamefont
  {Yu}},\ }\href {\doibase 10.1103/PhysRevD.26.3287} {\bibfield  {journal}
  {\bibinfo  {journal} {Phys. Rev. D}\ }\textbf {\bibinfo {volume} {26}},\
  \bibinfo {pages} {3287} (\bibinfo {year} {1982})}\BibitemShut {NoStop}%
\bibitem [{\citenamefont {Chivukula}\ and\ \citenamefont
  {Manohar}(1988)}]{Chivukula:1988gp}%
  \BibitemOpen
  \bibfield  {author} {\bibinfo {author} {\bibfnamefont {R.~S.}\ \bibnamefont
  {Chivukula}}\ and\ \bibinfo {author} {\bibfnamefont {A.~V.}\ \bibnamefont
  {Manohar}},\ }\href {\doibase 10.1016/0370-2693(88)90891-X} {\bibfield
  {journal} {\bibinfo  {journal} {Phys. Lett. B}\ }\textbf {\bibinfo {volume}
  {207}},\ \bibinfo {pages} {86} (\bibinfo {year} {1988})},\ \bibinfo {note}
  {[Erratum: Phys.Lett.B 217, 568 (1989)]}\BibitemShut {NoStop}%
\bibitem [{\citenamefont {Grinstein}\ \emph {et~al.}(1988)\citenamefont
  {Grinstein}, \citenamefont {Hall},\ and\ \citenamefont
  {Randall}}]{Grinstein:1988yu}%
  \BibitemOpen
  \bibfield  {author} {\bibinfo {author} {\bibfnamefont {B.}~\bibnamefont
  {Grinstein}}, \bibinfo {author} {\bibfnamefont {L.~J.}\ \bibnamefont {Hall}},
  \ and\ \bibinfo {author} {\bibfnamefont {L.}~\bibnamefont {Randall}},\ }\href
  {\doibase 10.1016/0370-2693(88)90916-1} {\bibfield  {journal} {\bibinfo
  {journal} {Phys. Lett. B}\ }\textbf {\bibinfo {volume} {211}},\ \bibinfo
  {pages} {363} (\bibinfo {year} {1988})}\BibitemShut {NoStop}%
\bibitem [{\citenamefont {Batell}\ \emph {et~al.}(2011)\citenamefont {Batell},
  \citenamefont {Pospelov},\ and\ \citenamefont {Ritz}}]{Batell:2009jf}%
  \BibitemOpen
  \bibfield  {author} {\bibinfo {author} {\bibfnamefont {B.}~\bibnamefont
  {Batell}}, \bibinfo {author} {\bibfnamefont {M.}~\bibnamefont {Pospelov}}, \
  and\ \bibinfo {author} {\bibfnamefont {A.}~\bibnamefont {Ritz}},\ }\href
  {\doibase 10.1103/PhysRevD.83.054005} {\bibfield  {journal} {\bibinfo
  {journal} {Phys. Rev. D}\ }\textbf {\bibinfo {volume} {83}},\ \bibinfo
  {pages} {054005} (\bibinfo {year} {2011})},\ \Eprint
  {http://arxiv.org/abs/0911.4938} {arXiv:0911.4938 [hep-ph]} \BibitemShut
  {NoStop}%
\bibitem [{\citenamefont {Ilten}\ \emph {et~al.}(2018)\citenamefont {Ilten},
  \citenamefont {Soreq}, \citenamefont {Williams},\ and\ \citenamefont
  {Xue}}]{Ilten:2018crw}%
  \BibitemOpen
  \bibfield  {author} {\bibinfo {author} {\bibfnamefont {P.}~\bibnamefont
  {Ilten}}, \bibinfo {author} {\bibfnamefont {Y.}~\bibnamefont {Soreq}},
  \bibinfo {author} {\bibfnamefont {M.}~\bibnamefont {Williams}}, \ and\
  \bibinfo {author} {\bibfnamefont {W.}~\bibnamefont {Xue}},\ }\href {\doibase
  10.1007/JHEP06(2018)004} {\bibfield  {journal} {\bibinfo  {journal} {JHEP}\
  }\textbf {\bibinfo {volume} {06}},\ \bibinfo {pages} {004} (\bibinfo {year}
  {2018})},\ \Eprint {http://arxiv.org/abs/1801.04847} {arXiv:1801.04847
  [hep-ph]} \BibitemShut {NoStop}%
\bibitem [{\citenamefont {Duerr}\ \emph {et~al.}(2021)\citenamefont {Duerr},
  \citenamefont {Ferber}, \citenamefont {Garcia-Cely}, \citenamefont {Hearty},\
  and\ \citenamefont {Schmidt-Hoberg}}]{Duerr:2020muu}%
  \BibitemOpen
  \bibfield  {author} {\bibinfo {author} {\bibfnamefont {M.}~\bibnamefont
  {Duerr}}, \bibinfo {author} {\bibfnamefont {T.}~\bibnamefont {Ferber}},
  \bibinfo {author} {\bibfnamefont {C.}~\bibnamefont {Garcia-Cely}}, \bibinfo
  {author} {\bibfnamefont {C.}~\bibnamefont {Hearty}}, \ and\ \bibinfo {author}
  {\bibfnamefont {K.}~\bibnamefont {Schmidt-Hoberg}},\ }\href {\doibase
  10.1007/JHEP04(2021)146} {\bibfield  {journal} {\bibinfo  {journal} {JHEP}\
  }\textbf {\bibinfo {volume} {04}},\ \bibinfo {pages} {146} (\bibinfo {year}
  {2021})},\ \Eprint {http://arxiv.org/abs/2012.08595} {arXiv:2012.08595
  [hep-ph]} \BibitemShut {NoStop}%
\bibitem [{\citenamefont {Brivio}\ \emph {et~al.}(2017)\citenamefont {Brivio},
  \citenamefont {Gavela}, \citenamefont {Merlo}, \citenamefont {Mimasu},
  \citenamefont {No}, \citenamefont {del Rey},\ and\ \citenamefont
  {Sanz}}]{Brivio:2017ije}%
  \BibitemOpen
  \bibfield  {author} {\bibinfo {author} {\bibfnamefont {I.}~\bibnamefont
  {Brivio}}, \bibinfo {author} {\bibfnamefont {M.~B.}\ \bibnamefont {Gavela}},
  \bibinfo {author} {\bibfnamefont {L.}~\bibnamefont {Merlo}}, \bibinfo
  {author} {\bibfnamefont {K.}~\bibnamefont {Mimasu}}, \bibinfo {author}
  {\bibfnamefont {J.~M.}\ \bibnamefont {No}}, \bibinfo {author} {\bibfnamefont
  {R.}~\bibnamefont {del Rey}}, \ and\ \bibinfo {author} {\bibfnamefont
  {V.}~\bibnamefont {Sanz}},\ }\href {\doibase 10.1140/epjc/s10052-017-5111-3}
  {\bibfield  {journal} {\bibinfo  {journal} {Eur. Phys. J. C}\ }\textbf
  {\bibinfo {volume} {77}},\ \bibinfo {pages} {572} (\bibinfo {year} {2017})},\
  \Eprint {http://arxiv.org/abs/1701.05379} {arXiv:1701.05379 [hep-ph]}
  \BibitemShut {NoStop}%
\bibitem [{\citenamefont {Aloni}\ \emph
  {et~al.}(2019{\natexlab{a}})\citenamefont {Aloni}, \citenamefont {Soreq},\
  and\ \citenamefont {Williams}}]{Aloni:2018vki}%
  \BibitemOpen
  \bibfield  {author} {\bibinfo {author} {\bibfnamefont {D.}~\bibnamefont
  {Aloni}}, \bibinfo {author} {\bibfnamefont {Y.}~\bibnamefont {Soreq}}, \ and\
  \bibinfo {author} {\bibfnamefont {M.}~\bibnamefont {Williams}},\ }\href
  {\doibase 10.1103/PhysRevLett.123.031803} {\bibfield  {journal} {\bibinfo
  {journal} {Phys. Rev. Lett.}\ }\textbf {\bibinfo {volume} {123}},\ \bibinfo
  {pages} {031803} (\bibinfo {year} {2019}{\natexlab{a}})},\ \Eprint
  {http://arxiv.org/abs/1811.03474} {arXiv:1811.03474 [hep-ph]} \BibitemShut
  {NoStop}%
\bibitem [{\citenamefont {Strassler}\ and\ \citenamefont
  {Zurek}(2007)}]{Strassler:2006im}%
  \BibitemOpen
  \bibfield  {author} {\bibinfo {author} {\bibfnamefont {M.~J.}\ \bibnamefont
  {Strassler}}\ and\ \bibinfo {author} {\bibfnamefont {K.~M.}\ \bibnamefont
  {Zurek}},\ }\href {\doibase 10.1016/j.physletb.2007.06.055} {\bibfield
  {journal} {\bibinfo  {journal} {Phys. Lett. B}\ }\textbf {\bibinfo {volume}
  {651}},\ \bibinfo {pages} {374} (\bibinfo {year} {2007})},\ \Eprint
  {http://arxiv.org/abs/hep-ph/0604261} {arXiv:hep-ph/0604261} \BibitemShut
  {NoStop}%
\bibitem [{\citenamefont {Knapen}\ \emph {et~al.}(2021)\citenamefont {Knapen},
  \citenamefont {Shelton},\ and\ \citenamefont {Xu}}]{Knapen:2021eip}%
  \BibitemOpen
  \bibfield  {author} {\bibinfo {author} {\bibfnamefont {S.}~\bibnamefont
  {Knapen}}, \bibinfo {author} {\bibfnamefont {J.}~\bibnamefont {Shelton}}, \
  and\ \bibinfo {author} {\bibfnamefont {D.}~\bibnamefont {Xu}},\ }\href
  {\doibase 10.1103/PhysRevD.103.115013} {\bibfield  {journal} {\bibinfo
  {journal} {Phys. Rev. D}\ }\textbf {\bibinfo {volume} {103}},\ \bibinfo
  {pages} {115013} (\bibinfo {year} {2021})},\ \Eprint
  {http://arxiv.org/abs/2103.01238} {arXiv:2103.01238 [hep-ph]} \BibitemShut
  {NoStop}%
\bibitem [{\citenamefont {Carloni}\ and\ \citenamefont
  {Sjostrand}(2010)}]{Carloni:2010tw}%
  \BibitemOpen
  \bibfield  {author} {\bibinfo {author} {\bibfnamefont {L.}~\bibnamefont
  {Carloni}}\ and\ \bibinfo {author} {\bibfnamefont {T.}~\bibnamefont
  {Sjostrand}},\ }\href {\doibase 10.1007/JHEP09(2010)105} {\bibfield
  {journal} {\bibinfo  {journal} {JHEP}\ }\textbf {\bibinfo {volume} {09}},\
  \bibinfo {pages} {105} (\bibinfo {year} {2010})},\ \Eprint
  {http://arxiv.org/abs/1006.2911} {arXiv:1006.2911 [hep-ph]} \BibitemShut
  {NoStop}%
\bibitem [{\citenamefont {Farchioni}\ \emph {et~al.}(2007)\citenamefont
  {Farchioni}, \citenamefont {Montvay}, \citenamefont {Munster}, \citenamefont
  {Scholz}, \citenamefont {Sudmann},\ and\ \citenamefont
  {Wuilloud}}]{Farchioni:2007dw}%
  \BibitemOpen
  \bibfield  {author} {\bibinfo {author} {\bibfnamefont {F.}~\bibnamefont
  {Farchioni}}, \bibinfo {author} {\bibfnamefont {I.}~\bibnamefont {Montvay}},
  \bibinfo {author} {\bibfnamefont {G.}~\bibnamefont {Munster}}, \bibinfo
  {author} {\bibfnamefont {E.~E.}\ \bibnamefont {Scholz}}, \bibinfo {author}
  {\bibfnamefont {T.}~\bibnamefont {Sudmann}}, \ and\ \bibinfo {author}
  {\bibfnamefont {J.}~\bibnamefont {Wuilloud}},\ }\href {\doibase
  10.1140/epjc/s10052-007-0394-4} {\bibfield  {journal} {\bibinfo  {journal}
  {Eur. Phys. J. C}\ }\textbf {\bibinfo {volume} {52}},\ \bibinfo {pages} {305}
  (\bibinfo {year} {2007})},\ \Eprint {http://arxiv.org/abs/0706.1131}
  {arXiv:0706.1131 [hep-lat]} \BibitemShut {NoStop}%
\bibitem [{\citenamefont {Creutz}(2007)}]{Creutz:2006ts}%
  \BibitemOpen
  \bibfield  {author} {\bibinfo {author} {\bibfnamefont {M.}~\bibnamefont
  {Creutz}},\ }\href {\doibase 10.1016/j.aop.2007.01.002} {\bibfield  {journal}
  {\bibinfo  {journal} {Annals Phys.}\ }\textbf {\bibinfo {volume} {322}},\
  \bibinfo {pages} {1518} (\bibinfo {year} {2007})},\ \Eprint
  {http://arxiv.org/abs/hep-th/0609187} {arXiv:hep-th/0609187} \BibitemShut
  {NoStop}%
\bibitem [{\citenamefont {Armoni}\ \emph {et~al.}(2003)\citenamefont {Armoni},
  \citenamefont {Shifman},\ and\ \citenamefont {Veneziano}}]{Armoni:2003fb}%
  \BibitemOpen
  \bibfield  {author} {\bibinfo {author} {\bibfnamefont {A.}~\bibnamefont
  {Armoni}}, \bibinfo {author} {\bibfnamefont {M.}~\bibnamefont {Shifman}}, \
  and\ \bibinfo {author} {\bibfnamefont {G.}~\bibnamefont {Veneziano}},\ }\href
  {\doibase 10.1103/PhysRevLett.91.191601} {\bibfield  {journal} {\bibinfo
  {journal} {Phys. Rev. Lett.}\ }\textbf {\bibinfo {volume} {91}},\ \bibinfo
  {pages} {191601} (\bibinfo {year} {2003})},\ \Eprint
  {http://arxiv.org/abs/hep-th/0307097} {arXiv:hep-th/0307097} \BibitemShut
  {NoStop}%
\bibitem [{\citenamefont {Tumasyan}\ \emph
  {et~al.}(2022{\natexlab{d}})\citenamefont {Tumasyan} \emph
  {et~al.}}]{CMS:2021rwb}%
  \BibitemOpen
  \bibfield  {author} {\bibinfo {author} {\bibfnamefont {A.}~\bibnamefont
  {Tumasyan}} \emph {et~al.} (\bibinfo {collaboration} {CMS}),\ }\href
  {\doibase 10.1140/epjc/s10052-022-10095-5} {\bibfield  {journal} {\bibinfo
  {journal} {Eur. Phys. J. C}\ }\textbf {\bibinfo {volume} {82}},\ \bibinfo
  {pages} {213} (\bibinfo {year} {2022}{\natexlab{d}})},\ \Eprint
  {http://arxiv.org/abs/2105.09178} {arXiv:2105.09178 [hep-ex]} \BibitemShut
  {NoStop}%
\bibitem [{\citenamefont {Aad}\ \emph {et~al.}(2021{\natexlab{b}})\citenamefont
  {Aad} \emph {et~al.}}]{ATLAS:2021kxv}%
  \BibitemOpen
  \bibfield  {author} {\bibinfo {author} {\bibfnamefont {G.}~\bibnamefont
  {Aad}} \emph {et~al.} (\bibinfo {collaboration} {ATLAS}),\ }\href {\doibase
  10.1103/PhysRevD.103.112006} {\bibfield  {journal} {\bibinfo  {journal}
  {Phys. Rev. D}\ }\textbf {\bibinfo {volume} {103}},\ \bibinfo {pages}
  {112006} (\bibinfo {year} {2021}{\natexlab{b}})},\ \Eprint
  {http://arxiv.org/abs/2102.10874} {arXiv:2102.10874 [hep-ex]} \BibitemShut
  {NoStop}%
\bibitem [{\citenamefont {Bergsma}\ \emph {et~al.}(1985)\citenamefont {Bergsma}
  \emph {et~al.}}]{CHARM:1985anb}%
  \BibitemOpen
  \bibfield  {author} {\bibinfo {author} {\bibfnamefont {F.}~\bibnamefont
  {Bergsma}} \emph {et~al.} (\bibinfo {collaboration} {CHARM}),\ }\href
  {\doibase 10.1016/0370-2693(85)90400-9} {\bibfield  {journal} {\bibinfo
  {journal} {Phys. Lett. B}\ }\textbf {\bibinfo {volume} {157}},\ \bibinfo
  {pages} {458} (\bibinfo {year} {1985})}\BibitemShut {NoStop}%
\bibitem [{\citenamefont {Schwaller}\ \emph {et~al.}(2015)\citenamefont
  {Schwaller}, \citenamefont {Stolarski},\ and\ \citenamefont
  {Weiler}}]{Schwaller:2015gea}%
  \BibitemOpen
  \bibfield  {author} {\bibinfo {author} {\bibfnamefont {P.}~\bibnamefont
  {Schwaller}}, \bibinfo {author} {\bibfnamefont {D.}~\bibnamefont
  {Stolarski}}, \ and\ \bibinfo {author} {\bibfnamefont {A.}~\bibnamefont
  {Weiler}},\ }\href {\doibase 10.1007/JHEP05(2015)059} {\bibfield  {journal}
  {\bibinfo  {journal} {JHEP}\ }\textbf {\bibinfo {volume} {05}},\ \bibinfo
  {pages} {059} (\bibinfo {year} {2015})},\ \Eprint
  {http://arxiv.org/abs/1502.05409} {arXiv:1502.05409 [hep-ph]} \BibitemShut
  {NoStop}%
\bibitem [{wor()}]{workshop}%
  \BibitemOpen
  \href@noop {} {\enquote {\bibinfo {title} {{New} ideas in detecting
  long-lived particles at the {LHC Workshop}},}\ }\bibinfo {howpublished}
  {\url{https://indico.physics.lbl.gov/event/633/}}\BibitemShut {NoStop}%
\bibitem [{\citenamefont {Aaij}\ \emph
  {et~al.}(2017{\natexlab{b}})\citenamefont {Aaij} \emph
  {et~al.}}]{LHCb:2016awg}%
  \BibitemOpen
  \bibfield  {author} {\bibinfo {author} {\bibfnamefont {R.}~\bibnamefont
  {Aaij}} \emph {et~al.} (\bibinfo {collaboration} {LHCb}),\ }\href {\doibase
  10.1103/PhysRevD.95.071101} {\bibfield  {journal} {\bibinfo  {journal} {Phys.
  Rev. D}\ }\textbf {\bibinfo {volume} {95}},\ \bibinfo {pages} {071101}
  (\bibinfo {year} {2017}{\natexlab{b}})},\ \Eprint
  {http://arxiv.org/abs/1612.07818} {arXiv:1612.07818 [hep-ex]} \BibitemShut
  {NoStop}%
\bibitem [{\citenamefont {Foroughi-Abari}\ and\ \citenamefont
  {Ritz}(2020)}]{Foroughi-Abari:2020gju}%
  \BibitemOpen
  \bibfield  {author} {\bibinfo {author} {\bibfnamefont {S.}~\bibnamefont
  {Foroughi-Abari}}\ and\ \bibinfo {author} {\bibfnamefont {A.}~\bibnamefont
  {Ritz}},\ }\href {\doibase 10.1103/PhysRevD.102.035015} {\bibfield  {journal}
  {\bibinfo  {journal} {Phys. Rev. D}\ }\textbf {\bibinfo {volume} {102}},\
  \bibinfo {pages} {035015} (\bibinfo {year} {2020})},\ \Eprint
  {http://arxiv.org/abs/2004.14515} {arXiv:2004.14515 [hep-ph]} \BibitemShut
  {NoStop}%
\bibitem [{\citenamefont {Aaboud}\ \emph
  {et~al.}(2019{\natexlab{b}})\citenamefont {Aaboud} \emph
  {et~al.}}]{ATLAS:2018tup}%
  \BibitemOpen
  \bibfield  {author} {\bibinfo {author} {\bibfnamefont {M.}~\bibnamefont
  {Aaboud}} \emph {et~al.} (\bibinfo {collaboration} {ATLAS}),\ }\href
  {\doibase 10.1103/PhysRevD.99.052005} {\bibfield  {journal} {\bibinfo
  {journal} {Phys. Rev. D}\ }\textbf {\bibinfo {volume} {99}},\ \bibinfo
  {pages} {052005} (\bibinfo {year} {2019}{\natexlab{b}})},\ \Eprint
  {http://arxiv.org/abs/1811.07370} {arXiv:1811.07370 [hep-ex]} \BibitemShut
  {NoStop}%
\bibitem [{\citenamefont {Alpigiani}\ \emph {et~al.}(2020)\citenamefont
  {Alpigiani} \emph {et~al.}}]{Alpigiani:2020tva}%
  \BibitemOpen
  \bibfield  {author} {\bibinfo {author} {\bibfnamefont {C.}~\bibnamefont
  {Alpigiani}} \emph {et~al.} (\bibinfo {collaboration} {MATHUSLA}),\
  }\href@noop {} {\  (\bibinfo {year} {2020})},\ \Eprint
  {http://arxiv.org/abs/2009.01693} {arXiv:2009.01693 [physics.ins-det]}
  \BibitemShut {NoStop}%
\bibitem [{\citenamefont {Feng}\ \emph
  {et~al.}(2018{\natexlab{b}})\citenamefont {Feng}, \citenamefont {Galon},
  \citenamefont {Kling},\ and\ \citenamefont {Trojanowski}}]{Feng:2017vli}%
  \BibitemOpen
  \bibfield  {author} {\bibinfo {author} {\bibfnamefont {J.~L.}\ \bibnamefont
  {Feng}}, \bibinfo {author} {\bibfnamefont {I.}~\bibnamefont {Galon}},
  \bibinfo {author} {\bibfnamefont {F.}~\bibnamefont {Kling}}, \ and\ \bibinfo
  {author} {\bibfnamefont {S.}~\bibnamefont {Trojanowski}},\ }\href {\doibase
  10.1103/PhysRevD.97.055034} {\bibfield  {journal} {\bibinfo  {journal} {Phys.
  Rev. D}\ }\textbf {\bibinfo {volume} {97}},\ \bibinfo {pages} {055034}
  (\bibinfo {year} {2018}{\natexlab{b}})},\ \Eprint
  {http://arxiv.org/abs/1710.09387} {arXiv:1710.09387 [hep-ph]} \BibitemShut
  {NoStop}%
\bibitem [{\citenamefont {Aaij}\ \emph
  {et~al.}(2017{\natexlab{c}})\citenamefont {Aaij} \emph
  {et~al.}}]{Aaij:2016qsm}%
  \BibitemOpen
  \bibfield  {author} {\bibinfo {author} {\bibfnamefont {R.}~\bibnamefont
  {Aaij}} \emph {et~al.} (\bibinfo {collaboration} {LHCb}),\ }\href {\doibase
  10.1103/PhysRevD.95.071101} {\bibfield  {journal} {\bibinfo  {journal} {Phys.
  Rev. D}\ }\textbf {\bibinfo {volume} {95}},\ \bibinfo {pages} {071101}
  (\bibinfo {year} {2017}{\natexlab{c}})},\ \Eprint
  {http://arxiv.org/abs/1612.07818} {arXiv:1612.07818 [hep-ex]} \BibitemShut
  {NoStop}%
\bibitem [{\citenamefont {Lees}\ \emph {et~al.}(2014)\citenamefont {Lees} \emph
  {et~al.}}]{Lees:2014xha}%
  \BibitemOpen
  \bibfield  {author} {\bibinfo {author} {\bibfnamefont {J.~P.}\ \bibnamefont
  {Lees}} \emph {et~al.} (\bibinfo {collaboration} {BaBar}),\ }\href {\doibase
  10.1103/PhysRevLett.113.201801} {\bibfield  {journal} {\bibinfo  {journal}
  {Phys. Rev. Lett.}\ }\textbf {\bibinfo {volume} {113}},\ \bibinfo {pages}
  {201801} (\bibinfo {year} {2014})},\ \Eprint {http://arxiv.org/abs/1406.2980}
  {arXiv:1406.2980 [hep-ex]} \BibitemShut {NoStop}%
\bibitem [{\citenamefont {Anastasi}\ \emph {et~al.}(2018)\citenamefont
  {Anastasi} \emph {et~al.}}]{Anastasi:2018azp}%
  \BibitemOpen
  \bibfield  {author} {\bibinfo {author} {\bibfnamefont {A.}~\bibnamefont
  {Anastasi}} \emph {et~al.} (\bibinfo {collaboration} {KLOE-2}),\ }\href@noop
  {} {\bibfield  {journal} {\bibinfo  {journal} {Submitted to: Phys. Lett. B}\
  } (\bibinfo {year} {2018})},\ \Eprint {http://arxiv.org/abs/1807.02691}
  {arXiv:1807.02691 [hep-ex]} \BibitemShut {NoStop}%
\bibitem [{\citenamefont {Aaij}\ \emph {et~al.}(2020)\citenamefont {Aaij} \emph
  {et~al.}}]{LHCb:2019vmc}%
  \BibitemOpen
  \bibfield  {author} {\bibinfo {author} {\bibfnamefont {R.}~\bibnamefont
  {Aaij}} \emph {et~al.} (\bibinfo {collaboration} {LHCb}),\ }\href {\doibase
  10.1103/PhysRevLett.124.041801} {\bibfield  {journal} {\bibinfo  {journal}
  {Phys. Rev. Lett.}\ }\textbf {\bibinfo {volume} {124}},\ \bibinfo {pages}
  {041801} (\bibinfo {year} {2020})},\ \Eprint
  {http://arxiv.org/abs/1910.06926} {arXiv:1910.06926 [hep-ex]} \BibitemShut
  {NoStop}%
\bibitem [{\citenamefont {Batley}\ \emph {et~al.}(2015)\citenamefont {Batley}
  \emph {et~al.}}]{Batley:2015lha}%
  \BibitemOpen
  \bibfield  {author} {\bibinfo {author} {\bibfnamefont {J.~R.}\ \bibnamefont
  {Batley}} \emph {et~al.} (\bibinfo {collaboration} {NA48/2}),\ }\href
  {\doibase 10.1016/j.physletb.2015.04.068} {\bibfield  {journal} {\bibinfo
  {journal} {Phys. Lett.}\ }\textbf {\bibinfo {volume} {B746}},\ \bibinfo
  {pages} {178} (\bibinfo {year} {2015})},\ \Eprint
  {http://arxiv.org/abs/1504.00607} {arXiv:1504.00607 [hep-ex]} \BibitemShut
  {NoStop}%
\bibitem [{\citenamefont {{ATLAS
  Collaboration}}(2022{\natexlab{b}})}]{ATLAS:2022izj}%
  \BibitemOpen
  \bibfield  {author} {\bibinfo {author} {\bibnamefont {{ATLAS Collaboration}}}
  (\bibinfo {collaboration} {ATLAS}),\ }\href@noop {} {\  (\bibinfo {year}
  {2022}{\natexlab{b}})},\ \Eprint {http://arxiv.org/abs/2206.12181}
  {arXiv:2206.12181 [hep-ex]} \BibitemShut {NoStop}%
\bibitem [{\citenamefont {Blumlein}\ \emph {et~al.}(1992)\citenamefont
  {Blumlein} \emph {et~al.}}]{Blumlein:1991xh}%
  \BibitemOpen
  \bibfield  {author} {\bibinfo {author} {\bibfnamefont {J.}~\bibnamefont
  {Blumlein}} \emph {et~al.},\ }\href {\doibase 10.1142/S0217751X9200171X}
  {\bibfield  {journal} {\bibinfo  {journal} {Int. J. Mod. Phys.}\ }\textbf
  {\bibinfo {volume} {A7}},\ \bibinfo {pages} {3835} (\bibinfo {year}
  {1992})}\BibitemShut {NoStop}%
\bibitem [{\citenamefont {Riordan}\ \emph {et~al.}(1987)\citenamefont {Riordan}
  \emph {et~al.}}]{Riordan:1987aw}%
  \BibitemOpen
  \bibfield  {author} {\bibinfo {author} {\bibfnamefont {E.~M.}\ \bibnamefont
  {Riordan}} \emph {et~al.},\ }\href {\doibase 10.1103/PhysRevLett.59.755}
  {\bibfield  {journal} {\bibinfo  {journal} {Phys. Rev. Lett.}\ }\textbf
  {\bibinfo {volume} {59}},\ \bibinfo {pages} {755} (\bibinfo {year}
  {1987})}\BibitemShut {NoStop}%
\bibitem [{\citenamefont {Banerjee}\ \emph {et~al.}(2020)\citenamefont
  {Banerjee} \emph {et~al.}}]{NA64:2020qwq}%
  \BibitemOpen
  \bibfield  {author} {\bibinfo {author} {\bibfnamefont {D.}~\bibnamefont
  {Banerjee}} \emph {et~al.} (\bibinfo {collaboration} {NA64}),\ }\href
  {\doibase 10.1103/PhysRevLett.125.081801} {\bibfield  {journal} {\bibinfo
  {journal} {Phys. Rev. Lett.}\ }\textbf {\bibinfo {volume} {125}},\ \bibinfo
  {pages} {081801} (\bibinfo {year} {2020})},\ \Eprint
  {http://arxiv.org/abs/2005.02710} {arXiv:2005.02710 [hep-ex]} \BibitemShut
  {NoStop}%
\bibitem [{\citenamefont {Berlin}\ \emph {et~al.}(2018)\citenamefont {Berlin},
  \citenamefont {Gori}, \citenamefont {Schuster},\ and\ \citenamefont
  {Toro}}]{Berlin:2018pwi}%
  \BibitemOpen
  \bibfield  {author} {\bibinfo {author} {\bibfnamefont {A.}~\bibnamefont
  {Berlin}}, \bibinfo {author} {\bibfnamefont {S.}~\bibnamefont {Gori}},
  \bibinfo {author} {\bibfnamefont {P.}~\bibnamefont {Schuster}}, \ and\
  \bibinfo {author} {\bibfnamefont {N.}~\bibnamefont {Toro}},\ }\href {\doibase
  10.1103/PhysRevD.98.035011} {\bibfield  {journal} {\bibinfo  {journal} {Phys.
  Rev. D}\ }\textbf {\bibinfo {volume} {98}},\ \bibinfo {pages} {035011}
  (\bibinfo {year} {2018})},\ \Eprint {http://arxiv.org/abs/1804.00661}
  {arXiv:1804.00661 [hep-ph]} \BibitemShut {NoStop}%
\bibitem [{\citenamefont {{NA62 Collaboration}}(2019)}]{Collaboration:2691873}%
  \BibitemOpen
  \bibfield  {author} {\bibinfo {author} {\bibnamefont {{NA62
  Collaboration}}},\ }\href {https://cds.cern.ch/record/2691873} {\  (\bibinfo
  {year} {2019})}\BibitemShut {NoStop}%
\bibitem [{\citenamefont {Ilten}\ \emph {et~al.}(2016)\citenamefont {Ilten},
  \citenamefont {Soreq}, \citenamefont {Thaler}, \citenamefont {Williams},\
  and\ \citenamefont {Xue}}]{Ilten:2016tkc}%
  \BibitemOpen
  \bibfield  {author} {\bibinfo {author} {\bibfnamefont {P.}~\bibnamefont
  {Ilten}}, \bibinfo {author} {\bibfnamefont {Y.}~\bibnamefont {Soreq}},
  \bibinfo {author} {\bibfnamefont {J.}~\bibnamefont {Thaler}}, \bibinfo
  {author} {\bibfnamefont {M.}~\bibnamefont {Williams}}, \ and\ \bibinfo
  {author} {\bibfnamefont {W.}~\bibnamefont {Xue}},\ }\href {\doibase
  10.1103/PhysRevLett.116.251803} {\bibfield  {journal} {\bibinfo  {journal}
  {Phys. Rev. Lett.}\ }\textbf {\bibinfo {volume} {116}},\ \bibinfo {pages}
  {251803} (\bibinfo {year} {2016})},\ \Eprint
  {http://arxiv.org/abs/1603.08926} {arXiv:1603.08926 [hep-ph]} \BibitemShut
  {NoStop}%
\bibitem [{\citenamefont {Altmannshofer}\ \emph
  {et~al.}(2019{\natexlab{a}})\citenamefont {Altmannshofer} \emph
  {et~al.}}]{Kou:2018nap}%
  \BibitemOpen
  \bibfield  {author} {\bibinfo {author} {\bibfnamefont {W.}~\bibnamefont
  {Altmannshofer}} \emph {et~al.} (\bibinfo {collaboration} {Belle-II}),\
  }\href {\doibase 10.1093/ptep/ptz106} {\bibfield  {journal} {\bibinfo
  {journal} {PTEP}\ }\textbf {\bibinfo {volume} {2019}},\ \bibinfo {pages}
  {123C01} (\bibinfo {year} {2019}{\natexlab{a}})},\ \bibinfo {note} {[Erratum:
  PTEP 2020, 029201 (2020)]},\ \Eprint {http://arxiv.org/abs/1808.10567}
  {arXiv:1808.10567 [hep-ex]} \BibitemShut {NoStop}%
\bibitem [{\citenamefont {Chakraborty}\ \emph {et~al.}(2021)\citenamefont
  {Chakraborty}, \citenamefont {Kraus}, \citenamefont {Loladze}, \citenamefont
  {Okui},\ and\ \citenamefont {Tobioka}}]{Chakraborty:2021wda}%
  \BibitemOpen
  \bibfield  {author} {\bibinfo {author} {\bibfnamefont {S.}~\bibnamefont
  {Chakraborty}}, \bibinfo {author} {\bibfnamefont {M.}~\bibnamefont {Kraus}},
  \bibinfo {author} {\bibfnamefont {V.}~\bibnamefont {Loladze}}, \bibinfo
  {author} {\bibfnamefont {T.}~\bibnamefont {Okui}}, \ and\ \bibinfo {author}
  {\bibfnamefont {K.}~\bibnamefont {Tobioka}},\ }\href {\doibase
  10.1103/PhysRevD.104.055036} {\bibfield  {journal} {\bibinfo  {journal}
  {Phys. Rev. D}\ }\textbf {\bibinfo {volume} {104}},\ \bibinfo {pages}
  {055036} (\bibinfo {year} {2021})},\ \Eprint
  {http://arxiv.org/abs/2102.04474} {arXiv:2102.04474 [hep-ph]} \BibitemShut
  {NoStop}%
\bibitem [{\citenamefont {Zyla}\ \emph {et~al.}(2020)\citenamefont {Zyla} \emph
  {et~al.}}]{ParticleDataGroup:2020ssz}%
  \BibitemOpen
  \bibfield  {author} {\bibinfo {author} {\bibfnamefont {P.~A.}\ \bibnamefont
  {Zyla}} \emph {et~al.} (\bibinfo {collaboration} {Particle Data Group}),\
  }\href {\doibase 10.1093/ptep/ptaa104} {\bibfield  {journal} {\bibinfo
  {journal} {PTEP}\ }\textbf {\bibinfo {volume} {2020}},\ \bibinfo {pages}
  {083C01} (\bibinfo {year} {2020})}\BibitemShut {NoStop}%
\bibitem [{\citenamefont {Chobanova}\ \emph {et~al.}(2014)\citenamefont
  {Chobanova} \emph {et~al.}}]{Belle:2013nby}%
  \BibitemOpen
  \bibfield  {author} {\bibinfo {author} {\bibfnamefont {V.}~\bibnamefont
  {Chobanova}} \emph {et~al.} (\bibinfo {collaboration} {Belle}),\ }\href
  {\doibase 10.1103/PhysRevD.90.012002} {\bibfield  {journal} {\bibinfo
  {journal} {Phys. Rev. D}\ }\textbf {\bibinfo {volume} {90}},\ \bibinfo
  {pages} {012002} (\bibinfo {year} {2014})},\ \Eprint
  {http://arxiv.org/abs/1311.6666} {arXiv:1311.6666 [hep-ex]} \BibitemShut
  {NoStop}%
\bibitem [{\citenamefont {Aubert}\ \emph {et~al.}(2008)\citenamefont {Aubert}
  \emph {et~al.}}]{BaBar:2008rth}%
  \BibitemOpen
  \bibfield  {author} {\bibinfo {author} {\bibfnamefont {B.}~\bibnamefont
  {Aubert}} \emph {et~al.} (\bibinfo {collaboration} {BaBar}),\ }\href
  {\doibase 10.1103/PhysRevLett.101.091801} {\bibfield  {journal} {\bibinfo
  {journal} {Phys. Rev. Lett.}\ }\textbf {\bibinfo {volume} {101}},\ \bibinfo
  {pages} {091801} (\bibinfo {year} {2008})},\ \Eprint
  {http://arxiv.org/abs/0804.0411} {arXiv:0804.0411 [hep-ex]} \BibitemShut
  {NoStop}%
\bibitem [{\citenamefont {Lees}\ \emph {et~al.}(2011)\citenamefont {Lees} \emph
  {et~al.}}]{BaBar:2011vod}%
  \BibitemOpen
  \bibfield  {author} {\bibinfo {author} {\bibfnamefont {J.~P.}\ \bibnamefont
  {Lees}} \emph {et~al.} (\bibinfo {collaboration} {BaBar}),\ }\href {\doibase
  10.1103/PhysRevD.84.012001} {\bibfield  {journal} {\bibinfo  {journal} {Phys.
  Rev. D}\ }\textbf {\bibinfo {volume} {84}},\ \bibinfo {pages} {012001}
  (\bibinfo {year} {2011})},\ \Eprint {http://arxiv.org/abs/1105.5159}
  {arXiv:1105.5159 [hep-ex]} \BibitemShut {NoStop}%
\bibitem [{\citenamefont {Ertas}\ and\ \citenamefont
  {Kahlhoefer}(2020)}]{Ertas:2020xcc}%
  \BibitemOpen
  \bibfield  {author} {\bibinfo {author} {\bibfnamefont {F.}~\bibnamefont
  {Ertas}}\ and\ \bibinfo {author} {\bibfnamefont {F.}~\bibnamefont
  {Kahlhoefer}},\ }\href {\doibase 10.1007/JHEP07(2020)050} {\bibfield
  {journal} {\bibinfo  {journal} {JHEP}\ }\textbf {\bibinfo {volume} {07}},\
  \bibinfo {pages} {050} (\bibinfo {year} {2020})},\ \Eprint
  {http://arxiv.org/abs/2004.01193} {arXiv:2004.01193 [hep-ph]} \BibitemShut
  {NoStop}%
\bibitem [{\citenamefont {Dolan}\ \emph {et~al.}(2017)\citenamefont {Dolan},
  \citenamefont {Ferber}, \citenamefont {Hearty}, \citenamefont {Kahlhoefer},\
  and\ \citenamefont {Schmidt-Hoberg}}]{Dolan:2017osp}%
  \BibitemOpen
  \bibfield  {author} {\bibinfo {author} {\bibfnamefont {M.~J.}\ \bibnamefont
  {Dolan}}, \bibinfo {author} {\bibfnamefont {T.}~\bibnamefont {Ferber}},
  \bibinfo {author} {\bibfnamefont {C.}~\bibnamefont {Hearty}}, \bibinfo
  {author} {\bibfnamefont {F.}~\bibnamefont {Kahlhoefer}}, \ and\ \bibinfo
  {author} {\bibfnamefont {K.}~\bibnamefont {Schmidt-Hoberg}},\ }\href
  {\doibase 10.1007/JHEP12(2017)094} {\bibfield  {journal} {\bibinfo  {journal}
  {JHEP}\ }\textbf {\bibinfo {volume} {12}},\ \bibinfo {pages} {094} (\bibinfo
  {year} {2017})},\ \bibinfo {note} {[Erratum: JHEP 03, 190 (2021)]},\ \Eprint
  {http://arxiv.org/abs/1709.00009} {arXiv:1709.00009 [hep-ph]} \BibitemShut
  {NoStop}%
\bibitem [{\citenamefont {Acciarri}\ \emph {et~al.}(1997)\citenamefont
  {Acciarri} \emph {et~al.}}]{L3:1997exg}%
  \BibitemOpen
  \bibfield  {author} {\bibinfo {author} {\bibfnamefont {M.}~\bibnamefont
  {Acciarri}} \emph {et~al.} (\bibinfo {collaboration} {L3}),\ }\href {\doibase
  10.1016/S0370-2693(97)01003-4} {\bibfield  {journal} {\bibinfo  {journal}
  {Phys. Lett. B}\ }\textbf {\bibinfo {volume} {412}},\ \bibinfo {pages} {201}
  (\bibinfo {year} {1997})}\BibitemShut {NoStop}%
\bibitem [{\citenamefont {Jaeckel}\ and\ \citenamefont
  {Spannowsky}(2016)}]{Jaeckel:2015jla}%
  \BibitemOpen
  \bibfield  {author} {\bibinfo {author} {\bibfnamefont {J.}~\bibnamefont
  {Jaeckel}}\ and\ \bibinfo {author} {\bibfnamefont {M.}~\bibnamefont
  {Spannowsky}},\ }\href {\doibase 10.1016/j.physletb.2015.12.037} {\bibfield
  {journal} {\bibinfo  {journal} {Phys. Lett. B}\ }\textbf {\bibinfo {volume}
  {753}},\ \bibinfo {pages} {482} (\bibinfo {year} {2016})},\ \Eprint
  {http://arxiv.org/abs/1509.00476} {arXiv:1509.00476 [hep-ph]} \BibitemShut
  {NoStop}%
\bibitem [{\citenamefont {Anashkin}\ \emph {et~al.}(1999)\citenamefont
  {Anashkin} \emph {et~al.}}]{DELPHI:1999fgt}%
  \BibitemOpen
  \bibfield  {author} {\bibinfo {author} {\bibfnamefont {E.}~\bibnamefont
  {Anashkin}} \emph {et~al.} (\bibinfo {collaboration} {DELPHI}),\ }\href@noop
  {} {\  (\bibinfo {year} {1999})}\BibitemShut {NoStop}%
\bibitem [{\citenamefont {Acciarri}\ \emph {et~al.}(1995)\citenamefont
  {Acciarri} \emph {et~al.}}]{L3:1995nbq}%
  \BibitemOpen
  \bibfield  {author} {\bibinfo {author} {\bibfnamefont {M.}~\bibnamefont
  {Acciarri}} \emph {et~al.} (\bibinfo {collaboration} {L3}),\ }\href {\doibase
  10.1016/0370-2693(95)00527-R} {\bibfield  {journal} {\bibinfo  {journal}
  {Phys. Lett. B}\ }\textbf {\bibinfo {volume} {353}},\ \bibinfo {pages} {136}
  (\bibinfo {year} {1995})}\BibitemShut {NoStop}%
\bibitem [{\citenamefont {Aloni}\ \emph
  {et~al.}(2019{\natexlab{b}})\citenamefont {Aloni}, \citenamefont {Fanelli},
  \citenamefont {Soreq},\ and\ \citenamefont {Williams}}]{Aloni:2019ruo}%
  \BibitemOpen
  \bibfield  {author} {\bibinfo {author} {\bibfnamefont {D.}~\bibnamefont
  {Aloni}}, \bibinfo {author} {\bibfnamefont {C.}~\bibnamefont {Fanelli}},
  \bibinfo {author} {\bibfnamefont {Y.}~\bibnamefont {Soreq}}, \ and\ \bibinfo
  {author} {\bibfnamefont {M.}~\bibnamefont {Williams}},\ }\href {\doibase
  10.1103/PhysRevLett.123.071801} {\bibfield  {journal} {\bibinfo  {journal}
  {Phys. Rev. Lett.}\ }\textbf {\bibinfo {volume} {123}},\ \bibinfo {pages}
  {071801} (\bibinfo {year} {2019}{\natexlab{b}})},\ \Eprint
  {http://arxiv.org/abs/1903.03586} {arXiv:1903.03586 [hep-ph]} \BibitemShut
  {NoStop}%
\bibitem [{\citenamefont {Larin}\ \emph {et~al.}(2011)\citenamefont {Larin}
  \emph {et~al.}}]{PrimEx:2010fvg}%
  \BibitemOpen
  \bibfield  {author} {\bibinfo {author} {\bibfnamefont {I.}~\bibnamefont
  {Larin}} \emph {et~al.} (\bibinfo {collaboration} {PrimEx}),\ }\href
  {\doibase 10.1103/PhysRevLett.106.162303} {\bibfield  {journal} {\bibinfo
  {journal} {Phys. Rev. Lett.}\ }\textbf {\bibinfo {volume} {106}},\ \bibinfo
  {pages} {162303} (\bibinfo {year} {2011})},\ \Eprint
  {http://arxiv.org/abs/1009.1681} {arXiv:1009.1681 [nucl-ex]} \BibitemShut
  {NoStop}%
\bibitem [{\citenamefont {Abudin\'en}\ \emph {et~al.}(2020)\citenamefont
  {Abudin\'en} \emph {et~al.}}]{Belle-II:2020jti}%
  \BibitemOpen
  \bibfield  {author} {\bibinfo {author} {\bibfnamefont {F.}~\bibnamefont
  {Abudin\'en}} \emph {et~al.} (\bibinfo {collaboration} {Belle-II}),\ }\href
  {\doibase 10.1103/PhysRevLett.125.161806} {\bibfield  {journal} {\bibinfo
  {journal} {Phys. Rev. Lett.}\ }\textbf {\bibinfo {volume} {125}},\ \bibinfo
  {pages} {161806} (\bibinfo {year} {2020})},\ \Eprint
  {http://arxiv.org/abs/2007.13071} {arXiv:2007.13071 [hep-ex]} \BibitemShut
  {NoStop}%
\bibitem [{\citenamefont {Dugger}\ \emph {et~al.}(2012)\citenamefont {Dugger}
  \emph {et~al.}}]{GlueX:2012idx}%
  \BibitemOpen
  \bibfield  {author} {\bibinfo {author} {\bibfnamefont {M.}~\bibnamefont
  {Dugger}} \emph {et~al.} (\bibinfo {collaboration} {GlueX}),\ }\href@noop {}
  {\  (\bibinfo {year} {2012})},\ \Eprint {http://arxiv.org/abs/1210.4508}
  {arXiv:1210.4508 [hep-ex]} \BibitemShut {NoStop}%
\bibitem [{\citenamefont {Altmannshofer}\ \emph
  {et~al.}(2019{\natexlab{b}})\citenamefont {Altmannshofer} \emph
  {et~al.}}]{Belle-II:2018jsg}%
  \BibitemOpen
  \bibfield  {author} {\bibinfo {author} {\bibfnamefont {W.}~\bibnamefont
  {Altmannshofer}} \emph {et~al.} (\bibinfo {collaboration} {Belle-II}),\
  }\href {\doibase 10.1093/ptep/ptz106} {\bibfield  {journal} {\bibinfo
  {journal} {PTEP}\ }\textbf {\bibinfo {volume} {2019}},\ \bibinfo {pages}
  {123C01} (\bibinfo {year} {2019}{\natexlab{b}})},\ \bibinfo {note} {[Erratum:
  PTEP 2020, 029201 (2020)]},\ \Eprint {http://arxiv.org/abs/1808.10567}
  {arXiv:1808.10567 [hep-ex]} \BibitemShut {NoStop}%
\bibitem [{\citenamefont {Bjorken}\ \emph {et~al.}(1988)\citenamefont
  {Bjorken}, \citenamefont {Ecklund}, \citenamefont {Nelson}, \citenamefont
  {Abashian}, \citenamefont {Church}, \citenamefont {Lu}, \citenamefont {Mo},
  \citenamefont {Nunamaker},\ and\ \citenamefont {Rassmann}}]{Bjorken:1988as}%
  \BibitemOpen
  \bibfield  {author} {\bibinfo {author} {\bibfnamefont {J.~D.}\ \bibnamefont
  {Bjorken}}, \bibinfo {author} {\bibfnamefont {S.}~\bibnamefont {Ecklund}},
  \bibinfo {author} {\bibfnamefont {W.~R.}\ \bibnamefont {Nelson}}, \bibinfo
  {author} {\bibfnamefont {A.}~\bibnamefont {Abashian}}, \bibinfo {author}
  {\bibfnamefont {C.}~\bibnamefont {Church}}, \bibinfo {author} {\bibfnamefont
  {B.}~\bibnamefont {Lu}}, \bibinfo {author} {\bibfnamefont {L.~W.}\
  \bibnamefont {Mo}}, \bibinfo {author} {\bibfnamefont {T.~A.}\ \bibnamefont
  {Nunamaker}}, \ and\ \bibinfo {author} {\bibfnamefont {P.}~\bibnamefont
  {Rassmann}},\ }\href {\doibase 10.1103/PhysRevD.38.3375} {\bibfield
  {journal} {\bibinfo  {journal} {Phys. Rev. D}\ }\textbf {\bibinfo {volume}
  {38}},\ \bibinfo {pages} {3375} (\bibinfo {year} {1988})}\BibitemShut
  {NoStop}%
\bibitem [{\citenamefont {Blumlein}\ \emph {et~al.}(1991)\citenamefont
  {Blumlein} \emph {et~al.}}]{Blumlein:1990ay}%
  \BibitemOpen
  \bibfield  {author} {\bibinfo {author} {\bibfnamefont {J.}~\bibnamefont
  {Blumlein}} \emph {et~al.},\ }\href {\doibase 10.1007/BF01548556} {\bibfield
  {journal} {\bibinfo  {journal} {Z. Phys. C}\ }\textbf {\bibinfo {volume}
  {51}},\ \bibinfo {pages} {341} (\bibinfo {year} {1991})}\BibitemShut
  {NoStop}%
\bibitem [{\citenamefont {Abdallah}\ \emph {et~al.}(2009)\citenamefont
  {Abdallah} \emph {et~al.}}]{DELPHI:2008uka}%
  \BibitemOpen
  \bibfield  {author} {\bibinfo {author} {\bibfnamefont {J.}~\bibnamefont
  {Abdallah}} \emph {et~al.} (\bibinfo {collaboration} {DELPHI}),\ }\href
  {\doibase 10.1140/epjc/s10052-009-0874-9} {\bibfield  {journal} {\bibinfo
  {journal} {Eur. Phys. J. C}\ }\textbf {\bibinfo {volume} {60}},\ \bibinfo
  {pages} {17} (\bibinfo {year} {2009})},\ \Eprint
  {http://arxiv.org/abs/0901.4486} {arXiv:0901.4486 [hep-ex]} \BibitemShut
  {NoStop}%
\bibitem [{\citenamefont {Berlin}\ and\ \citenamefont
  {Kling}(2019)}]{Berlin:2018jbm}%
  \BibitemOpen
  \bibfield  {author} {\bibinfo {author} {\bibfnamefont {A.}~\bibnamefont
  {Berlin}}\ and\ \bibinfo {author} {\bibfnamefont {F.}~\bibnamefont {Kling}},\
  }\href {\doibase 10.1103/PhysRevD.99.015021} {\bibfield  {journal} {\bibinfo
  {journal} {Phys. Rev. D}\ }\textbf {\bibinfo {volume} {99}},\ \bibinfo
  {pages} {015021} (\bibinfo {year} {2019})},\ \Eprint
  {http://arxiv.org/abs/1810.01879} {arXiv:1810.01879 [hep-ph]} \BibitemShut
  {NoStop}%
\bibitem [{\citenamefont {Lees}\ \emph {et~al.}(2017)\citenamefont {Lees} \emph
  {et~al.}}]{BaBar:2017tiz}%
  \BibitemOpen
  \bibfield  {author} {\bibinfo {author} {\bibfnamefont {J.~P.}\ \bibnamefont
  {Lees}} \emph {et~al.} (\bibinfo {collaboration} {BaBar}),\ }\href {\doibase
  10.1103/PhysRevLett.119.131804} {\bibfield  {journal} {\bibinfo  {journal}
  {Phys. Rev. Lett.}\ }\textbf {\bibinfo {volume} {119}},\ \bibinfo {pages}
  {131804} (\bibinfo {year} {2017})},\ \Eprint
  {http://arxiv.org/abs/1702.03327} {arXiv:1702.03327 [hep-ex]} \BibitemShut
  {NoStop}%
\bibitem [{\citenamefont {Hook}\ \emph {et~al.}(2011)\citenamefont {Hook},
  \citenamefont {Izaguirre},\ and\ \citenamefont {Wacker}}]{Hook:2010tw}%
  \BibitemOpen
  \bibfield  {author} {\bibinfo {author} {\bibfnamefont {A.}~\bibnamefont
  {Hook}}, \bibinfo {author} {\bibfnamefont {E.}~\bibnamefont {Izaguirre}}, \
  and\ \bibinfo {author} {\bibfnamefont {J.~G.}\ \bibnamefont {Wacker}},\
  }\href {\doibase 10.1155/2011/859762} {\bibfield  {journal} {\bibinfo
  {journal} {Adv. High Energy Phys.}\ }\textbf {\bibinfo {volume} {2011}},\
  \bibinfo {pages} {859762} (\bibinfo {year} {2011})},\ \Eprint
  {http://arxiv.org/abs/1006.0973} {arXiv:1006.0973 [hep-ph]} \BibitemShut
  {NoStop}%
\bibitem [{\citenamefont {Pierce}\ \emph {et~al.}(2018)\citenamefont {Pierce},
  \citenamefont {Shakya}, \citenamefont {Tsai},\ and\ \citenamefont
  {Zhao}}]{Pierce:2017taw}%
  \BibitemOpen
  \bibfield  {author} {\bibinfo {author} {\bibfnamefont {A.}~\bibnamefont
  {Pierce}}, \bibinfo {author} {\bibfnamefont {B.}~\bibnamefont {Shakya}},
  \bibinfo {author} {\bibfnamefont {Y.}~\bibnamefont {Tsai}}, \ and\ \bibinfo
  {author} {\bibfnamefont {Y.}~\bibnamefont {Zhao}},\ }\href {\doibase
  10.1103/PhysRevD.97.095033} {\bibfield  {journal} {\bibinfo  {journal} {Phys.
  Rev. D}\ }\textbf {\bibinfo {volume} {97}},\ \bibinfo {pages} {095033}
  (\bibinfo {year} {2018})},\ \Eprint {http://arxiv.org/abs/1708.05389}
  {arXiv:1708.05389 [hep-ph]} \BibitemShut {NoStop}%
\bibitem [{\citenamefont {Izaguirre}\ \emph {et~al.}(2016)\citenamefont
  {Izaguirre}, \citenamefont {Krnjaic},\ and\ \citenamefont
  {Shuve}}]{Izaguirre:2015zva}%
  \BibitemOpen
  \bibfield  {author} {\bibinfo {author} {\bibfnamefont {E.}~\bibnamefont
  {Izaguirre}}, \bibinfo {author} {\bibfnamefont {G.}~\bibnamefont {Krnjaic}},
  \ and\ \bibinfo {author} {\bibfnamefont {B.}~\bibnamefont {Shuve}},\ }\href
  {\doibase 10.1103/PhysRevD.93.063523} {\bibfield  {journal} {\bibinfo
  {journal} {Phys. Rev. D}\ }\textbf {\bibinfo {volume} {93}},\ \bibinfo
  {pages} {063523} (\bibinfo {year} {2016})},\ \Eprint
  {http://arxiv.org/abs/1508.03050} {arXiv:1508.03050 [hep-ph]} \BibitemShut
  {NoStop}%
\bibitem [{\citenamefont {Liu}\ \emph {et~al.}(2019)\citenamefont {Liu},
  \citenamefont {Liu},\ and\ \citenamefont {Wang}}]{Liu:2018wte}%
  \BibitemOpen
  \bibfield  {author} {\bibinfo {author} {\bibfnamefont {J.}~\bibnamefont
  {Liu}}, \bibinfo {author} {\bibfnamefont {Z.}~\bibnamefont {Liu}}, \ and\
  \bibinfo {author} {\bibfnamefont {L.-T.}\ \bibnamefont {Wang}},\ }\href
  {\doibase 10.1103/PhysRevLett.122.131801} {\bibfield  {journal} {\bibinfo
  {journal} {Phys. Rev. Lett.}\ }\textbf {\bibinfo {volume} {122}},\ \bibinfo
  {pages} {131801} (\bibinfo {year} {2019})},\ \Eprint
  {http://arxiv.org/abs/1805.05957} {arXiv:1805.05957 [hep-ph]} \BibitemShut
  {NoStop}%
\bibitem [{\citenamefont {Aaij}\ \emph {et~al.}(2018)\citenamefont {Aaij} \emph
  {et~al.}}]{LHCb:2017trq}%
  \BibitemOpen
  \bibfield  {author} {\bibinfo {author} {\bibfnamefont {R.}~\bibnamefont
  {Aaij}} \emph {et~al.} (\bibinfo {collaboration} {LHCb}),\ }\href {\doibase
  10.1103/PhysRevLett.120.061801} {\bibfield  {journal} {\bibinfo  {journal}
  {Phys. Rev. Lett.}\ }\textbf {\bibinfo {volume} {120}},\ \bibinfo {pages}
  {061801} (\bibinfo {year} {2018})},\ \Eprint
  {http://arxiv.org/abs/1710.02867} {arXiv:1710.02867 [hep-ex]} \BibitemShut
  {NoStop}%
\end{thebibliography}%

\end{document}